\RequirePackage{ifpdf}
\documentclass[a4paper,11pt]{JHEP3}

\pdfoutput=1
\usepackage{epsfig}
\usepackage{cite}
\usepackage{graphicx}
\usepackage{subfigure}
\usepackage{inputenc}
\usepackage{amsmath}
\usepackage{slashed}
\usepackage{cancel}
\usepackage{slashed}

\newcommand{\beq}{\begin{equation}}
\newcommand{\eeq}{\end{equation}}
\newcommand{\beqn}{\begin{eqnarray}}
\newcommand{\eeqn}{\end{eqnarray}}
\def\df{{\rm d}}
\newcommand{\bn}{{\bar n}}
\def\nslash{n\!\!\!\slash}

\newcommand{\mcdot}{\!\cdot\!}
\newcommand{\eq}[1]{Eq.~\eqref{eq:#1}}
\newcommand{\eqs}[2]{Eqs.~\eqref{eq:#1} and \eqref{eq:#2}}

\newcommand{\fullF}{\mathcal{F}}

\newcommand{\nPC}{$n$PC  }

\newcommand{\Vfull}{V}

\def\eps{\delta}

\def\nn{\nonumber\\}

\def\pslash{p\!\!\!\slash}
\def\ppslash{p^{\,\prime}\!\!\!\!\!\slash}

\title{A numerical formulation of resummation in effective field theory}

\author{Christian W.~Bauer$^{a}$ and Pier Francesco Monni$^b$\\
   $^a$Ernest Orlando Lawrence Berkeley National Laboratory, University of California, Berkeley, CA 94720, USA\\
   $^b$CERN, Theoretical Physics Department, CH-1211 Geneva 23, Switzerland\\   
      E-mail: \email {cwbauer@lbl.gov}, \email{pier.monni@cern.ch}
        }

\received{\today}               
\accepted{\today}               

\preprint{CERN-TH-2018-055}

\abstract{In this article we show how the resummation of infrared and collinear logarithms within Soft-Collinear Effective Theory (SCET) can be formulated in a way that makes it suitable for a Monte-Carlo implementation. This is done by applying the techniques developed for automated resummation using the branching formalism, which have resulted in the general resummation approach {\tt CAESAR/ARES}. This work builds a connection between the two resummation approaches, and paves the way to automated resummation in SCET. As a case study we consider the resummation of the thrust distribution in electron-positron collisions at next-to-leading logarithm (NLL). However, the results presented here are easily generalizable to more complicated observables and processes as well as to higher orders in the logarithmic accuracy.}

\begin{document} 

\section{Introduction}
\label{sec:introduction}

A well known fact of perturbation theory is the presence of logarithmic terms, sensitive to the ratio of scales in the problem, whose power grows with the perturbative order. For most processes of interest at high energy colliders, two powers of such logarithms ($L$) arise for every power of the strong coupling constant, and the numerical size of these logarithms can be of order $L \sim 1 / \alpha_s$. This makes fixed order (FO) perturbation theory for such processes ill behaved, requiring a rearrangement of the perturbative expansion, in which these large logarithms are resummed to all orders in perturbation theory. Instead of simply counting the powers of the strong coupling constant, where N$^n$LO refers to a calculation satisfying $ \sigma^{\rm exact}/\sigma^{{\rm N}^n{\rm LO}}  -1\sim \alpha_s^{n+1}$, one instead performs a logarithmic resummation, in which N$^n$LL implies that $\ln[ \sigma^{\rm exact}/\sigma^{{\rm N}^n{\rm LL}} ] \sim \alpha_s^m L^{m-n}$ for any $m \ge n$. As long as the ratio of scales in the logarithm is large ($L \gg 1$), this reorganization of perturbation theory provides a sensible expansion.

In this paper we will study the 2-jet cross section at lepton colliders (equivalent considerations apply to the 0-jet cross sections at hadron colliders), that we denote as
\begin{align}
\label{eq:SigmaDef}
\Sigma(v) = \frac{1}{\sigma_B}\sum_n \int \! \df \Phi_n \, \frac{\df \sigma}{\df \Phi_n} \Theta\left[V(\Phi_n) < v\right]
\,,\end{align}
where $\sigma_B$ denotes the Born cross section and $V(\Phi_n)$ an observable that goes to zero in the Born kinematics. The radiation phase space $\Phi_n$ has the property that in the limit $v \to 0$ each strongly interacting particle is either infinitely soft or it is collinear to either of the two Born legs (either the directions of the 2-jets in lepton colliders or the beam directions for hadron colliders). The resummed expression commonly takes the form
\begin{align}
\label{eq:SigmaGeneralForm}
\Sigma(v) = \sigma_B \left[1 + \alpha_s \, C^{(1)} +  \alpha_s^2 \, C^{(2)} + \ldots\right] \exp\left[ L g_1(\alpha_s L) + g_2(\alpha_s L) + \ldots\right]
\,.
\end{align}
Resummation to order N$^k$LL requires knowledge of the functions $g_{m}(\alpha_s L)$ with $m \le k+1$ and the coefficients $C^{(m)}$ with $m \le k-1$. So for example, for NLL resummation one needs $g_1(\alpha_s L)$ and $g_2(\alpha_s L)$, but only the leading order coefficient $C^{(0)} = 1$. 

There are two main approaches to calculate the resummed expressions. The first is based on deriving a factorization theorem for the specific cross section under consideration, and then using evolution equations to resum the logarithms of the various ingredients of the factorization theorem. This approach was started by Collins, Soper and Sterman~\cite{Collins:1984kg}, where the transverse momentum distribution of the vector boson in $W$ and $Z$ production was studied. The development of soft-collinear effective theory (SCET)~\cite{Bauer:2000ew,Bauer:2000yr,Bauer:2001ct,Bauer:2001yt} has formulated this approach in the framework of effective field theories. In SCET, the interactions between particles are already factorized at the level of the Lagrangian~\cite{Bauer:2001ct}. As long as the measurement function can be shown to factorize~\cite{Bauer:2008dt,Bauer:2008jx}, the relevant interactions can be separated as arising from collinear and soft particles, which do not interact with one another directly. Once a factorization theorem for a given observable has been derived, the important property of SCET is that in dimensional regularization each element of the factorization theorem only depends on a single scale through the ratio to the renormalization scale~\cite{Bauer:2000ew,Bauer:2000yr}. This means that all logarithmic terms can be obtained from the renormalization (RG) group evolution of each element in the factorization theorem, resulting in analytic results for resummed cross sections.

An alternate approach to resummation is based on the branching formalism\cite{Catani:1990rr,Catani:1991kz}, which is built on the factorization properties of the QCD squared amplitudes. It relies on slicing the radiation phase space by means of a resolution scale $q_0$ and then describing the radiation of particles above and below that scale separately. At lowest order this approach is the same as what is used in a parton shower, but by not requiring to produce fully exclusive final states which satisfy momentum conservation, one can systematically improve the branching formalism to any logarithmic accuracy desired. The logarithmic dependence on the resolution scale $q_0$ cancels for infrared and collinear (IRC) safe observables, leaving only a power suppressed dependence on $q_0$, which allows one to take the limit for $q_0\to 0$. The key difference between the branching formalism and the resummation from factorization is that in the factorization approach the resummation is obtained by solving a set of differential equations (RG equations in the case of SCET), while in the branching formalism resummation is usually performed through an all order calculation in perturbation theory using a Monte-Carlo (MC) algorithm. For sufficiently simple observables one can rewrite the branching formalism in terms of differential equations, reproducing the results from the factorization approach.

Each of these two approaches has its advantages and disadvantages. The factorization approach has the advantage that to perform resummation only requires to obtain the desired differential RG equations that, in particular using SCET, can be obtained by computing well defined loop diagrams. On the other hand, the formalism only works for observables for which a factorization theorem is known. While these factorization theorems can easily be obtained for the simplest observables, for more complicated observables the factorization formulae are not known. In fact, the most difficult part to obtain resummation in SCET is often the derivation of the appropriate factorization theorem. 
Since the branching formalism does not rely on a specific factorization theorem, it can be applied to a rather wide class of observables. In fact, one can show that for any continuously global, recursively infrared and collinear (rIRC) safe observable~\cite{Banfi:2004yd}, resummed results can be obtained using an appropriate MC algorithm. The downside of the branching formalism is that going to higher logarithmic accuracy requires a careful definition of the necessary squared amplitudes and phase space constraints, making extensions to arbitrarily high accuracy less systematic. Besides the formulation of resummation, the two approaches differ in a number of important aspects. For instance, while in the branching formalism the uncertainties associated with higher-order terms are estimated by varying the renormalization (and factorization in the hadronic case) and resummation scales, in SCET each of the ingredients of the factorization theorem has a characteristic scale that can be separately varied (commonly done through profile functions~\cite{Abbate:2010xh}) to estimate the perturbative uncertainty. Another important difference is in the way the two methods are sensitive to non-perturbative effects. While in the branching formalism this sensitivity comes from the scale at which the running coupling is evaluated, in a factorized approach one has an operator definition of the factorization ingredients in the non-perturbative regime.

As was shown in refs~\cite{Banfi:2001bz,Banfi:2004yd,Banfi:2014sua}, one can reduce the requirements of the accuracy of the ingredients in the MC algorithm by one order if one has analytic knowledge of the resummation of any other observable that has the same double logarithmic structure as the observable one is interested in resumming. In other words, if one wants to resum $\Sigma(v)$ for a given observable $v$ to order N$^n$LL, and one knows the resummmation analytically for $\Sigma(v_s)$, where $v_s$ is a different (simpler) observable that has the same double logarithmic structure as $\Sigma(v)$, then the ratio $\Sigma(v) / \Sigma(v_s)$ can be computed using a MC that only requires ingredients at N$^{n-1}$LL accuracy. This was used in~\cite{Banfi:2004yd} to construct a completely generic method capable of computing any IRC safe observable to NLL accuracy. The simplified observable was chosen such that the branching formalism could be solved analytically. This approach was later reformulated to NNLL for generic rIRC safe observables in~\cite{Banfi:2014sua,Banfi:2016zlc} and to even higher orders for specific observables~\cite{Bizon:2017rah}.

Both approaches can be applied to obtain high-order resummations for a multitude of final-state observables for scattering processes in $e^+e^-$~\cite{Becher:2008cf,Chien:2010kc,Becher:2012qc,Hoang:2014wka,Becher:2015lmy,Banfi:2014sua,Frye:2016aiz,Banfi:2016zlc,Moult:2018jzp,Baron:2018nfz} as well as in hadronic collisions~\cite{Bauer:2002nz,Banfi:2004nk,Bozzi:2005wk,Becher:2010tm,Stewart:2010pd, Banfi:2011dx,Berger:2010xi,Jouttenus:2011wh ,Becher:2012qa,Zhu:2012ts,Banfi:2012jm,Becher:2013xia,Stewart:2013faa,Procura:2014cba,Li:2016ctv,Monni:2016ktx,Bizon:2017rah}.

By combining the two approaches with one another one keeps the advantages of each, while removing the main obstacle. If one can find for any observable $v$ a simplified observable $v_s$ with a simple factorization formula, one can use SCET to obtain the analytical resummation of this simplified observable, while using the branching formalism to relate this analytical result to the resummed result for the more complicated observable, for which a factorization formula might be difficult or impossible to obtain. This is in spirit very similar to the {\tt CAESAR/ARES} approach, but instead of finding a simplified observable for which the branching formalism can be solved analytically, one chooses the simplified observable such that a factorization theorem is easily derived and its resummation performed in SCET. In fact, there is a large body of observables for which high logarithmic accuracy is known in SCET (see references above). By combining this with the branching formalism allows one to obtain results for all related observables to the same level of accuracy.

It is this combination of SCET results with the branching formalism that we will address in this paper. A major part of this discussion will explain how to deal with UV divergent phase space regions that are crucial in the SCET approach (since it is the UV divergences from these regions that give rise to the anomalous dimensions leading to the RG equations in SCET), but can not be present in a MC approach which has to integrate over physical regions of phase space. We will explain this combination using the thrust distribution (using $\tau = 1 - T$) as an explicit example. Although the ingredients for a N$^3$LL resummation are currently known~\cite{Becher:2008cf,Abbate:2010xh,Becher:2006mr,Kelley:2011ng,Monni:2011gb} (with the sole exception of the four-loop cusp anomalous dimension), in this paper we limit ourselves to NLL for the sake of simplicity. However our results are easily generalizable to more complicated observables of interest as well as to the computation of higher-order corrections. We leave this to a forthcoming publication~\cite{futureWork}.

This paper is organized as follows: We review the resummation using the branching formalism in Section \ref{sec:reviewCAESAR} and the SCET approach to resummation in Section~\ref{sec:reviewSCET}. The main part of the paper is contained in Section~\ref{sec:automatedSCET}, where we discuss how to combine the two approaches to obtain a numerical approach to resummation in SCET. Conclusions and an outlook to future work is presented in Section~\ref{sec:conclusions}.

\section{Review of QCD resummation in the {\tt CAESAR/ARES} framework}
\label{sec:reviewCAESAR}

In this section we briefly review how resummation is carried out in the approach of refs.~\cite{Banfi:2004yd,Banfi:2014sua}. We begin by discussing the basic setup of the formalism for a general observable $v$ and to arbitrary order in the resummation, but then restrict ourselves to the specific case of the thrust distribution when deriving the NLL result in more detail. We consider observables that vanish in the 2-jet limit, and when considering the thrust distribution we use $\tau\equiv 1-T$. 

At Born level, the final state consists of two back-to-back particles  along 
the thrust axis with center of mass energy $Q$. 
Beyond Born level, further radiation is present and the final state consists in general of $n$
secondary emissions, $k_1,\dots,k_n$, and of the primary quark and
antiquark which recoil against these additional emissions.
We denote the value of the observable by $\Vfull(\Phi_B; k_1,\dots,k_n)$, where the Born phase space $\Phi_B$ contains the dependence on the two Born momenta.

In order to single out the dependence on the Born phase space, we write
\begin{equation}
\Sigma(v) =  \frac{1}{\sigma_{ B}} \int \!\df\Phi_B  \, \Sigma(\Phi_B; v)
\,,
\end{equation}
and we will work with the expression $\Sigma(\Phi_B; v)$ for most of this paper. This means that the $\Phi_B$ phase space integral needs to be performed at the end, and the final result needs to be divided by the Born cross section $\sigma_{B}$.

In the infrared and collinear limit, $\Sigma(\Phi_B; v)$ receives contributions from either virtual, or soft and/or collinear real corrections.  In the following, we denote by $ {\cal V}(\Phi_B)$ the quark form factor at all orders (see e.g.~\cite{Dixon:2008gr}). Therefore we can write
\begin{equation}
  \label{eq:Sigma-2}
  \Sigma(\Phi_B; v) = {\cal V}(\Phi_B) \sum_{n=0}^{\infty}
  \int\!\prod_{i=1}^n [\df k_i]
  |M(\Phi_B; k_1,\dots ,k_n)|^2\,\Theta\left[V(\Phi_B; k_1,\dots,k_n)<v\right]\,,
\end{equation}
where $M$ is the matrix element for $n$ real emissions (the case with $n=0$ reduces to the Born matrix element), and $[\df k_i]$ denotes the phase space for the emission $k_i$. Each matrix element in Eq.~\eqref{eq:Sigma-2} receives higher-order virtual corrections, while the number of real emissions is fixed by the index of the sum. The $\Theta$ function represents the measurement function for the observable under consideration.

\subsection{The simplified observable and the transfer function}
The general strategy of the {\tt CAESAR/ARES} approach is to write the cross section for a rIRC safe observable $v$ into the cross section of a simpler observable $v_s$ which 
has the same logarithmic structure as $v$ at
lowest order\footnote{This last requirement is not, strictly
speaking, necessary, although it will lead to important
simplifications in formulating a Monte Carlo solution.}, and a transfer function that accounts for the difference between the two observables $v$ and $v_s$. The latter is formulated in such a way that it can be evaluated efficiently using Monte Carlo methods. 

As we will discuss in a little while, a good choice for such a simple observable is $v_s=v_{\max}$, where the 
simple observable is defined by taking a suitably defined maximum of the observable calculated 
for independent emissions. We use the following notation
\begin{align}
\Sigma_{\rm max}(\Phi_B; v) \equiv \Sigma(\Phi_B; v_{\max})
\end{align}
from now on. A detailed definition of this observable will follow shortly.

Using some trivial manipulations, one can write
\begin{align}
\label{eq:Sigma-0}
\Sigma(\Phi_B; v) &= \Sigma_{\rm max}(\Phi_B; v) \frac{\Sigma_{\rm max}(\Phi_B; \eps v)}{\Sigma_{\rm
  max}(\Phi_B; v)} \frac{\Sigma(\Phi_B; v)}{\Sigma_{\rm max}(\Phi_B; \eps v)} 
\nn
&\equiv \Sigma_{\rm max}(\Phi_B; v) \, \fullF(\Phi_B; v)
\,,
\end{align}
where we introduced the small positive parameter $\eps\ll 1$ that is independent of the observable's value $v$.
An important comment at this stage is in order. Eq.~\eqref{eq:Sigma-0} is strictly valid only for observables that admit a resummed cross section of the type~\eqref{eq:SigmaGeneralForm}. This is not the case for observables which can vanish also in the presence of resolved emissions due to kinematic cancellations (for instance for the transverse momentum of a color singlet in $pp$ collisions). In this case Eq.~\eqref{eq:Sigma-0} takes the form of a convolution between $\Sigma_{\rm max}$ and $\fullF$. In this article we limit ourselves to observables that behave like Eq.~\eqref{eq:SigmaGeneralForm} in the $v\to 0$ limit, and leave the study of the above observables for a future publication.

The product of ratios gives the relation between the cross section of the desired observable
$\Sigma(\Phi_B; v)$ and the cross section of the simplified observable $\Sigma_{\max}(\Phi_B; v)$, and it accounts for
the exact behavior of the observable in the presence of
radiation. For this reason it is normally referred to as
multiple-emissions functions. For the sake of brevity, we have dubbed it
{\it transfer} function
\begin{equation}
\label{eq:fullFDef}
\fullF(\Phi_B; v) \equiv \frac{\Sigma_{\rm max}(\Phi_B; \eps v)}{\Sigma_{\rm
  max}(\Phi_B; v)} \frac{\Sigma(\Phi_B; v)}{\Sigma_{\rm max}(\Phi_B; \eps v)}
  \,.
\end{equation}

An important property of this transfer function is that it is IRC and UV finite, and as long as $\Sigma_{\max}(\Phi_B; v)$ has the same LL structure as $\Sigma(\Phi_B; v)$ (as we are assuming), its contribution starts at NLL (as will be shown later). The small parameter $\eps$ was introduced in order to allow for the transfer function to be easily calculable in a MC framework, as we will discuss below. The basic idea is that in the second ratio, $\Sigma(\Phi_B; v) / \Sigma_{\rm max}(\Phi_B; \eps v)$, the denominator removes the unresolved emissions with $V < \delta v$, such that this ratio is IRC finite. The resulting dependence on the resolution parameter $\delta$ is cancelled against the first ratio, which can be calculated analytically once $\Sigma_{\rm  max}$ is known.

The first step in computing a resummed expression for $\Sigma(\Phi_B; v)$ is to build an explicit logarithmic
counting for the squared matrix elements. 
To achieve this, one introduces the $n$-particle correlated ($n$PC) squared matrix elements
$|\tilde{M}(k_1, \dots,k_n)|^2$, which are defined recursively as
\begin{align}
|\tilde{M}(k_a)|^2 &=\frac{|M(\Phi_B; k_a)|^2}{|M(\Phi_B)|^2} \equiv |M(k_a)|^2\,,\\
|\tilde{M}(k_a,k_b)|^2 &= \frac{|M(\Phi_B; k_a,k_b)|^2}{|M(\Phi_B)|^2}-|M(k_a)|^2|M(k_b)|^2\,,\nn
|\tilde{M}(k_a,k_b,k_c)|^2 &= \frac{|M(\Phi_B; k_a,k_b,k_c)|^2}{|M(\Phi_B)|^2}-|M(k_a)|^2|M(k_b)|^2|M(k_c)|^2\nn
&~-|\tilde{M}(k_a,k_b)|^2|M(k_c)|^2-|\tilde{M}(k_a,k_c)|^2|M(k_b)|^2-|\tilde{M}(k_b,k_c)|^2|M(k_a)|^2
\nn
& \ldots
\notag
\end{align}
These denote the parts of the squared amplitudes with $n$ real emissions that can not be obtained 
by multiplying together squared amplitudes with less than $n$ real emissions, and therefore represent
the fully correlated part. With these definitions, the renormalized squared
amplitude for $n$ real emissions can be
 decomposed as
\begin{align}
\label{eq:nPC}
&\frac{|M(\Phi_B; k_1,\dots ,k_n)|^2}{|M_B(\Phi_B)|^2} = 
\prod_{{\substack{i=1\\\phantom{x}}}}^{n} \left|M(k_i)\right|^2
+\sum_{a > b}\left|\tilde{M}(k_a, k_b)\right|^2\prod_{\substack{i=1\\ i\neq a,b}}^{n} \left|M(k_i)\right|^2
\nn
& \qquad
+ \sum_{a > b > c}\left|\tilde{M}(k_a, k_b,k_c)\right|^2\prod_{\substack{i=1\\ i\neq a,b,c}}^{n} \left|M(k_i)\right|^2
\nn
& \qquad
+\sum_{a > b}\sum_{\substack{ c > d\\ c,d\neq a,b}}\left|\tilde{M}(k_a, k_b)\right|^2 \left|\tilde{M}(k_c, k_d)\right|^2\prod_{\substack{i=1\\ i\neq a,b,c,d}}^{n} \left|M(k_i)\right|^2
+ \dots 
\end{align}
Each of the correlated squared amplitudes admits a perturbative
expansion
\begin{equation}
\label{eq:nPC-def}
|\tilde{M}(k_a, \dots,k_n)|^2~\equiv~ \sum_{j=0}^{\infty}\left(\frac{\alpha_s}{2\pi}\right)^{n+j}n\mbox{PC}^{(j)}(k_a, \dots,k_n)
\,,
\end{equation}
where the index $j$ denotes the order of virtual corrections to the squared amplitude with $n$ real emissions.
The rIRC safety of the observables considered here guarantees a hierarchy between the different blocks in the decomposition~\eqref{eq:nPC}, in the sense that correlated blocks with $n$ particles generally start contributing at one logarithmic order higher than correlated blocks with $n-1$ particles~\cite{Banfi:2004yd,Banfi:2014sua}. 

Having introduced the \nPC decomposition of the squared matrix elements allows one to precisely define the simplified observable $V_{\max}$. It is defined to be the maximum value of the full observable $V$ calculated on the sum of momenta in each correlated block. In equations, this becomes
\begin{equation}
\label{eq:Sigma-2a}
\Sigma_{\max}(\Phi_B; v) =  {\cal V}(\Phi_B) \, \sum_{n=0}^{\infty}
\int\!\prod_{i=1}^n [\df k_i] \, 
|M(\Phi_B; k_1,\dots ,k_n)|^2\,\Theta\left[V_{\max}(\Phi_B; k_1,\dots,k_n)<v\right]\,,
\end{equation}
where $V_{\max}(\Phi_B; k_1,\dots,k_n)$ is defined through its action on the $n$-particle correlated blocks as
\begin{align}
\label{eq:VMaxDef}
&\int\! \prod_{i=1}^n [\df k_i] \, \left|\tilde M(k_1, \ldots, k_{m_1}) \right|^2 \ldots \left|\tilde M(k_{m_k+1}, \ldots, k_{n}) \right|^2 \,\Theta\left[V_{\max}(\Phi_B; k_1,\dots,k_n)<v\right]
\\
& \qquad = \int\! \prod_{i=1}^n [\df k_i] \int \! [\df q_1] \delta(q_1 - k_1 - \ldots - k_{m_1})\ldots  \int\![\df q_k] \delta(q_k - k_{m_k+1} - \ldots - k_{n})
\nn
& \qquad \times \left|\tilde M(k_1, \ldots, k_{m_1}) \right|^2 \ldots \left|\tilde M(k_{m_k+1}, \ldots, k_{n}) \right|^2 
\Theta\left[\max\{\tilde V(\Phi_B; q_1), \dots, \tilde V(\Phi_B; q_k)\}<v\right]
\,.
\notag
\end{align}
where $\tilde V(\Phi_B; q)$ denotes the functional form of the observable on a single emission. It is important that the content of each correlated block is treated inclusively when computing the relative $\tilde V$ so that the above definition is collinear safe and can be extended at all orders. It is obvious that, in general, the cross section $\Sigma_{\rm max}$ has no physical meaning, but it simply defines one ingredient for our resummation approach.

It is worth stressing that one has some freedom in choosing the form of $\tilde V$. Conceptually, the simplest choice is to set $\tilde V = V$ evaluated on the inclusive content of each block. In general, however, the only important feature is that it shares the same leading logarithms with the observable we are ultimately interested in. It is therefore very useful to define the simple observable such that the corresponding $\Sigma_{\rm max}$ can be used for a whole class of more complicated observables. This can be achieved, for instance, by using the soft-collinear approximation of the full observable $\tilde V = V_{\rm sc}$ instead of its full form $V$. This indeed, besides simplifying further the computation of $\Sigma_{\rm max}$, guarantees that this ingredient can be directly used for the resummation of all observables that share the same soft-collinear limit for a single emission, which defines a much broader class than the first definition given above. For the sake of simplicity, however, we avoid this technical complication in the rest of this article, and refer the interested reader to refs.~\cite{Banfi:2004yd,Banfi:2014sua} for more details. 

We now continue with the derivation of the master formula. Using \eq{Sigma-2a} together with \eqs{nPC}{VMaxDef}, $\Sigma_{\max}(\Phi_B; v)$ can be written in terms of the nPC$^{(j)}$ blocks. Since the observable acts separately on each \nPC block, the expression in \eq{Sigma-2a} can easily be shown to exponentiate. To perform a N$^k$LL resummation for global, rIRC-safe observables, one needs to include $n$PC$^{(j)}$ blocks with $n+j \leq k+1$, but additional simplifications might be made on each $n$PC$^{(j)}$ block. For example, to LL accuracy, only the $1$PC$^{(0)}$ block is required, and one only needs to keep the soft-collinear limit of it. Thus, one obtains
\begin{align}
  \label{eq:SigmaMaxLL}
\Sigma^{\rm LL}_{\max}(\Phi_B; v) &= |M_B(\Phi_B)|^2{\cal V}(\Phi_B) \frac{1}{n!}
\prod_{i=1}^n \left\{ \int
[\df k_i] |M^{(0)}_{\rm sc}(k_i)|^2\Theta\left[\tilde V(\Phi_B; k_i)<v\right]\right\}
\nn
&=  |M_B(\Phi_B)|^2 \, e^{-R_{\rm LL}(\Phi_B; v)}
\,,
\end{align}
where the $1/n!$ prefactor accounts for $n$ identical gluons in the final state.
The LL radiator function $R_{\rm LL}(\Phi_B; v)$ is the combination of the virtual and real contribution which at this order simply reads
\begin{align}
\label{eq:RLLDef}
R_{\rm LL}(\Phi_B; v) = \int\!
[\df k] |M^{(0)}_{\rm sc}(k)|^2\,\Theta\left[\tilde V(\Phi_B; k)>v\right]
\,.
\end{align}
The definition of $V_{\max}$ ensures the exponentiation at higher orders as well, such that one can always write
\begin{align}
\Sigma_{\max}(\Phi_B; v)  &=  |M_B(\Phi_B)|^2 \, \Sigma_{\max}^{0} e^{-R(\Phi_B; v)}
\end{align}
where $R(\Phi_B; v)$ is called the radiator function, and $\Sigma_{\max}^{0}$ denote constant terms. $\Sigma_{\max}^{0}$ differs from one starting at NNLL order.

Using the expression for $\Sigma_{\max}(\Phi_B; v)$ obtained just above, the first ratio in \eq{fullFDef} can easily be computed, and we give the explicit expression when deriving the results at NLL accuracy below. 

To compute the second ratio in \eq{fullFDef} we need to carefully define the notion of resolved (unresolved) momenta, by demanding that the value $V_{\max}$ evaluated on the set of momenta is above (below) a resolution scale $\eps v$
\begin{align}
\overline \Theta_{\eps v}\left[\{k_i\}\right] &\equiv \Theta\left[V_{\max}(\Phi_B;\{k_i\}) > \eps v\right]
\nn
\Theta_{\eps v}\left[\{k_i\}\right] &\equiv \Theta\left[V_{\max}(\Phi_B;\{k_i\}) < \eps v\right]
\,.
\end{align}
A key point now is to notice that for $\eps \to 0$ one can neglect
the unresolved real emissions in the observable measurement function
as they are much softer and/or more collinear than any other resolved
emission in the final state. rIRC safety then guarantees that
\begin{align}
& \Theta\left[V(\Phi_B; k_1,\dots,k_n)<v\right] \, \Theta_{\eps v}\left[ \{k_{1}, \ldots , k_{l}\}\right] 
\nn
&\qquad 
=  \Theta\left[V(\Phi_B; k_{l+1},\dots,k_n)<v\right] \, \Theta_{\eps v}\left[ \{k_{1}, \ldots , k_{l}\}\right] + v\,\eps ^p
\,,
\end{align}
where $p$ is a positive and real constant, independent of $v$.
This allows one to split the total momentum $q$ of each \nPC block in $\Sigma(\Phi_B; v)$ into a resolved and unresolved component (depending on whether the value of $\tilde V(\Phi_B;q)$ is greater or less than $\eps v$).\footnote{A second comment is in order. Once again, for observables that can vanish because of kinematic cancellations (a primary example being the transverse momentum of a color singlet in $pp$ collisions), our choice of the simple observable can have issues when the above cancellations occur. A more appropriate, and more general, prescription would be to use $\eps V(q_1)$ as a resolution scale, where $q_1$ is the total four momentum of the {\it hardest} correlated block. In this case Eq.~\eqref{eq:Sigma-0} takes the form of a convolution as discussed in Refs.~\cite{Banfi:2001bz,Monni:2016ktx}.  We will however proceed with the initial choice in the rest of this article for the sake of simplicity, since all of the other considerations made here are fully general.} \\
Thus, for each \nPC block we use
\begin{align}
\int \! [\df k] \, |M(k)|^2 &= \int [\df k] \, |M(k)|^2\,\left[ \Theta_{\eps v}\left( k\right) + \overline \Theta_{\eps v}\left( k\right)\right]
\nn
& \equiv \int^{\eps v} \!\!  [\df k] \, |M(k)|^2 + \int_{\eps v} [\df k] \, |M(k)|^2\,,
\nn
\int
\! [\df k_1][\df k_2] \, |\tilde{M}(k_1,k_2)|^2 &= \int
\! [\df k_1][\df k_2] \, |\tilde{M}(k_1,k_2)|^2 \,\left[\Theta_{\eps v}\left( k_1, k_2\right) + \overline \Theta_{\eps v}\left(  k_1, k_2\right)\right] 
  \nn
&\equiv \int^{\eps v}
  \! [\df k_1][\df k_2] \, |\tilde{M}(k_1,k_2)|^2+\int_{\eps v}
  \!\!  [\df k_1][\df k_2] \, |\tilde{M}(k_1,k_2)|^2,
\end{align}
and so on. Up to power corrections in the small parameter $\eps$, this allows us to separate $\Sigma(\Phi_B; v)$ into a resolved component (where all momenta are resolved) and an unresolved component,
\begin{align}
\Sigma(\Phi_B; v) & = |M_B(\Phi_B)|^2 {\cal V}(\Phi_B)
\left[\sum_{n=0}^{\infty}
\int^{\eps v}\!\prod_{i=1}^n [\df k_i]
\frac{|M(\Phi_B; k_1,\dots ,k_n)|^2}{|M_B(\Phi_B)|^2} \right]\notag
\\
& \qquad \times \left[\sum_{n=0}^{\infty}
\int_{\eps v}\!\prod_{i=1}^n [\df k_i]
\frac{|M(\Phi_B; k_1,\dots ,k_n)|^2}{|M_B(\Phi_B)|^2}\,\Theta\left[V(\Phi_B; k_1,\dots,k_n)<v\right]\right]
+ {\cal O}(v \, \eps^p)
\nn
&= \Sigma_{\max}(\Phi_B; \eps v) \left[\sum_{n=0}^{\infty}
\int_{\eps v}\!\prod_{i=1}^n [\df k_i]
\frac{|M(\Phi_B; k_1,\dots ,k_n)|^2}{|M_B(\Phi_B)|^2}\,\Theta\left[V(\Phi_B; k_1,\dots,k_n)<v\right]\right]
\nn
& \qquad \qquad + {\cal O}(v\,\eps^p)
\,.
\end{align}
One therefore finds for the second ratio in \eq{fullFDef}
\begin{align}
\label{eq:ratio2}
\frac{\Sigma(\Phi_B; v)}{\Sigma_{\rm max}(\Phi_B; \eps v)} &=\sum_{n=0}^{\infty}
\int_{\eps v}\prod_{i=1}^n [\df k_i]
\frac{|M(\Phi_B; k_1,\dots ,k_n)|^2}{|M_B(\Phi_B)|^2}\,\Theta\left[V(\Phi_B;  k_1,\dots,k_n)<v\right]
+ {\cal O}(v\,\eps^p)
\,.
\end{align}

Note that the above discussion holds to any logarithmic accuracy. To go to a given order in the resummation of $\Sigma(v)$ or $\Sigma_{\rm max}(v)$ one needs to rewrite the full matrix element in terms of the nPC$^{(j)}$ blocks, and only keep the blocks that are relevant at the desired logarithmic order. 

For the two ratios required in the transfer function \eq{fullFDef}, the argument of the numerator and denominator scale with the observable $v$. This implies that to compute the ratio to a given logarithmic accuracy, one needs the numerator and denominator at one logarithmic order lower~\cite{Banfi:2004yd,Banfi:2014sua}. 
To understand this fact better, let us consider the first ratio in Eq.~\eqref{eq:fullFDef} as an example. At NLL order, we can write $\Sigma_{\rm max}(v)=\exp[L_v g_1(\alpha_sL_v) + g_2(\alpha_s L_v)]$, where $L_x=\ln(1/x)$. We find
\begin{align}
\label{eq:onlyLLArgument}
\frac{\Sigma_{\max}(\eps v)}{\Sigma_{\max}(v)} &= \exp\left\{L_{\eps v} g_1(\alpha_s L_{\eps v}) - L_v g_1(\alpha_s L_v) + g_2(\alpha_s L_{\eps v}) - g_2(\alpha_s L_{v})\right\}
\nn
&= \exp \left\{ L_\eps \left[ g_1(\alpha_s L_v) + \alpha_s L_v g_1'(\alpha_s L_v) \right] + \ldots\right\},
\end{align} 
where we have dropped all terms contributing beyond NLL. One can clearly see that the result depends only on the LL function $g_1(\alpha_s L_v)$, such that each term is only required to LL accuracy.

Furthermore, one can perform additional kinematical expansions to simplify the expressions of the nPC$^{(j)}$ blocks, and we decompose each block $n{\rm PC}^{(j)}$ by singling out its most singular (hence leading) term $[ n{\rm PC}^{(j)} ]_{\rm sc}$, that is obtained by taking the soft and collinear limit of all emissions, i.e.
\begin{equation}
n{\rm PC}^{(j)}=[ n{\rm PC}^{(j)} ]_{\rm sc} + [ n{\rm PC}^{(j)} ]_{\rm \cancel{sc}}.
\end{equation}
In summary, the ingredients required to a given order in logarithmic counting are summarized in Table~\ref{tab:LogCounting}. \begin{table}
\begin{center}
\begin{tabular}{|c|c|c|c|c|}
\hline
 & \multicolumn{2}{c |}{$\Sigma_{\rm max}(v)$} & \multicolumn{2}{c |}{$\fullF(v)$}
 \\\hline
 & $[ n{\rm PC}^{(j)} ]_{\rm sc}$ &  $[ n{\rm PC}^{(j)} ]_{\rm \cancel{sc}}$ & $[ n{\rm PC}^{(j)} ]_{\rm sc}$ &  $[ n{\rm PC}^{(j)} ]_{\rm \cancel{sc}}$
 \\\hline\hline
LL & $n+j \le 1$ & -- & --  & --
\\ \hline
NLL & $n+j \le 2$ & $n+j \le 1$ & $n+j \le 1$  & --
\\ \hline
NNLL & $n+j \le 3$ & $n+j \le 2$ & $n+j \le 2$  & $n+j \le 1$
\\ \hline\hline
N$^k$LL & $n+j \le k+1$ & $n+j \le k$ & $n+j \le k$  & $n+j \le k-1$
\\ \hline\end{tabular}
\end{center}
\caption{The order at which the various $n$PC$^{(j)}$ are required for the computation of $\Sigma_{\rm max}(v)$ and $\fullF(v)$. 
\label{tab:LogCounting}}
\end{table}
In the next section we perform the calculation at NLL for the thrust event shape.

%
\subsection{Resumming the thrust distribution to NLL accuracy}
\label{sec:simplifiedResummation}

In this section we compute all ingredients necessary to obtain $\Sigma(\Phi_B; \tau)$ for the thrust distribution to NLL accuracy, using \eq{Sigma-0}. The thrust distribution is an additive observable, which satisfies
\begin{align}
V(\Phi_B; k_1,\dots,k_n) = \sum_{i=1}^n \tau_i \,, \qquad {\rm with} \qquad \tau_i \equiv V(\Phi_B; k_i)
\,.
\end{align}

The first ingredient is the NLL resummation of the simplified observable. To NLL accuracy one obtains [using the obvious extension of the LL result given in \eq{SigmaMaxLL}]
\begin{align}
  \label{eq:SigmaMaxNLL}
\Sigma^{\rm NLL}_{\max}(\Phi_B; \tau)  = |M_B(\Phi_B)|^2 \, e^{-R_{\rm NLL}(\Phi_B; \tau)}
\,,
\end{align}
with
\begin{align}
R_{\rm NLL}(\Phi_B; \tau)&= \int\!
[\df k]\left[ |M^{(0)}(k)|^2 + |M_{\rm sc}^{(1)}(k)|^2 + \int [\df k_a][\df k_b] |\tilde M_{\rm sc}^{(0)}(k_a, k_b)|^2 \delta(k - k_a - k_b)\right]
\nn
& \qquad \qquad \times
\Theta\left[\tilde V(\Phi_B; k)>\tau\right]
\,,
\end{align}
where one keeps the full kinematical dependence in the tree level contribution of $|M(k_i)|^2$, but only the soft-collinear limit of the one-loop contribution to  $|M(k_i)|^2$ and of $|\tilde M^{(0)}(k_a, k_b)|^2$. 
One can evaluate the integrals involving $ |M_{\rm sc}^{(1)}(k)|^2$ and $|\tilde M_{\rm sc}^{(0)}(k_a, k_b)|^2$ in dimensional regularization, and neglecting NNLL corrections one finds 
\begin{align}
\label{eq:RNLL_w_K}
R_{\rm NLL}(\Phi_B; \tau)&= \int\!
[\df k]\left[ |M^{(0)}(k)|^2 + \frac{\alpha_s(k_t)}{2\pi} |M_{\rm sc}^{(0)}(k)|^2 K\right]
\Theta\left[\tilde V(\Phi_B; k)>\tau\right]
\,,
\end{align}
where
\begin{equation}
\label{eq:K}
K = \left(\frac{67}{18}-\frac{\pi^2}{6}\right)C_A - \frac{5}{9} n_f\,.
\end{equation}

For the computation of the transfer function one only needs to keep the 1PC$^{(0)}$ block in its soft-collinear limit.
Therefore, the first ratio in $\fullF^{\rm NLL}(\tau)$ can be written as
\begin{align}
\label{eq:Sigma_maxRatio_NLL}
\frac{\Sigma^{\rm LL}_{\max}(\Phi_B; \eps \tau)}{\Sigma^{\rm LL}_{\max}(\Phi_B; \tau)} &= e^{R_{\rm LL}(\Phi_B; \tau) - R_{\rm LL}(\Phi_B; \eps \tau)}\equiv\Delta_{\rm LL}(\Phi_B; \tau, \eps \tau)
\,,\end{align}
where $R_{\rm LL}(\Phi_B; v)$ was given in \eq{RLLDef}.

To compute the second ratio of the transfer function to NLL accuracy, one uses \eq{ratio2}, which leads to (up to power corrections in $\eps$)
\begin{align}
\label{eq:Sigma_resolved_NLL}
\frac{\Sigma^{\rm LL}(\Phi_B; \tau)}{\Sigma^{\rm LL}_{\rm max}(\Phi_B; \eps \tau)} &= \sum_{n=0}^{\infty}\frac{1}{n!}
\prod_{i=1}^n \left\{ \int_{\eps \tau}
[\df k_i] |M^{(0)}_{\rm sc}(k_i)|^2\right\}\,\Theta\left[\sum_i \tau_i  <\tau\right]
\,.
\end{align}
Combining \eqs{Sigma_maxRatio_NLL}{Sigma_resolved_NLL}, we obtain the final expression for the transfer function
\begin{align}
\label{eq:FNLL}
\fullF_{\rm NLL}(\Phi_B; \tau) =\Delta_{\rm LL}(\Phi_B; \tau, \eps \tau) \sum_{n=0}^{\infty}
  \left(\frac{1}{n!}\int_{\eps \tau} \prod_{{\substack{i=1\\\phantom{x}}}}^{n} [\df k_i]|M^{(0)}_{\rm sc}(k_i)|^2\right)\Theta\left[\sum_i \tau_i  <\tau\right]
\,.
\end{align}
Eqs.~\eqref{eq:SigmaMaxNLL}, \eqref{eq:RNLL_w_K}, \eqref{eq:Sigma_maxRatio_NLL} and \eqref{eq:FNLL} provide all the ingredients to calculate $\Sigma(v)$ to NLL accuracy. 

The transfer function in \eq{FNLL} can easily be computed using an MC approach. 
Using \eq{RLLDef} one can write (recall that we choose $\tilde V = V$)
\begin{align}
\label{eq:rp_explicit}
R'_{\rm LL}(\Phi_B; \tau) & \equiv \frac{\df  R_{\rm LL}(\Phi_B; \tau) }{\df \ln(1/\tau)}= \tau\int [\df k] |M^{(0)}_{\rm sc}(k)|^2 \delta(V(\Phi_B; k) - \tau)\notag\\
& = \int \frac{\df k_t}{k_t}\int_0^{\ln\frac{Q}{k_t}} \df\eta\, \frac{\df\phi}{2\pi}4 C_F \frac{\alpha_s(k_t)}{\pi}\delta\left[\ln (V(\Phi_B; k)) - \ln(\tau)\right]
\,.\end{align}
Trading the $1/n!$ in Eq.~\eqref{eq:FNLL} with an ordering in $v_i$, this allows us to write the transfer function in the form
\begin{align}
\label{eq:FNLL_PS}
\fullF_{\rm NLL}(\Phi_B; \tau) &= \Delta_{\rm LL}(\Phi_B; \tau, \eps \tau)\Bigg[1 + \int_{\eps \tau}^\tau \! \frac{\df \tau_1}{\tau_1} \, R'_{\rm LL}(\Phi_B; \tau_1)
\nn
& \qquad+ \int_{\eps \tau}^\tau \! \frac{\df \tau_1}{\tau_1} \, R'_{\rm LL}(\Phi_B; \tau_1)\int_{\eps \tau}^{\tau_1}\!  \frac{\df \tau_2}{\tau_2} \, R'_{\rm LL}(\Phi_B; \tau_2) + \ldots\Bigg]
\Theta\left[\sum_i \tau_i  <\tau\right]
\nn
&= \Bigg[\Delta_{\rm LL}(\Phi_B; \tau, \eps \tau) + \int_{\eps \tau}^\tau \! \frac{\df \tau_1}{\tau_1} \, \Delta_{\rm LL}(\Phi_B; \tau, \tau_1)R'_{\rm LL}(\Phi_B; \tau_1)\Delta_{\rm LL}(\Phi_B; \tau_1, \eps \tau) 
\nn
& \qquad \qquad+  \ldots\Bigg] \Theta\Big[\sum_i \tau_i  <\tau\Big]
\,,\end{align}
where to obtain the second identity we have used 
\begin{align}
\Delta_{\rm LL}(\Phi_B; \tau_1, \tau_2) = \Delta_{\rm LL}(\Phi_B; \tau_1, \tau') \Delta_{\rm LL}(\Phi_B; \tau', \tau_2)
\,.\end{align}
Since $\Delta_{\rm LL}(\Phi_B; \tau, \tau')$ and $R'_{\rm LL}(\Phi_B; \tau)$ satisfy
\begin{align}
\tau' \frac{\df}{\df \tau'} \, \Delta_{\rm LL}(\Phi_B; \tau, \tau') = R'_{\rm LL}(\Phi_B; \tau')\,\Delta_{\rm LL}(\Phi_B; \tau, \tau')
\,,\end{align}
this strongly resembles the standard parton shower evolution. It is therefore solved using the usual parton shower algorithm:
\begin{enumerate}
\item Start with $i=0$ and $\tau_0 = \tau$
\item Increase $i$ by one
\item Generate $\tau_i$ randomly according to\footnote{This is done by generating a random number $r$ and then solving $\Delta_{\rm LL}(\Phi_B; \tau_{i-1}, \tau_i) = \frac{\Delta_{\rm LL}(\Phi_B; \tau_{i-1}, \eps \tau)}{\Delta_{\rm LL}(\Phi_B; \tau_{i}, \delta \tau)} = r$ for $\tau_i$} $\Delta_{\rm LL}(\Phi_B; \tau_{i-1}, \tau_i) R'_{\rm LL}(\Phi_B; \tau_i)$
\item If $\tau_i < \eps \tau$ exit the algorithm, otherwise go back to step 2
\end{enumerate}
If the sum over all generated $\tau_i$ is less than $\tau$, accept the event, otherwise reject it. The value of $\fullF(\tau)$ is equal to the fraction of the accepted events.

\subsection{Neglecting subleading effects: the {\tt CAESAR} formula}
The expression for the transfer function obtained in the previous section contains effects beyond NLL. It is often useful to be able to neglect all subleading effects and hence have a pure NLL answer. This can be done through a set of simplifications that we briefly summarize below. We stress that the operations performed in the present section are not, strictly speaking, necessary, but they can simplify considerably the numerical evaluation of the transfer function, and allow for an analytic solution in some cases.

There are two important sources of subleading corrections in the treatment presented in the previous section. First, since the relevant squared amplitudes in the transfer function $1$PC$^{(0)}$ are taken in the soft-collinear limit, it is natural to also approximate the observable $V$ in the same limit. It is convenient to parametrize the emissions' momenta as
\begin{equation}
  \label{eq:Sudakov}
  k_i = z_i^{(1)}p_1 +  z_i^{(2)}p_2 + \kappa_{t,i}\,,
\end{equation}
where $\kappa_{t,i}$ is a space-like four-vector $\kappa_{t,i}=(0,\vec k_{t,i})$, orthogonal to the two reference momenta $p_1$ and $p_2$ that are aligned with the thrust axis $\vec{n}_T$
\begin{equation}
\label{eq:com-frame}
p_1 = \frac{Q}{2}(1,\vec n_T)\,,\qquad p_2 = \frac{Q}{2}(1,-\vec n_T)\,.
\end{equation}
Finally, we introduce the emission's rapidity $\eta_i$ with respect to
the thrust axis, which is given by
\begin{equation}
  \label{eq:rapidity}
  \eta_i = \frac{1}{2}\ln \frac{z_i^{(1)}}{z_i^{(2)}}\,,\quad {\rm with}\quad 
  |\eta_i| < \ln\frac{Q}{k_{t,i}}\,,
\end{equation}
where the boundary for $\eta_i$ is obtained by imposing
$z_i^{(\ell)}<1$ for any leg $\ell=1,2$.

Using the additivity of thrust one finds
\begin{align}
\label{eq:Vsc_add}
V_{\rm sc}(\Phi_B; k_1,\dots,k_n) = \sum_{i=1}^n \tau_i \,, \qquad {\rm with} \qquad \tau_i \equiv V_{\rm sc}(\Phi_B; k_i) = \frac{k_{ti}}{Q}e^{-|\eta_i|}
\,.
\end{align}
The above expression for the observable can be used in the evaluation of both the Sudakov radiator and the transfer function, neglecting terms beyond NLL order. Starting again from $\Sigma^{\rm NLL}_{\max}$, we can evaluate the integral~\eqref{eq:RNLL_w_K} by parametrizing the phase space $[\df q]$ in terms of the transverse momentum $q_t$ and rapidity $\eta$ of the emission $q$ in the centre-of-mass frame of the emitting dipole. For the soft-collinear contribution one finds
\begin{align}
\label{eq:1PC0S}
\int\!
[\df k]&\left[1+ \frac{\alpha_s(k_t)}{2\pi} K \right]|M_{\rm sc}^{(0)}(k)|^2
\Theta\left[V_{\rm sc}(\Phi_B; k)>\tau\right] \notag\\
&=\sum_{\ell=1}^2\int^Q_0 \frac{\df k_t}{k_t}\int_{0}^{\ln\frac{Q}{k_t}}\df \eta^{(\ell)}\int_{-\pi}^{\pi}\frac{\df\phi}{2\pi}2 C_F\frac{\alpha_s(k_t)}{\pi}\left[1+ \frac{\alpha_s(k_t)}{2\pi} K \right]\Theta\left[\frac{k_t}{Q} e^{-|\eta^{(\ell)}|}> \tau\right]\,,
\end{align}
where the sum runs over the two Born emitters.
The remaining hard-collinear contribution can be recast as~\cite{Banfi:2004yd,Banfi:2014sua}
\begin{align}
\int\!
[\df k]|M_{hc}^{(0)}(k)|^2
\Theta\left[V_{\rm sc}(\Phi_B; k)>\tau\right] =  -\frac{3}{2}C_F\sum_{\ell=1}^2\int_0^Q \frac{\df k_t}{k_t}\frac{\alpha_s(k_t)}{\pi} \, \Theta\left[\frac{k_t^2}{Q^2}>\tau\right]
\,.\end{align}
Using the above integrals, one can express $R_{\rm NLL}(\Phi_B; v)$ as
\begin{equation}
R_{\rm NLL}(\Phi_B; \tau) = - L g_1(\alpha_s L) - g_2(\alpha_s L)
\,,
\end{equation}
where $L=\ln\frac{1}{\tau} $ and the expressions of $g_i$ have been long known~\cite{Catani:1991kz} and are summarized in Appendix~\ref{app:g_functions}. The derivative of the LL radiator, required in the transfer function, is then given by
\begin{equation}
R'_{\rm LL}(\Phi_B; \tau) = - \alpha_s L g'_1(\alpha_s L) - g_1(\alpha_s L)
\,.
\end{equation}

The second source of subleading corrections in the formulation of Section~\ref{sec:simplifiedResummation} has to do with the phase space bounds of the resolved radiation. In particular, we see from Eq.~\eqref{eq:1PC0S} that 
\begin{equation}
|\eta_i^{(\ell)}| < \ln\frac{Q}{k_{ti}} = \frac{1}{2}\ln\frac{1}{\tau_i}
\,,
\end{equation}
where in the last step we used Eq.~\eqref{eq:Vsc_add}. At NLL the upper rapidity bound can be approximated as
\begin{equation}
\frac{1}{2}\ln\frac{1}{\tau_i} = \frac{1}{2}\ln\frac{1}{\tau} + {\cal O}\left(\ln\frac{\tau}{\tau_i}\right)
\,,
\end{equation}
which is then common to all resolved emissions. In our notation, this operation amounts to Taylor expanding the functions $R'_{\rm LL}(\Phi_B; v_i)$ in the resolved radiation as 
\begin{equation}
R'_{\rm LL}(\Phi_B; \tau_i) = R'_{\rm LL}(\Phi_B; \tau) + R''_{\rm LL}(\Phi_B; \tau) \ln\frac{\tau}{\tau_i} + {\cal O}\left(R'''\right)
\,,
\end{equation}
where all terms in the r.h.s. beyond the first one are logarithmically subleading (each extra derivative suppresses the contribution by one logarithmic order). Similarly, the first ratio in the transfer function can be expanded about $\tau$, in order to retain only the actual NLL terms necessary to cancel the $\eps$ dependence of the resolved radiation
\begin{align}
\frac{\Sigma^{\rm LL}_{\max}(\Phi_B; \eps \tau)}{\Sigma^{\rm LL}_{\max}(\Phi_B; \tau)} &= e^{R_{\rm LL}(\Phi_B; \tau) - R_{\rm LL}(\Phi_B; \eps \tau)}\simeq e^{-R'_{\rm LL}(\Phi_B; \tau)\ln\frac{1}{\eps}}
\,.\end{align}
With these simplifications, we can recast the transfer function as 
\begin{align}
\label{eq:FNLL_CAESAR}
\fullF_{\rm NLL}(\Phi_B; \tau) 
& =\delta^{R'_{\rm LL}(\Phi_B; \tau)} \sum_{n=0}^{\infty}
  \left(\frac{1}{n!}\prod_{{\substack{i=1\\\phantom{x}}}}^{n} \int_{\eps \tau}^\tau\frac{\df \tau_i}{\tau_i} \, R'_{\rm LL}(\Phi_B; \tau)\right)\Theta\left[\sum_i \tau_i  <\tau\right]
\,,
\end{align}
which can be written as
\begin{align}
\label{eq:FNLL_CAESAR2}
\fullF_{\rm NLL}(\Phi_B; \tau) 
& = \Bigg[ \left(\frac{\tau}{\eps \tau}\right)^{-R'_{\rm LL}(\Phi_B; \tau)} + \int_{\eps \tau}^\tau \! \frac{\df \tau_1}{\tau_1} R'_{\rm LL}(\Phi_B; \tau) \left(\frac{\tau}{\tau_1}\right)^{-R'_{\rm LL}(\Phi_B; \tau)} \left(\frac{\tau_1}{\eps \tau}\right)^{-R'_{\rm LL}(\Phi_B; \tau)}
\nn
& \qquad + \ldots \Bigg]\Theta\left[\sum_i \tau_i  <\tau\right]
\,.
\end{align}

Eq.~\eqref{eq:FNLL_CAESAR} is purely NLL, and does not contain any correction of subleading logarithmic nature, but still has the same general form as \eq{FNLL_PS}. The algorithm to compute it simplifies considerably:
\begin{enumerate}
\item Start with $i=0$ and $\tau_0 = \tau$
\item Increase $i$ by one
\item Generate $\tau_i$ randomly according to $(\tau_{i-1}/\tau_i)^{-R'_{\rm LL}(\Phi_B; \tau)} = r$, with $r \in [0,1]$
\item If $\tau_i < \eps \tau$ exit the algorithm, otherwise go back to step 2
\end{enumerate}
If the sum over all generated $\tau_i$ is less than $\tau$, accept the event, otherwise reject it. The value of $\fullF(v)$ is equal to the fraction of the accepted events. The form of the transfer function can be manipulated further in order to make its numerical evaluation more efficient by getting rid of the $\Theta$ function in Eq.~\eqref{eq:FNLL_CAESAR2}, as shown in refs.~\cite{Banfi:2004yd,Banfi:2014sua}.

There is a second advantage of using \eq{FNLL_CAESAR2} rather than \eq{FNLL_PS}. We notice that the starting equation~\eqref{eq:FNLL_PS} involves the function $R'$ (and hence the running coupling) evaluated at scales $Q\tau_i$ that can get as small as $Q\eps \tau$. When $\eps\to 0$ the above scale hits the Landau pole of the theory, which requires a prescription to deal with the non-perturbative region (e.g. a cutoff or a non-perturbative model) if this equation is implemented in a Monte Carlo method. On the other hand, the final equation~\eqref{eq:FNLL_CAESAR2} does not have this issue since we expanded the arguments of the couplings about $Q\tau \gg \Lambda_{\rm QCD}$, hence avoiding the Landau pole as long as the observable $\tau$ is sufficiently large.

For an additive observable such as thrust considered here, further manipulations are possible to obtain an analytic solution which reads
\begin{align}
\label{eq:FNLL_CAESAR_analytic}
\fullF_{\rm NLL}(\Phi_B; \tau) = \frac{e^{-\gamma_E R'_{\rm LL}(\Phi_B; \tau)}}{\Gamma[1+R'_{\rm LL}(\Phi_B; \tau)]}
\,,
\end{align}
which leads to the following NLL formula for the thrust cumulative distribution
\begin{equation}
\label{eq:Sigma_NLL_thrust_CAESAR}
\Sigma(\tau) = e^{ L g_1(\alpha_s L) + g_2(\alpha_s L)}\frac{e^{-\gamma_E R'_{\rm LL}(\Phi_B; \tau)}}{\Gamma[1+R'_{\rm LL}(\Phi_B; \tau)]}
\,.
\end{equation}

\section{Review of resummation in SCET}
\label{sec:reviewSCET}
SCET~\cite{Bauer:2000ew,Bauer:2000yr,Bauer:2001ct,Bauer:2001yt} is an effective field theory of QCD constructed to capture the long distance physics arising from soft and collinear radiation. To describe these different types of long distance effects requires two separate types of fields in the effective theory: soft and collinear fields. All short distance physics is integrated out of the theory, and contributes only via short distance matching coefficients. 

\begin{figure}[t!]
	\begin{eqnarray}
	%
	%
	&& \begin{picture}(20,10)(20,0)
	\mbox{\includegraphics[width=3.0truecm]{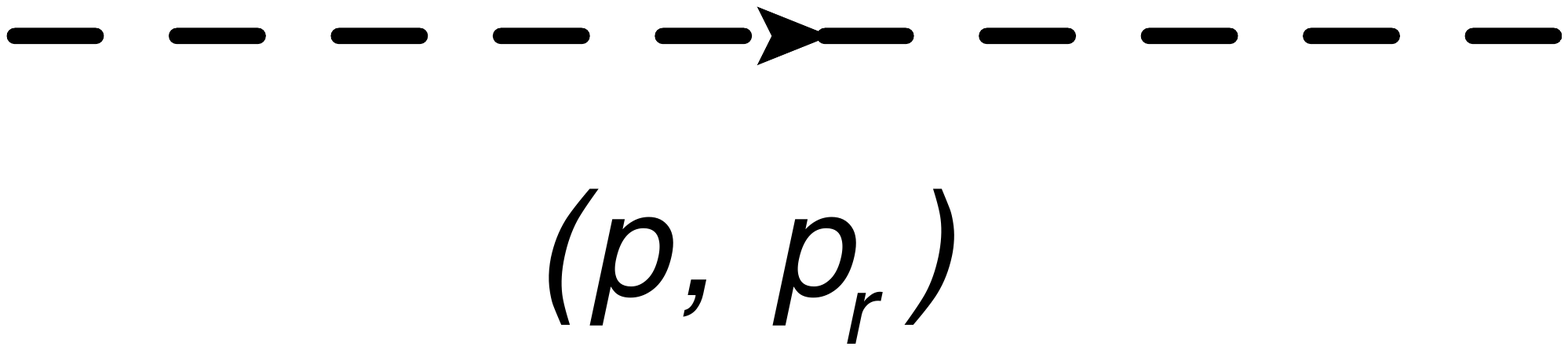} } 
	\end{picture} \qquad\qquad\quad \Large 
	\raise59pt \hbox{$  = \ \mbox{\normalsize $i$}\, \frac{\nslash}{2}\: 
		{\bn\cdot p \over n\cdot p_r\, \bn\cdot p\: +\: p_\perp^2 +i0}$ } 
	\nn[-40pt]
	%
	%
	&& \begin{picture}(20,10)(20,0)
	\mbox{\includegraphics[width=3.0truecm]{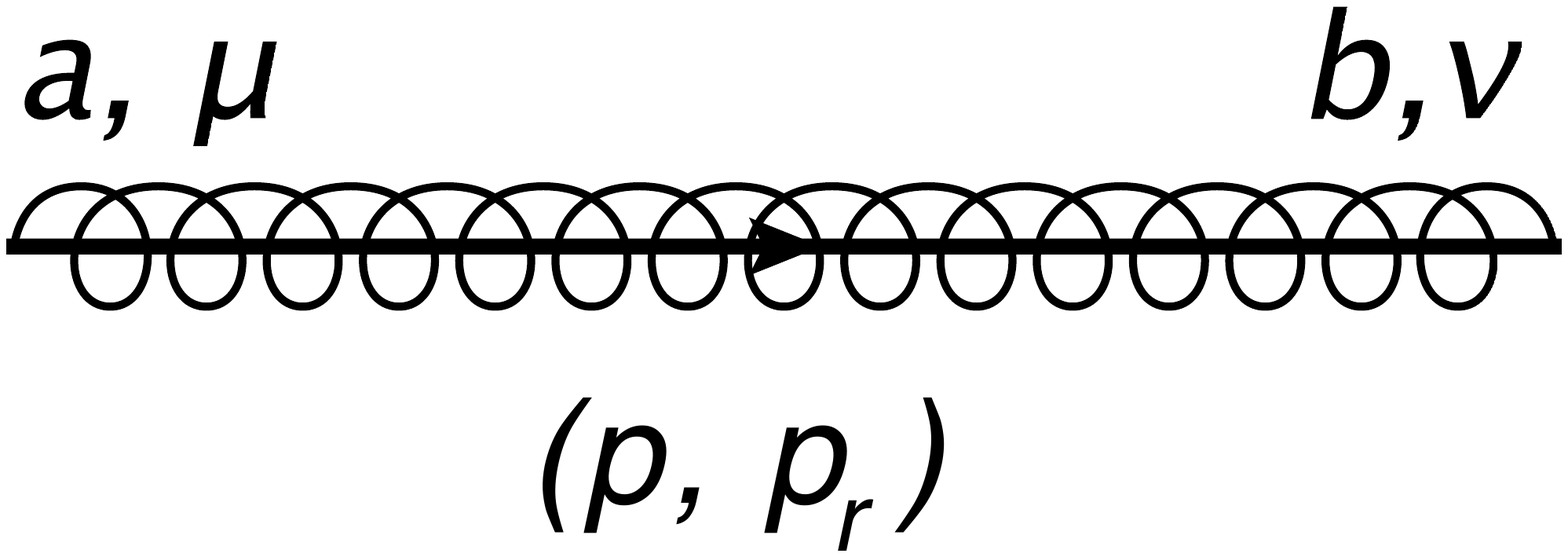} } 
	\end{picture} \qquad\qquad\quad \Large 
	\raise53pt \hbox{$  = \: 
		{(-i) g^{\mu\nu} \delta_{a,b} \over n\cdot p_r\, \bn\cdot p\: +\: p_\perp^2 +i0}$ } 
	\nn[-35pt]
	%
	%
	&& \begin{picture}(20,10)(20,0)
	\mbox{\includegraphics[width=3.0truecm]{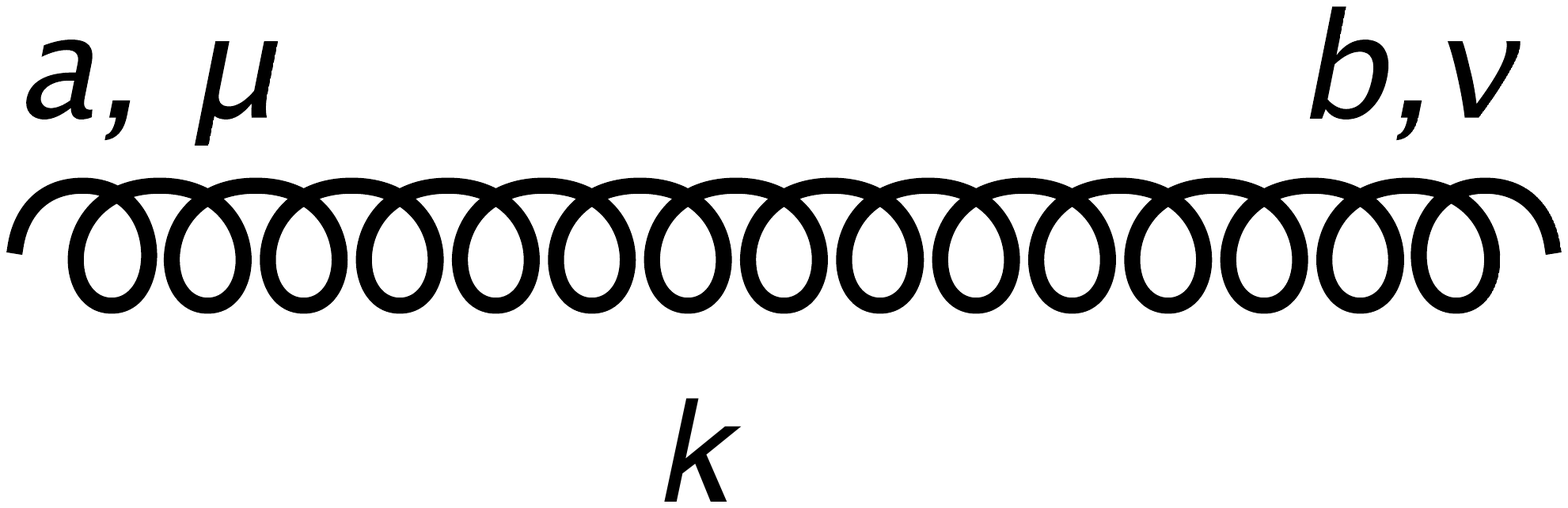} } 
	\end{picture} \qquad\qquad\quad \Large 
	\raise52pt \hbox{$  = \: 
		{(-i) g^{\mu\nu} \delta_{a,b} \over k^2 +i0}$ } 
	\nn[-40pt]
	%
	%
	&& \begin{picture}(20,80)(20,0)
	\mbox{\includegraphics[width=3.0truecm]{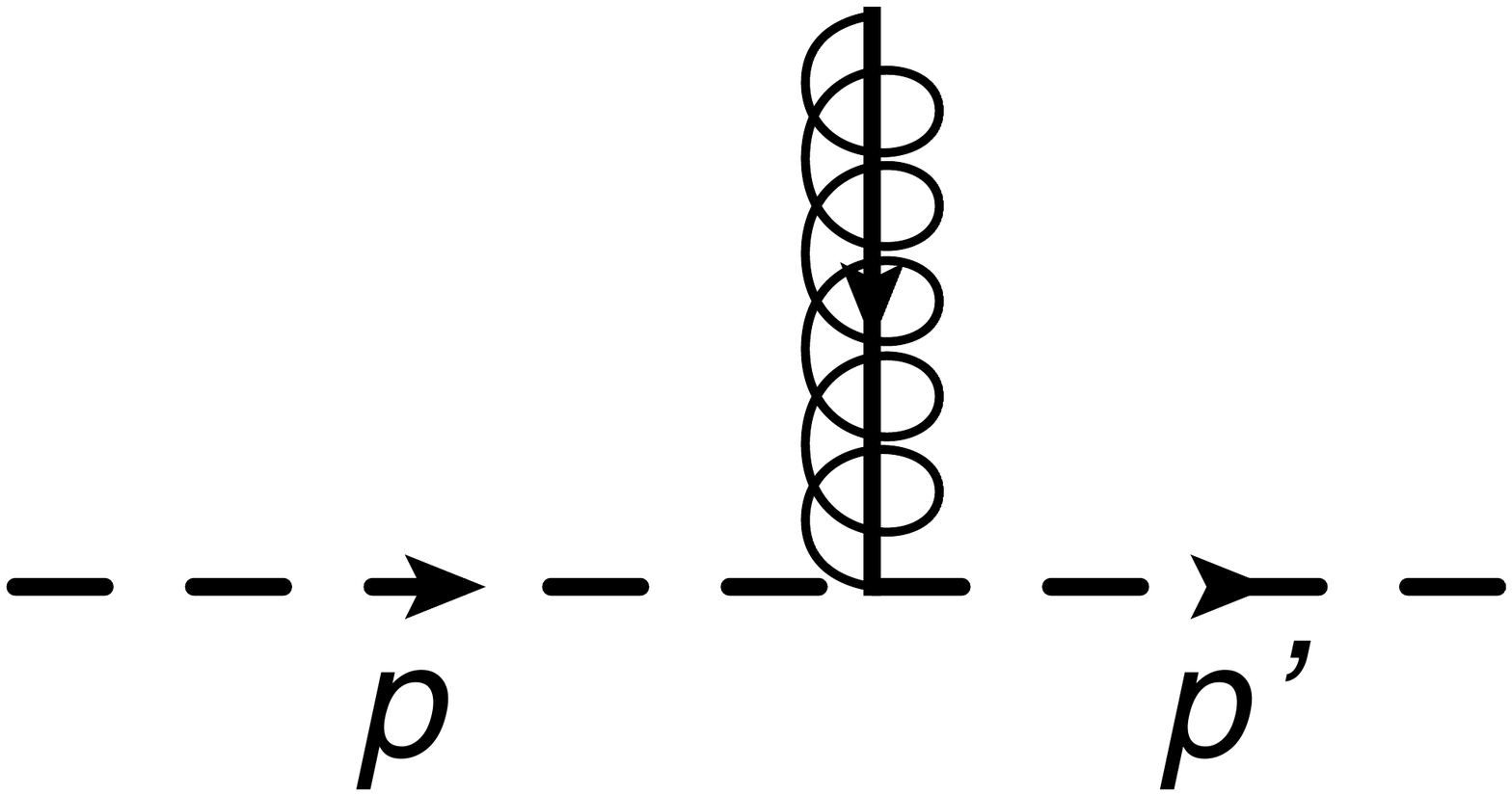}  }
	\end{picture} \qquad\qquad\quad \Large 
	\raise53pt \hbox{$  = \mbox{\normalsize $i g\,T^a$} \, 
		\bigg[ n_\mu + \frac{ \gamma^\perp_\mu \pslash_\perp }{\bn \cdot p} 
		+ \frac{ \ppslash_\perp \gamma^\perp_\mu }{\bn \cdot p^{\,\prime}}
		- \frac{\ppslash_\perp \pslash_\perp   }{\bn \cdot p\:  
			\bn \cdot p^{\,\prime}}\bar n_\mu \bigg]\, \frac{\bar\nslash}{2}  $ } 
	\nn[-45pt]
	%
	%
	&& \begin{picture}(20,80)(20,0) 
	\mbox{\includegraphics[width=3.0truecm]{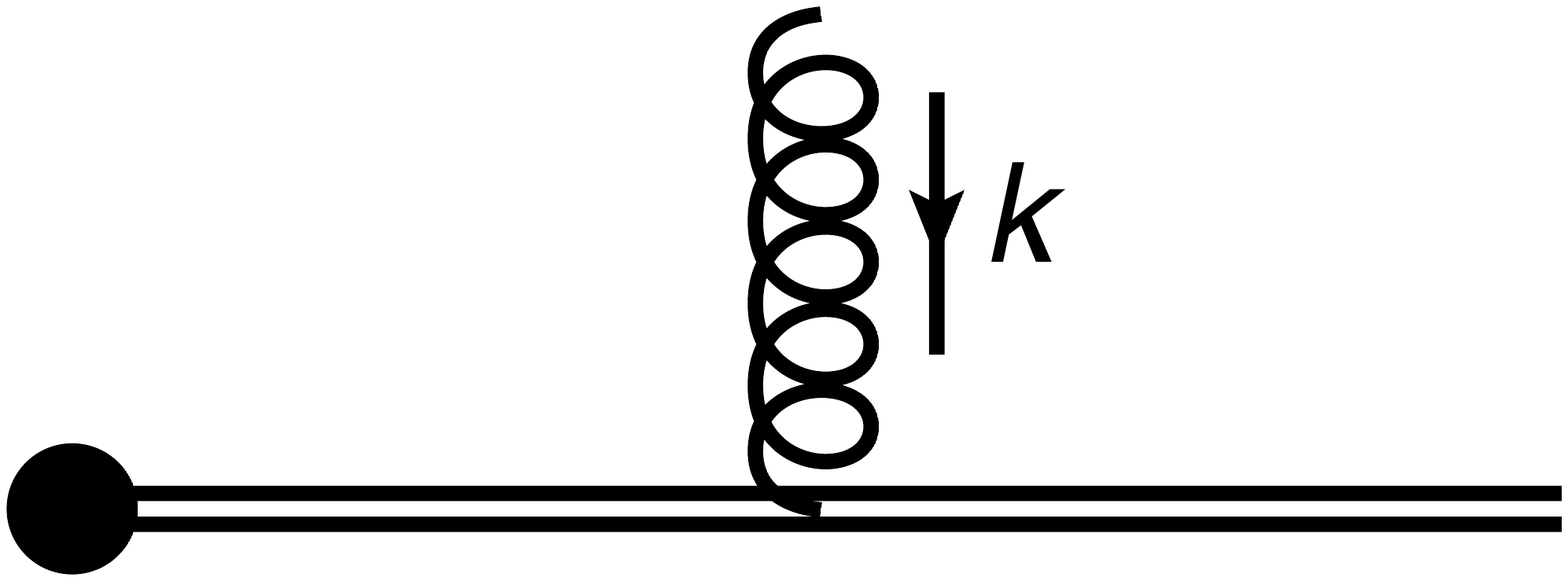}  }
	\end{picture} \qquad\qquad\quad \Large 
	\raise50pt \hbox{$  = \mbox{\normalsize $-g\,T^a$} 
		\: {n_\mu \over n\cdot k}  $ } 
	\nn[-45pt]
	%
	%
	&& \begin{picture}(20,80)(20,0) 
	\mbox{\includegraphics[width=3.0truecm]{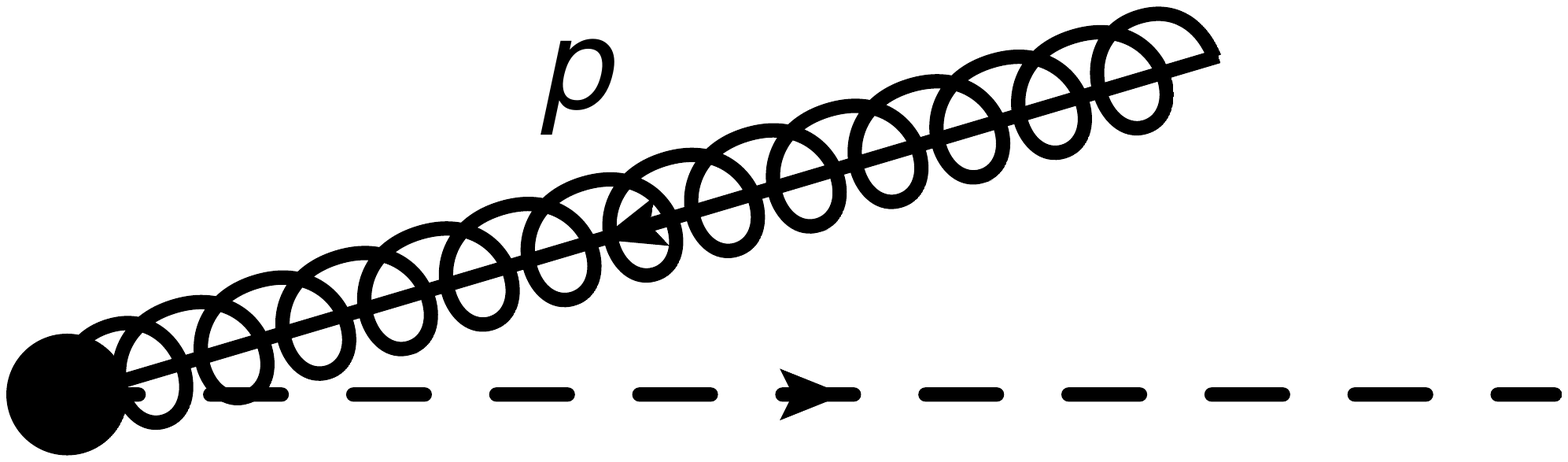}  }
	\end{picture} \qquad\qquad\quad \Large 
	\raise50pt \hbox{$  = \mbox{\normalsize $-g\,T^a$} 
		\:{\bar n_\mu \over \bar n\cdot p}   $ } 
	\notag
	\end{eqnarray}
	\vskip-1cm
	{\caption[1]{Feynman rules for the leading-power SCET$_{\rm I}$ Lagrangian and operators: collinear quark and
	                 gluon propagator with label $p$ and residual momentum $p_r$, soft gluon propagator, coupling of 
	                 collinear quark and gluon, emission from a soft ($Y_n$) and collinear ($W_n$) Wilson line, 
	                 respectively. When cutting a propagator, we replace the denominator $a$ of the propagator with 
	                 $(-2\pi i) \delta(a)$. }
		\label{Fig:Feynman} }
\end{figure}

Given that SCET has several degrees of freedom and exhibits a rich gauge structure, a detailed derivation of it is beyond the scope of this work and we refer the reader to the original literature\cite{Bauer:2000ew,Bauer:2000yr,Bauer:2001ct,Bauer:2001yt} for details. One important feature, however, is that by defining the collinear fields in an appropriate way~\cite{Bauer:2001yt}, the SCET Lagrangian can be written in a way that at leading power the collinear and soft degrees of freedom can be completely separated, giving
\begin{align} 
\label{eq:LSCET}
{\cal L}_{\rm SCET} = {\cal L}_s + \sum_i {\cal L}_{n_i}
\end{align}
where the soft Lagrangian is identical to the full QCD Lagrangian. In the following we are going to use the collinear fermionic Lagrangian which can be written as
\begin{align} 
\label{eq:Lxixi0}
{\cal L}^{f}_{n} = 
	\bar{\xi}_{n}\Big( i n\cdot D_{n}+ i \slashed{D}_{n\perp}\frac{1}{i\bn\cdot D_n}
	i \slashed{D}_{n\perp}\Big) \frac{\slashed{\bn}}{2}\,\xi_{n} \,.
\end{align}
Here $\xi_n$ denotes a collinear fermion field after a so-called BPS~\cite{Bauer:2001yt} field redefinition, and the derivatives $D_n$ are covariant with respect to collinear gauge transformations and therefore only include collinear gluons fields. Note that a single collinear fermion field $\xi_n$ can be made invariant under collinear gauge transformations by combining it with a collinear Wilson line $W_n$ to define
\begin{align}
\chi=W_n^\dagger\xi_n
\,.
\end{align}
Operators in SCET are typically constructed out these gauge invariant fields. The Feynman rules that are obtained from the SCET Lagrangian are given in Fig.~\ref{Fig:Feynman}.

The starting point for resummation in SCET is the derivation of a factorization theorem that expresses the cross section as a combination of contributions arising from three different singular sectors: hard, soft and collinear. Although such a type of separation is already performed at the level of the SCET Lagrangian~\cite{Bauer:2001ct}, the observable under consideration mixes the various soft and collinear modes in its definition. Therefore, in order to derive a factorization theorem one must decompose the observable into soft and collinear contributions~\cite{Bauer:2008dt,Bauer:2008jx}, which can be treated separately and then combined to give the final value of the observable. This allows for a separation between the phase space of the soft and collinear sectors, and therefore makes the factorization manifest. A clear complication arises for complex observables, for which the separation between soft and collinear modes in the measurement function can become quite cumbersome. 

We express a generic factorization theorem as
\begin{align}
\label{eq:SCETFactorzationGeneral}
\Sigma(v) = H_{n_1 n_2}(\mu) \, J_{n_1}(v, \ldots; \mu) \otimes J_{n_2}(v, \ldots; \mu) \otimes S_{n_1 n_2}(v, \ldots; \mu)
\,.
\end{align}
The hard function $H_{n_1 n_2}$ only depends on the directions $n_{i}$, but is independent of the observable. The jet functions describe the evolution of the radiation collinear to the directions $n_i$, and the soft function describes the soft interaction between the two jet functions. The precise definition of the jet and soft functions, as well as of the convolution $\otimes$ in Eq.~\eqref{eq:SCETFactorzationGeneral}, depend on the definition of the observable whose value is required to be less than $v$ in the integrated cross section $\Sigma(v)$.

In this paper we will need two types of observables. The first is the thrust observable we intend to resum, which is an additive observable for which the total value of the observable is the sum over the contributions from each particle. The factorization formula for such an additive observable takes the form~\cite{Korchemsky:1999kt,Fleming:2007qr,Schwartz:2007ib}
\begin{align}
\label{eq:SigmaAddSCET}
\Sigma(v) &= H_{n_1 n_2}(\mu) \left[\prod_{i=1}^2 \int \! \df v_i \, J_{n_i}(v_i; \mu)\right] \int \! \df v_s\,S_{n_1 n_2}(v_s; \mu) \, \, \Theta[v > \sum_i v_i + v_s]
\,.\end{align}
The two collinear directions $n_1$ and $n_2$ are back to back along the thrust axis $t$ such that $n_1 = n$, $n_2 = \bar n$ with $n = (1, \hat t)$, $\bar n = (1, -\hat t)$ and $n \!\cdot\! \bar n = 2$. Suppressing the dependence of the hard and soft function on the directions $n$ and $\bar n$, we write 
\begin{align}
  \Sigma(\tau) = H(\mu) \int \! \df \tau_n \, J_{n}(\tau_n; \mu)\int \! \df \tau_{\bar n} \, J_{\bar n}(\tau_{\bar n} ;\mu) \int \! \df \tau_s\,S(\tau_s; \mu) \, \, \Theta[\tau > \tau_n + \tau_{\bar n} + \tau_s]
\,.
\end{align}

We will also need an expression for the simple observable used to define $\Sigma_{\rm max}$ in the previous section. This is defined by first grouping the various collinear and soft emissions separately into clusters in an infrared and collinear safe manner, computing the observable from each cluster and taking the maximum value of those. Such an observable factorizes in a multiplicative way, such that no convolutions are required
\begin{align}
\label{eq:SigmaMaxSCET}
\Sigma_{\rm max}(v) &= H(\mu) \left[\prod_{i=1}^2J^{\rm max}_{n_i}(v; \mu)\right] S^{\rm max}(v; \mu) 
\,.\end{align}
We start by discussing the resummation for the additive observables described by the factorization formula~\eqref{eq:SigmaAddSCET}, and then we comment in more detail on the definition of $\Sigma_{\rm max}$ in SCET.\\

The soft and jet functions that appear in the above factorization theorems have the operator 
definition~\cite{Bauer:2002ie,Bauer:2003di,Fleming:2007qr,Becher:2006qw,Sterman:1986aj}
\begin{align}
S(\tau_s;\mu) &= \frac{1}{N_c} {\rm
  Tr}\langle 0| \bar{Y}_{\bn}^{\dagger}(0)Y_{n}^{\dagger}(0)\delta(\tau_s -V_{\rm soft})Y_{n}(0)  \bar{Y}_{\bn}(0)|0\rangle
  \,,\notag\\
J_n(\tau_n;\mu) &= \int \frac{\df l^+}{2\pi}{\cal J}_n(\tau_n,l^+;\mu)
\,,\notag\\
J_\bn(\tau_\bn;\mu) &= \int \frac{\df l^-}{2\pi}{\cal J}_\bn(\tau_\bn,l^-;\mu)
\,,
\end{align}
where
\begin{align}
{\cal J}_n(\tau_n,l^+;\mu) 
  \frac{\slashed{n}_{\alpha \beta}}{2}&=\frac{1}{N_c} {\rm 
  Tr}\int \df^4x \, e^{il\cdot x}\langle 0| \chi_{n,\alpha}(x)\delta(\tau_n -V_{ n}){\bar 
                                        \chi}_{n,\beta}(0)|0\rangle
                                        \,,\notag\\
{\cal J}_{\bar n}(\tau_\bn,l^-;\mu) 
  \frac{\slashed{\bar n}_{\alpha \beta}}{2}&=\frac{1}{N_c} {\rm 
  Tr}\int \df^4x \, e^{il\cdot x}\langle 0| {\bar \chi}_{\bar{n},\beta}(x)\delta(\tau_\bn -V_{\bn})\chi_{\bar{n},\alpha}(0)|0\rangle
  \,,
\end{align}
and $Y_n(x)$ denotes a soft Wilson line along the $n$ direction. $V_{\rm soft}$, $V_{n}$ and $V_{\bn}$ denote the expression of either thrust $V$ or the simple observable $V_{\rm max}$  as function of the final state momenta in the soft and collinear approximations, respectively. For notational simplicity, from now on we will omit the trace operation as well as the $1/N_c$ prefactor in the color average of the above expressions, which will be understood in the rest of this article.

\subsection{Resummation via Renormalization group equations}
\label{sec:SCET_Overview_RGE}
Once a factorization theorem has been obtained, one can use the renormalization group equations to resum the logarithmic dependence in the various contributions to the factorized cross sections. For this to work, it is crucial that each contribution depends kinematically on only a single scale $\mu_F$. This ensures that the logarithmic dependence in each contribution is directly tied to the dependence on the renormalization scale, since it can only occur in the form $\ln(\mu / \mu_F)$. It immediately follows that each contribution is free from logarithmic dependence if one chooses $\mu = \mu_F$ (the initial condition), and that the logarithms can be resummed using the RG equations. 

Before we discuss this in more detail, we take a short digression and discuss a feature of SCET that will be important later. 
In SCET, both the physical phase space and the observable's measurement function are expanded out according to the scaling of soft and collinear modes, since it ensures that each ingredient in the factorization formula depends on only a single scale\footnote{An exception is given by some observables which require the introduction of an additional regulator to handle the rapidity divergences, which are classified as SCET$_{\rm II}$ problems~\cite{Bauer:2002aj}. In this case soft and jet functions will generally depend on two scales. This fact does not affect the treatment we present in the rest of this article, as our final formulation of the resummation in Section~\ref{sec:automatedSCET} equally applies to both cases.}. Written in terms of the invariants $y_{qg} = s_{qg} / Q^2$ and $y_{\bar q g} = s_{\bar q g} / Q^2$, the matrix element squared of the real radiation behaves as $1 / (y_{q g} y_{\bar q g})$, such that divergences arise both in the IR ($y \to 0$) or UV ($y \to \infty$) limit. To understand the consequences of this, we investigate the phase space boundary of a single emission, which are given in full QCD as
\begin{align}
\label{eq:QCD_bounds}
{\rm QCD:} \quad \int \! \df y_{qg} \, \df y_{\bar q g} \,\, \Theta[\min( y_{qg},  y_{\bar q g}, 1 - y_{qg} -  y_{\bar q g}) < \tau] \, \Theta[0<y_{ij}<1]
\,,\end{align}
where $y_{ij}$ denotes both $y_{qg}$ and $y_{\bar q g}$. They are shown graphically in Fig~\ref{fig:phaseSpace}a). Clearly, neither of the two Mandelstam variables can exceed the physical bound set by the total energy in the event $Q^2$, and therefore the phase space integration over each variable is bounded from above. 
\FIGURE[h]{
 \centering
  \includegraphics[scale=0.18]{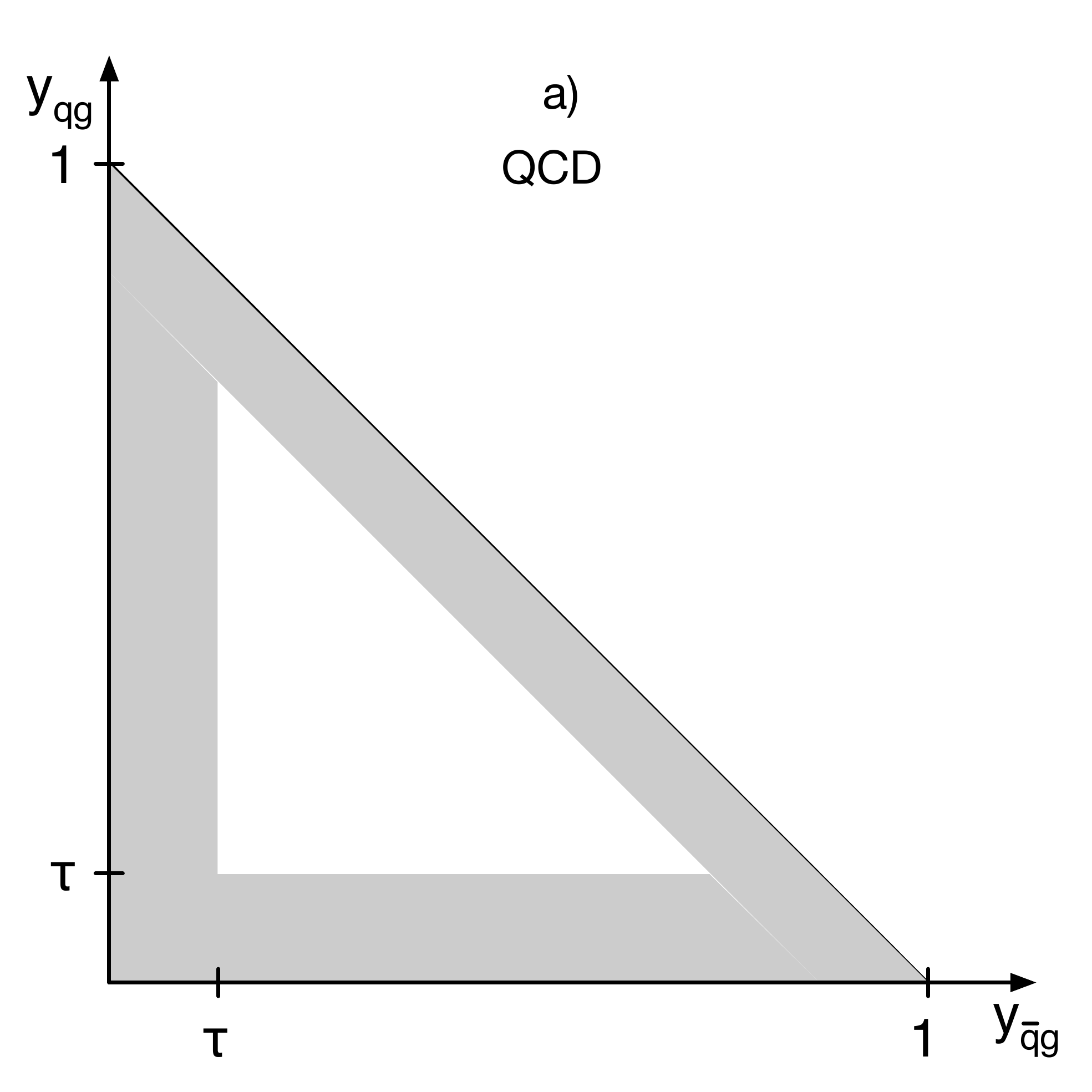}
  \includegraphics[scale=0.18]{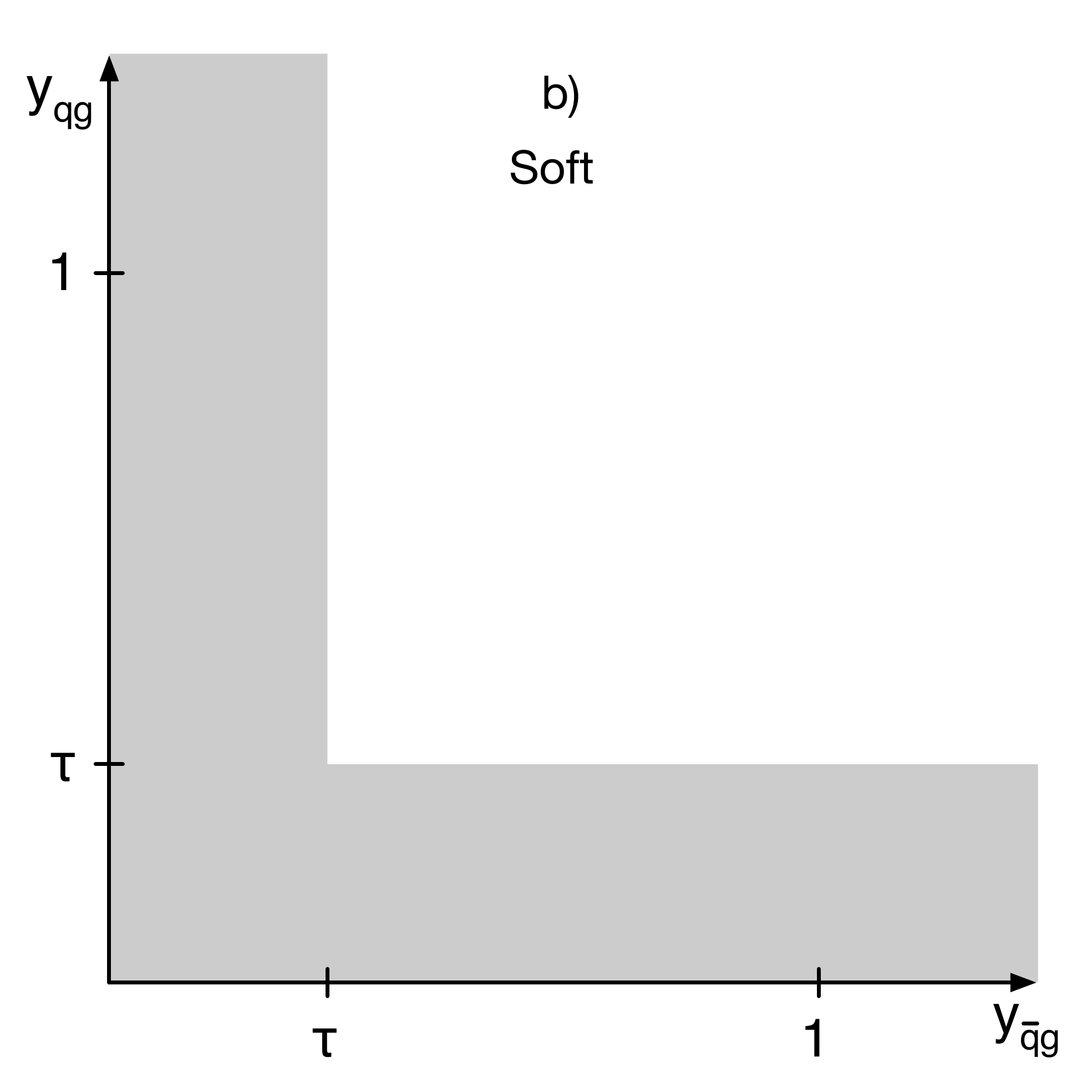}
  \includegraphics[scale=0.18]{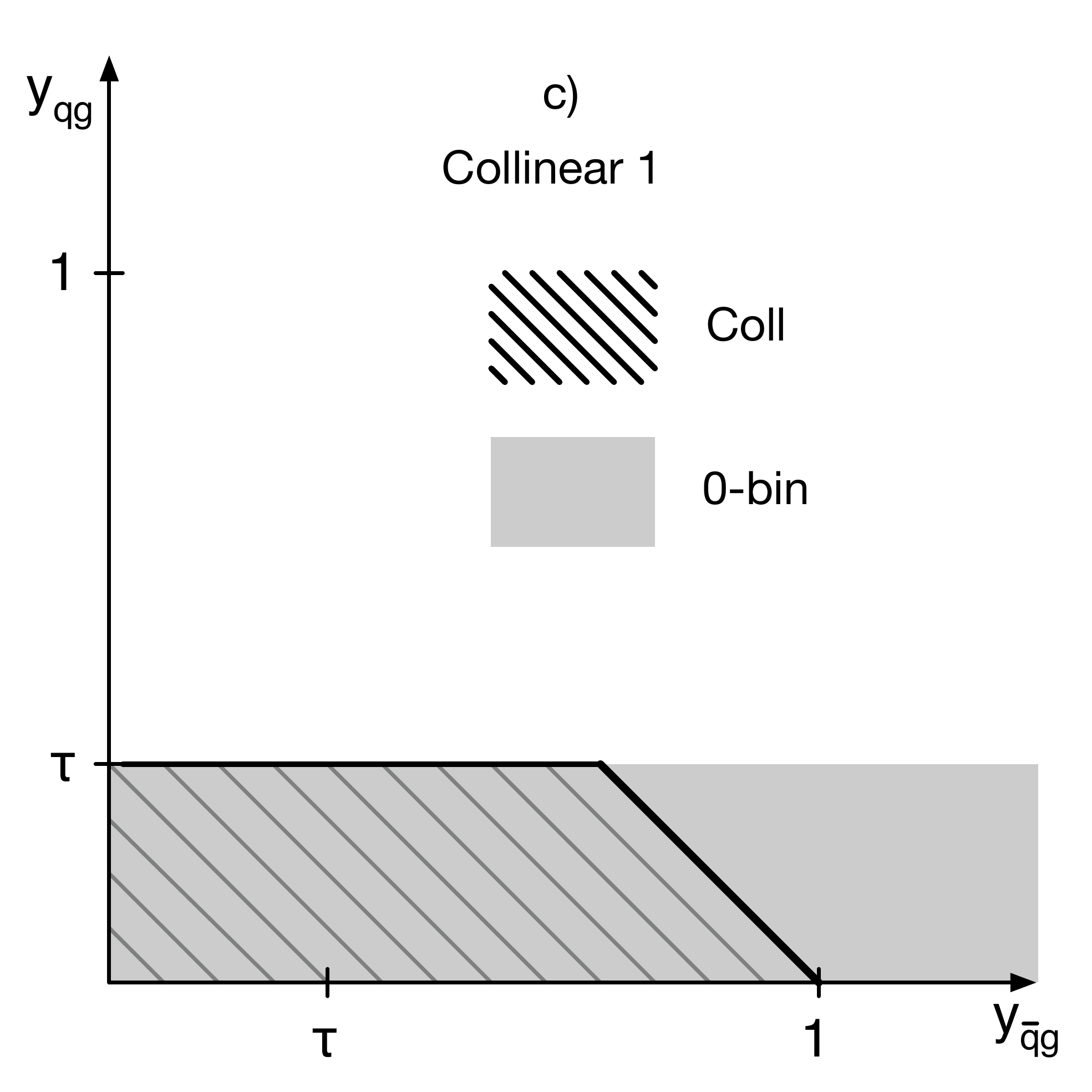}
  \includegraphics[scale=0.18]{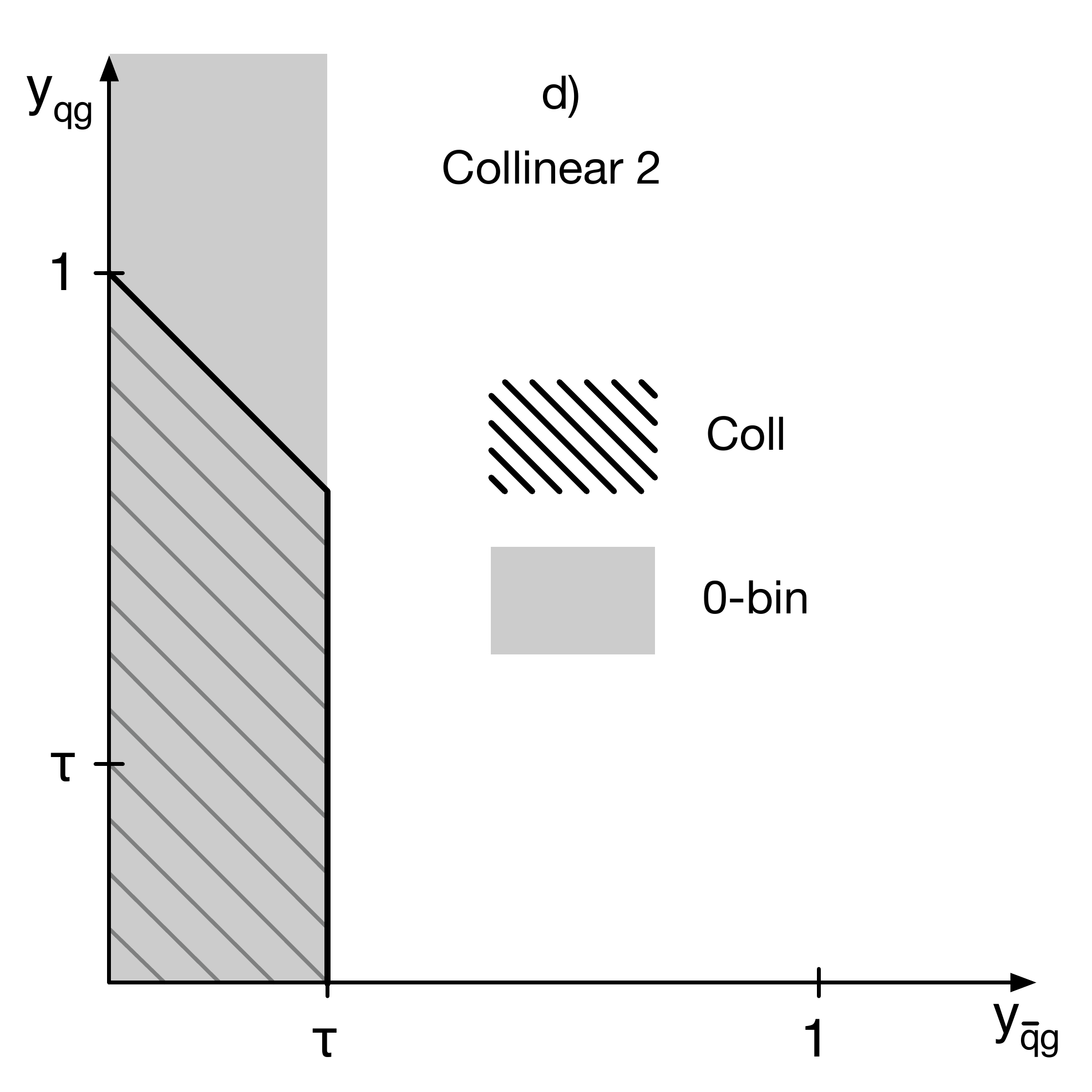}
\caption{\label{fig:phaseSpace}%
The regions of phase space contributing to the various pieces. In a) we show the phase space region of full QCD, in b) that of the soft function, and in c) and d) the region of the jet functions. 
}}

The phase space boundary of the soft function in SCET is obtained by expanding the full QCD phase space boundary about the limit $y_{q g}, y_{\bar q g} \ll 1$. This gives
\begin{align}
{\rm Soft}: \int \! \df y_{qg} \, \df y_{\bar q g} \,\, \Theta[\min( y_{qg},  y_{\bar q g}) < \tau] \, \Theta[0<y_{ij}]
\,,\end{align}
which is shown graphically in  Fig~\ref{fig:phaseSpace}b). This implies that the larger of the two Mandelstam variables $y_{q g}$ or $y_{\bar q g}$ is unbounded from above, leading to a UV divergence.  

The first collinear limit is obtained by taking the limit $y_{q g} \ll y_{\bar q g} \sim 1$ (the second is the same under the replacement $y_{q g} \leftrightarrow y_{\bar q g}$). This gives
\begin{align}
{\rm Coll}_1: \int \! \df y_{qg} \, \df y_{\bar q g} \,\, \Theta[\min( y_{qg}, 1-  y_{\bar q g}) < \tau] \,\Theta[0<y_{\bar q g}<1] \, \Theta[0<y_{q g}]
\,.\end{align}
The collinear regions are shown by the hatched region in  Fig~\ref{fig:phaseSpace} c) and d). In this case both variables are bounded from above, just as in the case of the full theory. However, adding the soft and collinear regions naively, leads to a double counting of the soft-collinear region~\cite{Manohar:2006nz}, which is handled in SCET by subtracting a 0-bin region from the collinear integrals, which is nothing but the soft limit of the collinear integral. The soft limit of the first collinear phase space region (with the obvious replacement to the obtain the soft limit of the second collinear phase space region) is given by
\begin{align}
0-{\rm bin}_1: \int \! \df y_{qg} \, \df y_{\bar q g} \,\, \Theta[0 < y_{qg} < \tau]\, \Theta[0 < y_{\bar qg}]
\,,\end{align}
such that the integral over $y_{\bar q g}$ is again unbounded from above, leading to a UV divergence. Diagrammatically, the 0-bin  regions are summarized by the gray region in Fig.~\ref{fig:phaseSpace} c) and d). 

While UV divergences are present in SCET as just discussed, each of the terms in the factorization formula~\eq{SigmaAddSCET} is IRC finite. Thus all divergences are of UV origin and are removed by renormalization. The renormalization of the UV divergences leads to renormalization group equations (RGE) for each component. As already discussed, each of the ingredients of the factorization theorem has its own characteristic scales that we denote by $\mu_H$, $\mu_J$, and $\mu_S$ for the hard, jet, and soft functions respectively. At these scales no logarithms are present to any order in perturbation theory. For the thrust observable considered in this work, the scales are~\cite{Schwartz:2007ib,Becher:2008cf}
\begin{align}
\mu_H = Q\,, \qquad \mu_J = Q \sqrt{\tau}\,, \qquad \mu_S = Q \tau
\,.
\end{align} 

The resummation in SCET is then performed by evolving the hard, soft and jet functions from their characteristic scales to a common renormalization scale $\mu$. The evolution is simply obtained by solving the corresponding RGE. The hard function is always multiplicatively renormalized, giving the following evolution equation
\begin{align}
\label{eq:RGE_H}
\mu \frac{\df}{\df \mu} H(\mu) = \left\{2\Gamma_{\rm cusp}[\alpha_s(\mu)] \ln\frac{Q^2}{\mu^2}+ 2 \gamma_{H}[\alpha_s(\mu)]\right\}H(\mu)
\,.
\end{align}
The precise form of the RGE for the soft and jet function depends on the observable under consideration. This dependence arises from the way the observable behaves in the presence of multiple soft or collinear emissions which make up the soft and jet functions. For instance, in the case of thrust, each new emission contributes to the observable additively, which implies the following non-local form for the RGEs~\cite{Bosch:2003fc,Bosch:2004th,Fleming:2007qr,Schwartz:2007ib,Becher:2008cf}
\begin{align}
\label{eq:RGE_J}
\mu \frac{\df}{\df \mu} J_{n_i}(\tau; \mu) &= \left\{-2\Gamma_{\rm cusp}[\alpha_s(\mu)] \ln\frac{\tau Q^2}{\mu^2}- 2 \gamma_{J}[\alpha_s(\mu)]\right\} J_{n}(\tau; \mu) \notag\\
&+ 2\Gamma_{\rm cusp}[\alpha_s(\mu)]\int_0^\tau\!  {\rm d}\tau' \, \frac{ J_{n_i}(\tau; \mu)- J_{n_i}(\tau'; \mu)}{\tau-\tau'}\,,
\\
\mu \frac{\df}{\df \mu} S(\tau; \mu) &= \left\{2\Gamma_{\rm cusp}[\alpha_s(\mu)] \ln\frac{\tau^2 Q^2}{\mu^2}- 2 \gamma_{S}[\alpha_s(\mu)]\right\} S(\tau; \mu) \notag\\
&- 4\Gamma_{\rm cusp}[\alpha_s(\mu)]\int_0^\tau\!  {\rm d}\tau' \, \frac{S(\tau; \mu)- S(\tau'; \mu)}{\tau-\tau'}
\,.
\label{eq:RGE_S}
\end{align}
Eqs.~\eqref{eq:RGE_J} and~\eqref{eq:RGE_S} are simplified in Laplace space, where the convolutions become simple products
\begin{align}
\label{eq:laplace_RGE_coll}
\mu \frac{\df}{\df \mu} \tilde{J}_{n_i}(u; \mu) &= \left\{-2\Gamma_{\rm cusp}[\alpha_s(\mu)] \ln\frac{u_0 \,Q^2}{u\,\mu^2}- 2 \gamma_{J}[\alpha_s(\mu)]\right\} \,\tilde{ J}_{n_i}(u;\mu)\,,
\\
\mu \frac{\df}{\df \mu} \tilde{S}(u; \mu) &= \left\{2\Gamma_{\rm cusp}[\alpha_s(\mu)] \ln\frac{u_0^2 Q^2}{u^2\mu^2}- 2 \gamma_{S}[\alpha_s(\mu)]\right\} \, \tilde{S}(u;\mu)
\,,
\label{eq:laplace_RGE_soft}
\end{align}
where $\tilde{ J}$ and $\tilde{S}$ denote the Laplace transform of the jet and soft functions, $u$ is the Laplace variable conjugate to $\tau$, and $u_0=e^{-\gamma_E}$.
Since the cross section $\Sigma(v)$ is independent of the renormalization scale, the anomalous dimensions of the various pieces satisfy the consistency condition
\begin{align}
\label{eq:anomalousDimensionConstraint}
  \gamma_H[\alpha_s(\mu)]=2 \gamma_{J}[\alpha_s(\mu)] + \gamma_{S}[\alpha_s(\mu)]
  \,,.\end{align}
An analogous condition, trivially satistfied, holds for the terms in the anomalous dimension proportional to $\Gamma_{\rm cusp}$.
We write the solution to the RGEs in Eqs.~\eqref{eq:RGE_H}, \eqref{eq:RGE_J} and~\eqref{eq:RGE_S} as
\begin{align}
H(\mu) &= H(\mu_H) \, U_H(\mu, \mu_H)
\,,
\nn
J(\tau; \mu) &=  \int \! \df \tau \, J(\tau; \mu_J) \, U_J(\tau - \tau'; \mu, \mu_J)
\,,
\nn
S(\tau; \mu) &=  \int \! \df \tau \, S(\tau; \mu_S) \, U_S(\tau - \tau'; \mu, \mu_S)
\,,\end{align}
where, as discussed above, all logarithms arise from the RG Kernels $U_F(\ldots; \mu, \mu_F)$. 
This leads to the final resummed $\Sigma(\tau)$, which takes the form
\begin{align}
  \Sigma(\tau) = & H(\mu_H) \, U_H(\mu, \mu_H) \int \! \df \tau_n\df \tau'_n \, J_{n}(\tau'_n; \mu_J)\, U_J(\tau_n- \tau'_n; \mu, \mu_J)
\nn
& \qquad \times  \int \! \df \tau_{\bar n}\df \tau'_{\bar n} \, J_{\bar n}(\tau'_{\bar n} ;\mu_J)\, U_J(\tau_{\bar n}- \tau'_{\bar n}; \mu, \mu_J) 
\nn
& \qquad \times \int \! \df \tau_s\df \tau'_s\,S(\tau_s; \mu_S)\, U_S(\tau_s- \tau'_s; \mu, \mu_S) \, \, \Theta[\tau - \tau_n - \tau_{\bar n} - \tau_s]
\,.
\end{align}
For an observable that is multiplicatively renormalized, such as $\Sigma_{\rm max}(\tau)$, one finds the simpler expression
\begin{align}
  \Sigma_{\max}(\tau) = & H(\mu_H) U_H(\mu, \mu_H) \, J_{n}(\tau; \mu_J)U_J(\tau; \mu, \mu_J) \, J_{\bar n}(\tau ;\mu_J)U_J(\tau; \mu, \mu_J) 
\nn
& \qquad \times \,S(\tau; \mu_S)U_S(\tau; \mu, \mu_S)
\,.
\end{align}

The boundary conditions $F(\ldots; \mu_F)$, as well as the anomalous dimensions $\Gamma_{\rm cusp}$ and $\gamma_F$ (for $F = H, J, S$) have a perturbative expansion
\begin{align}
\label{eq:anomalous_dimensions}
  \Gamma_{\rm cusp}[\alpha_s(\mu)]  &= \frac{\alpha_s(\mu)}{2\pi}\Gamma^{(1)}_{\rm cusp} + \left[\frac{\alpha_s(\mu)}{2\pi}\right]^2\Gamma^{(2)}_{\rm cusp} + \ldots
\nn
 \gamma_F[\alpha_s(\mu)]  &= \frac{\alpha_s(\mu)}{2\pi}\gamma^{(1)}_F+ \left[\frac{\alpha_s(\mu)}{2\pi}\right]^2\gamma^{(2)}_F+ \ldots\nn
F(\ldots; \mu_F) &= 1 + \frac{\alpha_s(\mu)}{2\pi}F^{(1)}+ \left[\frac{\alpha_s(\mu)}{2\pi}\right]^2F^{(2)}_F+ \ldots
\,.
\end{align}
The logarithmic accuracy is determined by the perturbative order with which the anomalous dimensions and boundary conditions are determined. For example, to achieve LL accuracy, one only needs $\Gamma_{\rm cusp}^{(1)}$, while for NLL accuracy $\Gamma_{\rm cusp}^{(n)}$ for $n \le 2$ and $\gamma_F^{(1)}$. For N$^k$LL accuracy, one needs $\Gamma_{\rm cusp}^{(n)}$ for $n \le k+1$, $\gamma_F^{(n)}$ for $n \le k$ and boundary condition $F^{(n)}$ with $n \le k-1$. The numerical values for $\Gamma_{\rm cusp}^{(1,2)}$ and $\gamma_F^{(1)}$, which are required for NLL resummation are given in Appendix~\ref{app:g_functions}. 
\begin{table}
\begin{center}
\begin{tabular}{|l|c|c|c|}
\hline
 &   $\Gamma_{\rm cusp}[\alpha_s]$ & $\gamma_F[\alpha_s]$ & $F(\mu_F)$ 
 \\\hline\hline
LL & 1 & -- & --
\\ \hline
NLL & 2 & 1 & -- 
\\ \hline
NNLL & 3 & 2 & 1 
\\ \hline\hline
N$^k$LL & k+1 & k & k-1
\\ \hline\end{tabular}
\end{center}
\caption{The loop order at which the various pieces ingredients to the RGE need to be computed to reach a given level in resummation accuracy.
\label{tab:LogCountingSCET}}
\end{table}

SCET and resummation based on factorization theorems in general is extremely powerful. Since higher logarithmic accuracy is achieved simply by computing anomalous dimensions and boundary conditions at higher perturbative accuracy, progress in our ability to perform fixed order calculations directly leads to higher logarithmic resummation, and some of the highest logarithmic accuracy has been achieved for several observables using this approach. The main drawback is that only observables for which a factorization theorem is known can be resummed using this approach. Deriving such a factorization theorem is often quite complicated, and for many observables it is not known.

\subsection{NLL resummation for thrust}
\label{sec:SCET_NLLThrust}
In this section we give the result for the thrust distribution at NLL accuracy, repeating the example of Section~\ref{sec:reviewCAESAR}.

\FIGURE[h]{
 \centering
  \includegraphics[scale=0.5]{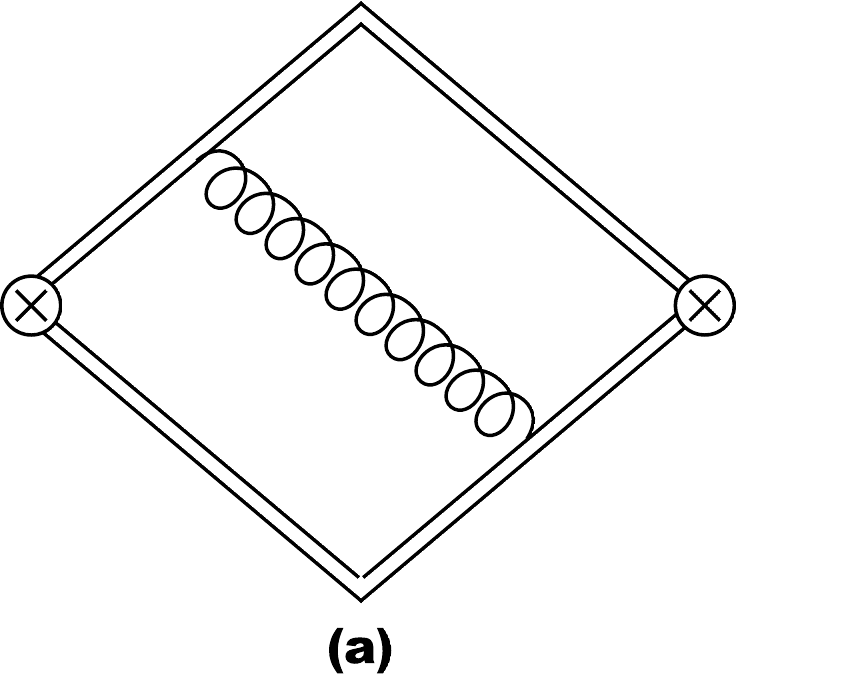}
  \includegraphics[scale=0.5]{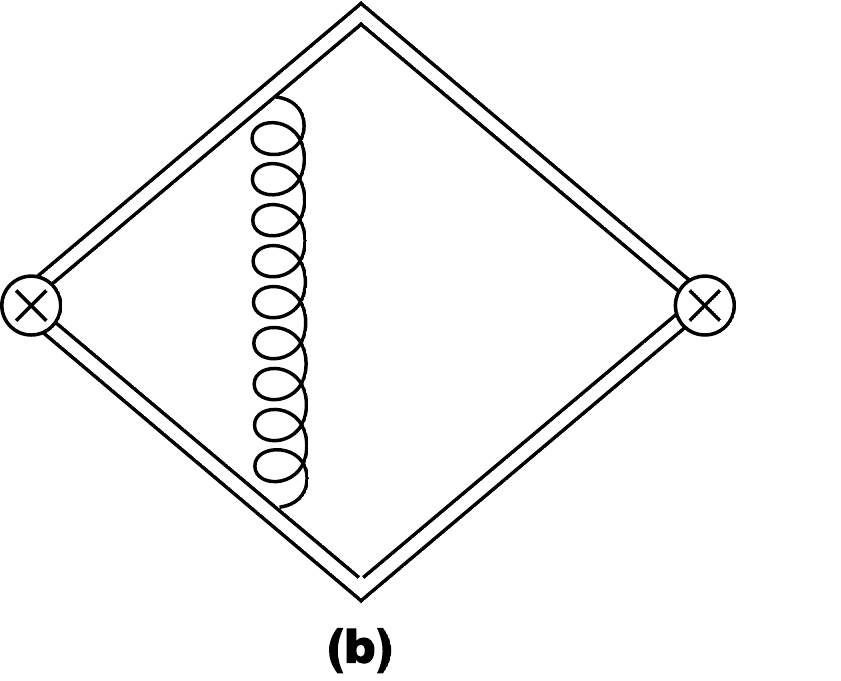}
  \caption{\label{fig:softfunction}%
Diagrams contributing to the one-loop soft function}}

\FIGURE[h]{
 \centering
   \includegraphics[scale=0.7]{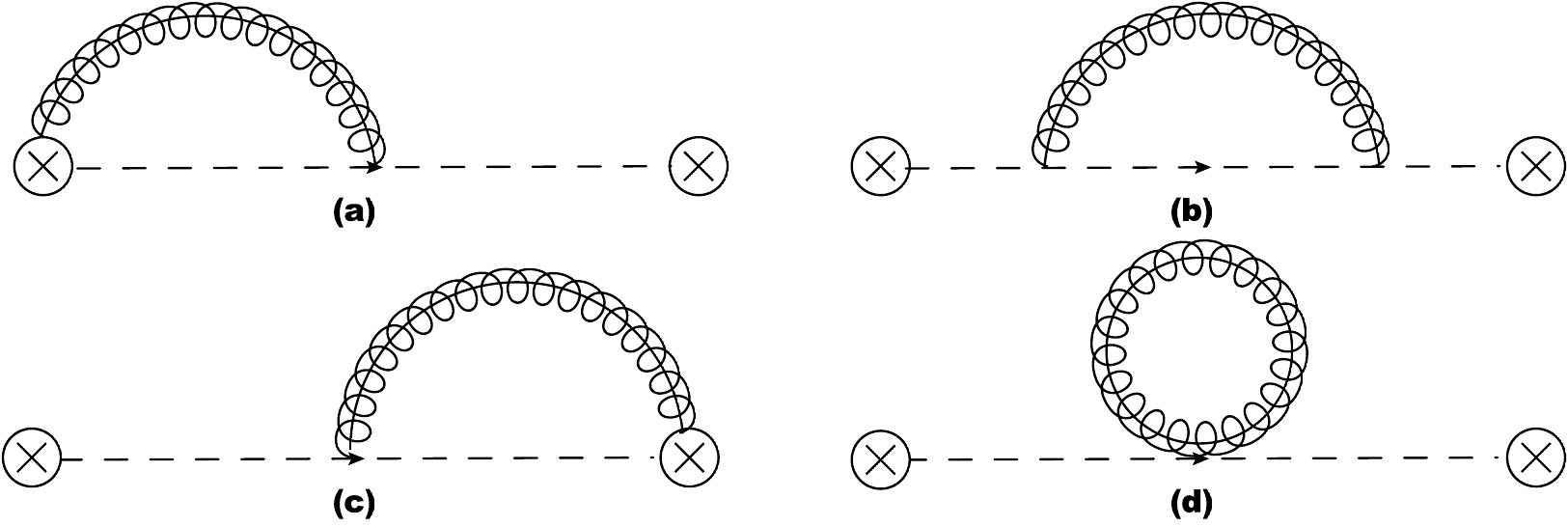}
  \caption{\label{fig:jetfunction}%
Diagrams contributing to the one-loop jet functions}}

We will perform all required calculations at 1-loop order, but will include the 2-loop cusp anomalous dimension when giving the final result. We parametrize the generic momentum $q$ as 
\begin{equation}
q^\mu = \frac{q\cdot n}{2}\bn^\mu + \frac{q\cdot \bn}{2} n^\mu + q^\mu_\perp = \frac{q^+}{2}\bn^\mu + \frac{q^-}{2} n^\mu + q^\mu_\perp
\,,
\end{equation}
and define
\begin{equation}
\df ^d k = \frac{1}{2} \df k^+ \df k^-\df^{d-2}k_\perp
\,.
\end{equation}
In the following we also use
\begin{equation}
\df^{d-2}k_\perp \delta(k^2)  = \frac{\pi^{1-\epsilon}}{\Gamma(1-\epsilon)} (k^+k^-)^{-\epsilon}\Theta(k^-) \Theta(k^+)
\,.
\end{equation}

We start with the computation of the soft function. The diagrams contributing to the one-loop corrections are reported in Figure~\ref{fig:softfunction}. The virtual correction is given by a scaleless integral, and hence vanishes in dimensional regularization, allowing us to set $\epsilon_{\rm IR}=\epsilon_{\rm UV}=\epsilon$. Conversely, the real correction (plus its conjugate) is obtained by cutting the gluon propagator. Using the Feynman rules given in Fig~\ref{Fig:Feynman} this gives
\begin{align}
\label{eq:SCET_Soft_bare}
S_{\rm bare}(\tau_s;\mu) &= \delta(\tau_s)+
2 g_s^2 \mu^{2\epsilon} C_F \, n \mcdot \bn\, Q \int \! \frac{{\rm d}^d k}{(2\pi)^d} \,(2 \pi) \delta(k^2) \,  \frac{1}{n \cdot k}\frac{1}{{\bar n} \cdot k}  \, \delta\left[{\rm min}(k^+, k^-) - Q \tau_s\right]
\nn
&=\delta(\tau_s)+C_F \, \frac{\alpha_s}{\pi} \, Q (\tau_s Q)^{-1-\epsilon} \mu^{2\epsilon} \, \frac{(4\pi)^\epsilon}{\Gamma(1-\epsilon)}\left[\int_{\tau_s Q}^{\infty} \frac{\df k^-}{(k^-)^{1+\epsilon}}  + \{k^-\to k^+\}\right]
\nn
&=\delta(\tau_s)+ 2 \, C_F \, \frac{\alpha_s}{\pi} \left(\frac{\mu}{Q}\right)^{2\epsilon} (\tau_s)^{-1-2\epsilon} \, \frac{(4\pi)^\epsilon}{\Gamma(1-\epsilon)} \, \frac{1}{\epsilon} 
\,.\end{align}
After renormalizing the coupling in the $\overline{\rm MS}$ scheme ($\alpha_s (4 \pi)^\epsilon\to \alpha_s(\mu) e^{\gamma_E \epsilon}$) we take the Laplace transform of the result and expand it in $\alpha_s(\mu)$ to obtain
\begin{align}
\tilde{S}_{\rm bare}(u;\mu) &= 1+ C_F \, \frac{\alpha_s(\mu)}{\pi} \left[ -\frac{1}{\epsilon^2} +\frac{2}{\epsilon}\ln\frac{Q\,u_0}{\mu\, u} - 2 \ln^2\frac{Q\,u_0}{\mu \,u} - \frac{\pi^2}{4}\right]
\,,
\end{align}
where $u_0=e^{-\gamma_E}$. 
One can see that this soft function does not contain any logarithms at the characteristic scale
\begin{align}
\label{eq:mus_laplace}
\mu_S = \frac{Q \, u_0}{u}
\,,
\end{align}
which corresponds to $\mu_S = Q \tau$ in thrust space. 

Next, we compute the jet function along the $n$ direction (analogous considerations apply to $J_\bn$), whose one-loop corrections are given by the diagrams of Figure~\ref{fig:jetfunction}.
Virtual corrections are again given by scaleless integrals, so the only non-vanishing contribution is obtained by cutting through the loop in the diagrams of Figure~\ref{fig:jetfunction}. The sum of the diagrams ${\rm (a)}$ and ${\rm (c)}$ can be obtained by using the SCET Feynman rules reported in Fig.~\ref{Fig:Feynman}. We obtain (using $l^-\simeq Q$)
\begin{align}
\label{eq:a+c}
J_{n \,{\rm bare}}^{(a)+(c)}(\tau_n;\mu) &= 2 g_s^2\mu^{2\epsilon} C_F n\cdot\bn \, Q \int \df l^+\int \frac{\df k^+ \df k^- \df^{d-2}k_\perp}{2(2\pi)^{d-1}}\frac{1}{l^+}\frac{Q - k^-}{k^-}\delta(k^2)\Theta[k^->0]
\nn
&\qquad \times\delta\left[(Q-k^-)(l^+-k^+)-k_\perp^2\right]\Theta(Q>k^-)\delta(l^+-\tau_n Q)
\nn
&=C_F\,\frac{\alpha_s}{\pi} (4 \pi)^\epsilon\left(\frac{\mu^2}{Q^2}\right)^{\epsilon} \tau_n^{-1-\epsilon} \frac{1}{\Gamma(1-\epsilon)}\int_0^1 \! \df x \, (1-x)^{1-\epsilon} \, x^{-1-\epsilon}
\nn
&=C_F\,\frac{\alpha_s}{\pi}(4 \pi)^\epsilon\left(\frac{\mu^2}{Q^2}\right)^{\epsilon} \tau_n^{-1-\epsilon} \frac{\Gamma(2-\epsilon)\Gamma(-\epsilon)}{\Gamma(2-2\epsilon) \Gamma(1-\epsilon)}
\,,
\end{align}
where we have defined $x = k^- / Q$. The calculation of the remaining two diagrams (${\rm (b)}$ and ${\rm (d)}$ in Figure~\ref{fig:jetfunction}) can be simplified further by noticing that their sum is related to the QCD wave function~\cite{Bauer:2000yr} as follows\\
\includegraphics[scale=0.29]{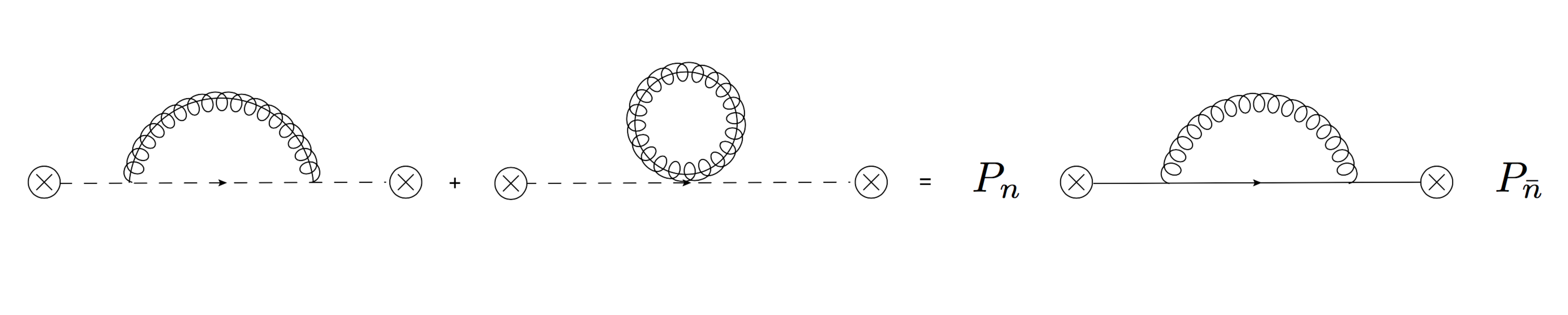}
\\
where the projectors $P_n$ and $P_\bn$ read
\begin{equation}
P_n=\frac{\slashed{n}\slashed{\bar n}}{4}
\,,\qquad P_\bn=\frac{\slashed{\bn}\slashed{ n}}{4}
\,.
\end{equation}
The result therefore reads
\begin{align}
\label{eq:b+d}
J_{n \,{\rm bare}}^{(b)+(d)}(\tau_n;\mu) &= g_s^2\mu^{2\epsilon} C_F \, n\cdot\bn \, \frac{d-2}{2} \, Q\int \frac{\df l^+}{l_+^2}\int \frac{\df k^+ \df k^- \df^{d-2}k_\perp}{2(2\pi)^{d-1}}\, (l^+-k^+) \delta(k^2)\Theta[k^->0]
\nn
&\qquad \times\delta\left[(Q-k^-)(l^+-k^+)-k_\perp^2\right]\Theta(Q>k^-)\, \delta(l^+-\tau_n Q)
\nn
&=C_F\,\frac{\alpha_s}{2\pi} (4 \pi)^\epsilon \left(\frac{\mu^2}{Q^2}\right)^{\epsilon} \tau_n^{-1-\epsilon}\frac{1}{\Gamma(1-\epsilon)} \frac{d-2}{2} \int_0^1 \! \df x \, x^{1-\epsilon}(1-x)^{-\epsilon}
\nn
&=C_F\,\frac{\alpha_s}{\pi} (4 \pi)^\epsilon \left(\frac{\mu^2}{Q^2}\right)^{\epsilon} \tau_n^{-1-\epsilon} \frac{1-\epsilon}{2} \frac{\Gamma(1-\epsilon)\Gamma(2-\epsilon)}{\Gamma(3-2 \epsilon) \Gamma(1-\epsilon)}
\,.
\end{align}

Since the integrals above include a contribution where the momentum $k^\mu$ becomes soft (and these effects have already been included in the soft function), this soft contribution needs to be subtracted. In SCET this procedure is called zero-bin subtraction~\cite{Manohar:2006nz}, but in this case is given by scaleless integrals and hence vanishes. 
Combining Eqs.~\eqref{eq:a+c} and~\eqref{eq:b+d} (after the usual $\overline{\rm MS}$ renormalization), performing the Laplace transform and expanding the result in $\alpha_s(\mu)$ one finds
\begin{align}
\label{eq:Jnbare}
\tilde{J}_{n \,{\rm bare}}(u;\mu) &= 1 + C_F\frac{\alpha_s(\mu)}{\pi}\left[ \frac{1}{\epsilon^2}+\frac{\frac{3}{4}+\ln\frac{\mu^2u}{Q^2u_0}}{\epsilon}+ \frac{1}{4}\left(3+2\ln\frac{\mu^2u}{Q^2u_0}\right)\ln\frac{\mu^2u}{Q^2u_0}+ \frac{7}{4}-\frac{\pi^2}{6}\right]
\,.
\end{align}
One can see that the jet function does not contain any logarithmically enhanced terms at the characteristic scale
\begin{align}
\label{eq:muj_laplace}
\mu_J = \frac{Q \sqrt{u_0}}{\sqrt{u}}
\,,
\end{align}
which corresponds to $\mu_S = Q \sqrt{\tau}$ in thrust space. 

The $1/\epsilon$ divergences are of UV origin, and in Laplace space can be renormalized with a multiplicative renormalization constant as follows
 \begin{align}
\tilde{S}= Z_S^{-1}\tilde{S}_{\rm bare}
\,,\notag\\
 \tilde{J}=Z_\psi Z_J^{-1}\tilde{J}_{\rm bare}
 \,,
\end{align}
where we defined
\begin{align}
Z_S &= 1+ C_F \, \frac{\alpha_s(\mu)}{\pi} \left( -\frac{1}{\epsilon^2} +\frac{2}{\epsilon}\ln\frac{Q\,u_0}{\mu\, u}\right)
\,,\notag\\
Z_J &= 1+C_F\frac{\alpha_s(\mu)}{\pi}\left(\frac{1}{\epsilon^2} + \frac{1}{2\epsilon} + \frac{1}{\epsilon} \ln\frac{\mu^2u}{Q^2u_0} \right)
\,,
\end{align}
and $Z_\psi$ is the wave-function renormalization
\begin{equation}
Z_\psi = 1-C_F\frac{\alpha_s(\mu)}{4\pi}\frac{1}{\epsilon}
\,.
\end{equation}

By imposing the RG invariance for the bare soft and jet functions one can obtain the RGE  of the renormalized ones
\begin{align}
\frac{d \ln \tilde{S}(u;\mu)  }{d\ln\mu} &= 4 C_F \frac{\alpha_s(\mu)}{\pi}\ln\frac{Q \, u_0}{\mu \, u}
\,,
\nn
\frac{d \ln \tilde{J}_n(u;\mu)  }{d\ln\mu} &= 2 C_F \frac{\alpha_s(\mu)}{\pi}\left( \ln\frac{\mu^2u}{Q^2u_0} + \frac{3}{4}\right)
\,,
\end{align}
which agrees with Eq.~\eqref{eq:laplace_RGE_soft}, with $\Gamma_{\rm cusp}^{(1)}=2C_F$ and $\gamma_J^{(1)}=-3 C_F$. One could directly renormalize Eqs.~\eqref{eq:SCET_Soft_bare},~\eqref{eq:Jnbare} in thrust space, which contain plus distributions. In this case the resulting RGEs take the form reported in Eqs.~\eqref{eq:RGE_J} and~\eqref{eq:RGE_S}.

Using only these one-loop results, the solution to the previous RGE reads (where $F = J, S$)
\begin{align}
\label{eq:solution-RGE-Laplace_soft}
\tilde{F}(u;\mu) = \tilde{F}\left(u;\mu_F\right) \tilde U_S\left(u; \mu,\mu_F \right) 
\,,
\end{align}
where at NLL the initial conditions read
\begin{equation}
\tilde{F}\left(u;\mu_F\right) = 1 + {\cal O}(\alpha_s)
\,.
\end{equation}
and
\begin{align}
\tilde U_S(u; \mu,\mu_S) &= \exp\left\{\int_{\mu_S}^{\mu} \frac{\df\mu'}{\mu'}4 C_F \frac{\alpha_s(\mu')}{\pi} \ln\frac{\mu_S}{\mu'}\right\}
\nn
\tilde U_J(u; \mu,\mu_0) &= \exp\left\{\int_{\mu_J}^{\mu} \frac{\df\mu'}{\mu'}2 C_F \frac{\alpha_s(\mu')}{\pi}\left( \ln\frac{\mu'^2}{\mu_J^2} + \frac{3}{4}\right) \right\}
\,.
\end{align}

Now Eq.~\eqref{eq:solution-RGE-Laplace_soft} can be inverted to thrust space. One can decide to set the scales as in Eqs.~\eqref{eq:mus_laplace},~\eqref{eq:muj_laplace} and perform the Laplace transform or, rather, to first perform the inverse Laplace transform with symbolic $\mu_S$ and $\mu_J$ and then set the scales to $\mu_S=\tau_s Q$,  $\mu_J=\sqrt{\tau_n} Q$ directly in thrust space. The difference between the two procedures is subleading, therefore we choose the latter which yields
\begin{align}
\label{eq:solution-RGE-thrust_soft}
S(\tau_s;\mu) = &\exp\left\{\int_{\mu_S}^{\mu} \frac{\df\mu}{\mu}4 C_F \frac{\alpha_s(\mu)}{\pi} \ln\frac{\mu_S}{\mu}\right\}\left[\frac{1}{\tau_s Q}\left(\frac{\tau_sQ}{\mu_S}\right)^{\eta_S}\frac{e^{-\gamma_E \eta_S}}{\Gamma(\eta_S)}\right]
\nn
J_n(\tau_n;\mu) = &\exp\left\{\int_{\mu_J}^{\mu} \frac{\df\mu}{\mu}C_F \frac{\alpha_s(\mu)}{\pi}\left( 2\ln\frac{\mu^2}{\mu_J^2} + \frac{3}{2}\right) \right\}\left[\frac{1}{\tau_n Q^2}\left(\frac{\tau_nQ^2}{\mu^2_J}\right)^{\eta_J}\frac{e^{-\gamma_E \eta_J}}{\Gamma(\eta_J)}\right],
\end{align}
where $\mu_S=\tau_s Q$,  $\mu_J=\sqrt{\tau_n} Q$ and 
\begin{align}
\eta_j=-\frac{\eta(\mu,\mu_J)}{2}\,, \qquad \eta_s=\eta(\mu,\mu_S)
\,,
\end{align}
with
\begin{equation}
\eta(\mu,\mu_F) = \int_{\mu_F}^{\mu} \frac{\df\mu'}{\mu'}4 C_F \frac{\alpha_s(\mu')}{\pi}
\,.
\end{equation}

Combining all results together, setting the common renormalization scale to $\mu = \mu_H = Q$, such that the hard function contains no logarithmically enhanced terms and to NLL order can be set to unity, and including the 2-loop cusp anomalous dimension, one obtains
\begin{align}
\label{eq:SigmaNLLResummedSCET}
  \Sigma_{\rm NLL}(\tau) &= \exp\left\{\int_{\sqrt{\tau}Q}^{Q} \frac{\df\mu}{\mu}\left( 4\Gamma_{\rm cusp}[\alpha_s(\mu)] \ln\frac{\mu^2}{\tau Q^2} - 4\gamma_J[\alpha_s(\mu)]\right) \right\}\notag\\
&\qquad \times\exp\left\{\int_{\tau Q}^{Q} \frac{\df\mu}{\mu}4 \Gamma_{\rm cusp}[\alpha_s(\mu)] \ln\frac{\tau Q}{\mu}\right\}\frac{e^{-\gamma_E (2\eta_J+\eta_S)}}{\Gamma(1+(2\eta_J+\eta_S))}
\,,
\end{align}
where the expressions for the anomalous dimensions are reported in Appendix~\ref{app:g_functions}.
After evaluating the integrals in the exponent, and neglecting terms beyond NLL, one finds
\begin{align}
\label{eq:SigmaResummedNLL}
 \Sigma_{\rm NLL}(\tau) =\exp\left\{ L g_1(\alpha_s L) + g_2(\alpha_s L)\right\}\frac{e^{-\gamma_E (2\eta_j+\eta_s)}}{\Gamma(1+(2\eta_j+\eta_s))}
 \,,
\end{align}
where the functions $g_i$ are reported in Appendix~\ref{app:g_functions}.
One can easily show that the above equation is equivalent to the QCD result of Eq~\eqref{eq:Sigma_NLL_thrust_CAESAR} by writing
\begin{align}
2\eta_J+\eta_S = \int_{\tau Q}^{\sqrt{\tau}Q}\frac{\df\mu}{\mu}4 C_F\frac{\alpha_s(\mu)}{\pi}
\,,
\end{align}
which is equal to $R'_{\rm LL}(\Phi_B; v)$ given in~\eqref{eq:rp_explicit}
\begin{align}
R'_{\rm LL}(\Phi_B; v)  &= \int \frac{\df k_t}{k_t}\int_0^{\ln\frac{Q}{k_t}} \df\eta\, \frac{\df\phi}{2\pi}4 C_F \frac{\alpha_s(k_t)}{\pi}\delta\left[\ln (k_t/Q) - \eta - \ln(\tau)\right] 
\nn
& = \int_{\tau Q}^{\sqrt{\tau}Q} \frac{\df k_t}{k_t}4 C_F \frac{\alpha_s(k_t)}{\pi}
\,.\end{align}

Before moving on, we report the result for $\Sigma_{\max}(\tau)$, which enters as an ingredient of the decomposition that will be used in Section~\ref{sec:automatedSCET}. The simple observable used to define $\Sigma_{\max}(\tau)$ is such that its UV divergences can be renormalized in a multiplicative way in thrust space, that is, the corresponding factorization theorem is multiplicative (see Eq.~\eqref{eq:SigmaMaxSCET}). To the order we are working, the resulting soft and jet functions are trivially obtained from the Laplace space results reported above by simply evaluating them directly in thrust space, i.e.
\begin{align}
\label{eq:thrustLaplaceRel_noLambda}
S^{\max}(\tau; \mu) &= \tilde S(u=u_0 / \tau; \mu)
\nn
J^{\max}(\tau; \mu) &= \tilde J(u=u_0 / \tau; \mu)
\,.
\end{align}
This gives
\begin{align}
  \Sigma_{\max}^{\rm NLL}(\tau) &= \exp\left\{\int_{\sqrt{\tau}Q}^{Q} \frac{\df\mu}{\mu}\left( 4\Gamma_{\rm cusp}[\alpha_s(\mu)] \ln\frac{\mu^2}{\tau Q^2} - 4\gamma_J[\alpha_s(\mu)]\right) \right\}\notag\\
&\qquad \times\exp\left\{\int_{\tau Q}^{Q} \frac{\df\mu}{\mu}4 \Gamma_{\rm cusp}[\alpha_s(\mu)] \ln\frac{\tau Q}{\mu}\right\}
\,.
\end{align}

At higher orders the initial conditions in Laplace space are different than they are in thrust space, such that the \eq{thrustLaplaceRel} is no longer exactly correct. To obtain the correct expression requires to perform the calculation of $S^{\max}$ and $J^{\max}$ directly in thrust space according to the factorization theorem~\eqref{eq:SigmaMaxSCET}.

\subsection{Neglecting subleading logarithmic effects}
\label{sec:SCET_subleading}
The exact definition of the logarithmic order in resummation is
somewhat convention dependent, and different prescriptions can be
found in the literature. The prescription given in the previous
section in Eq.~\eqref{eq:SigmaResummedNLL} includes in fact various
subleading logarithmic terms. For example, the cusp anomalous
dimensions at 2-loop order is only required for the contribution in
the first line of \eqs{RGE_J}{RGE_S}, while in the second line it is enough
to include the cusp anomalous dimension at 1-loop order. This implies
that, instead of using the full expression for
$\eta \equiv 2 \eta_j + \eta_s$ in the term
$e^{-\gamma_E \eta} / \Gamma(1+\eta)$, one can perform a Taylor
expansion of this result. For example, to NNLL accuracy one has
\begin{align}
\frac{e^{-\gamma_E \eta_{\rm NNLL}} }{ \Gamma(1+\eta_{\rm NNLL})} = \frac{e^{-\gamma_E \eta_{\rm NLL}} }{ \Gamma(1+\eta_{\rm NLL})} + \frac{\eta_{\rm NNLL} - \eta_{\rm NLL} }{\eta_{\rm NLL}} \frac{\df}{\df \eta_{\rm NLL}}\frac{e^{-\gamma_E \eta_{\rm NLL}} }{ \Gamma(1+\eta_{\rm NLL})} + \ldots
\end{align}
where
\begin{align}
\eta_{\rm NLL} &= 4  \int_{\tau Q}^{\sqrt{\tau}Q}\frac{\df\mu}{\mu} \left[ \frac{\alpha_s(\mu)}{2\pi}\Gamma^{(1)}_{\rm cusp}  \right]
\nn
\eta_{\rm NNLL} &= 4  \int_{\tau Q}^{\sqrt{\tau}Q}\frac{\df\mu}{\mu} \left[ \frac{\alpha_s(\mu)}{2\pi}\Gamma^{(1)}_{\rm cusp} + \left(\frac{\alpha_s(\mu)}{2\pi}\right)^2\Gamma^{(2)}_{\rm cusp}\right]
\,.\end{align}

Also, in general one finds differences depending on how the RG equations are solved. As already mentioned, performing resummation to a given order in Laplace space and then inverting the Laplace transform, gives results that differ beyond the order one is working compared to solving the RG equations directly in thrust space. A second example is that resumming the thrust distribution $\df \sigma / \df \tau'$ (by setting the scales to the characteristic scales of the distribution) and then computing $\Sigma(\tau)$ by integrating over $0 < \tau' < \tau$ yields results that again differ at higher logarithmic order from those obtained by directly resumming the distribution $\Sigma(\tau)$. For a detailed discussion of differences in logarithmic counting, see~\cite{Almeida:2014uva}.

This existence of different conventions needs to be kept in mind in the next section when comparing the results obtained from an automated SCET resummation with the analytical results. In particular, a consistent comparison between different approaches can be only performed up to formally subleading terms.

\section{Automated resummation in SCET}
\label{sec:automatedSCET}
The starting equation for the automated resummation in Section~\ref{sec:reviewCAESAR} was the separation of the desired cross section $\Sigma(v)$ into the product of the simplified cross section $\Sigma_{\max}(v)$ and the transfer function $\fullF(v)$ given in \eq{fullFDef}. The resummation of the simplified observable was computed analytically, while the transfer function could be obtained numerically. In this section we derive a similar result, but where all ingredients are defined within SCET.

To simplify the discussion, we consider here a factorizable observable (such as thrust) and perform
a similar decomposition at the level of the individual soft and jet functions. The SCET factorization theorem~\eqref{eq:SigmaAddSCET} for the thrust event shape that can be recast as (note that we drop the $\Phi_B$ dependence from now on)
\begin{align}
\label{eq:scet-F}
\Sigma(\tau) =  H\int \! \df \tau_n \, \Sigma'_{J_n}(\tau_n, \mu) \int \! \df \tau_\bn \, \Sigma'_{J_\bn}(\tau_\bn, \mu)\int \! \df \tau_s\, \Sigma'_{S}(\tau_s, \mu)\,\, \Theta[\tau > \tau_n + \tau_{\bar n} + \tau_s],
\end{align}
where we expressed the soft and jet functions as (with $F = S, J_n, J_\bn$)
\begin{align}
F(\tau_F, \mu) \equiv \Sigma'_{F}(\tau_F, \mu) = \frac{\df\Sigma_{F}(\tau_F)}{\df\tau_F}
\,.
\end{align}
Next, we define
\begin{align}
\label{eq:MC-soft-jet}
\Sigma_{F}(\tau_F, \mu) &\equiv \Sigma_{F}^{\rm max}(\tau, \mu)\fullF_F(\tau_F, \tau, \mu)
\,.
\end{align}
with
\begin{align}
\label{eq:transfer-soft-jet}
\fullF_{F}(\tau_F, \tau, \mu) &= \frac{\Sigma_{F}^{\rm max}(\eps\tau, \mu)}{\Sigma_{F}^{\rm max}(\tau, \mu)} \frac{\Sigma_{F}(\tau_F, \mu)}{\Sigma_{F}^{\rm max}(\eps\tau, \mu)}
\,.
\end{align}
This allows us to write
\begin{align}
\label{eq:Sigma_MC_SCET}
\Sigma(\tau) =  \Sigma_{\max}(\tau)\int \! \df \tau_n \, \fullF'_{J_n}(\tau_n, \tau, \mu) \int \! \df \tau_\bn \, \fullF'_{J_\bn}(\tau_\bn, \tau, \mu)\int \! \df \tau_s\, \fullF'_{S}(\tau_s, \tau, \mu)\,\, \Theta[\tau > \tau_n + \tau_{\bar n} + \tau_s]
\,,
\end{align}
where we defined $\fullF'_F \equiv \df \fullF'_F/\df\tau_F$ with $F = S, J_n, J_\bn$.

The goal is to compute each of the transfer functions through a MC algorithm defined uniquely in terms of either soft or collinear fields, in a way that is similar to Section~\ref{sec:reviewCAESAR}. We will show in Section~\ref{sec:MC_transfer} that in the framework of SCET one can compute each of the transfer functions $\fullF_J(\tau_n, \tau, \mu)$ and $\fullF_S(\tau_s, \tau, \mu)$ through a separate MC. This ensures that all observable dependence is restricted to the numerical MC algorithm. 

The computation of Eqs.~\eqref{eq:transfer-soft-jet} via MC methods requires that each can be obtained in 4 dimensions by recursively computing real emissions. This relies on two important facts: First, the transfer function has to be determined entirely through the real radiation, and second, each contribution needs to be finite in 4 dimensions. The first fact is trivially satisfied, since in the ratios $\Sigma_{F}(\tau)/\Sigma_{F}^{\rm max}(\eps\tau)$ the purely virtual corrections cancel exactly. The second requirement deserves some closer investigation.

The IRC divergences cancel quite trivially in the ratio $\Sigma_{F}(\tau)/\Sigma_{F}^{\rm max}(\eps\tau)$, since the numerator and denominator include the same unresolved real radiation (for rIRC safe observables). However, as we discussed in Section \ref{sec:SCET_Overview_RGE} and contrary to full QCD, in the standard formulation of SCET real radiation is UV divergent. 
The resulting UV divergences of the real radiation appear both in the soft and in the jet functions and they cancel entirely only in their combination to give the physical cross section. The existence of the above divergences is a feature of the effective theory formulation in which the UV bounds of the theory are completely integrated out into Wilson coefficients. This guarantees that each of the soft and jet functions only depends on a single characteristic scale, which allows for the resummation of the dominant logarithmic terms via RG equations.  In the usual formulation of SCET the UV divergences from the real radiation are regulated using dimensional regularization, and they constitute a crucial contribution to the anomalous dimensions which resummation is based on. However, the presence of the additional UV divergences prohibits a MC formulation of the problem that requires the phase-space integrals of the real radiation to be computable in 4 dimensions.

We solve this problem by introducing an explicit UV regulator for real phase space integrals into SCET. In this formulation of SCET, the UV divergences from virtual diagrams are regulated in dimensional regularization, just as before, while those from the real radiation are regulated with an alternative regulator, which can be chosen to be either a physical cutoff or an analytic regulator. This will give rise to a different RG structure in SCET, resulting in different logarithmic structures for the soft and jet functions individually. However, when soft and jet functions are combined into physical observables, one reproduces the same result as in the standard SCET formulation. By introducing such a regulator, we make sure that the UV divergences in the real radiation are now regulated in 4 dimensions, hence allowing for a formulation of the resummation through a MC algorithm.

In Section~\ref{sec:SCETNoUV} we discuss the standard resummation in SCET in the presence of this new UV regulator for real-emission phase space integrals. We perform an explicit computation of the relevant soft and jet functions at one loop, and we show how the resummation can be performed through RG evolution. In Section~\ref{sec:MC_transfer} we show how to formulate a MC solution to the corresponding RG equations. We briefly comment on the extension to other observables in the conclusions, while the detailed generalization will be treated in a future publication. An alternative interpretation of the results presented in this section in the context of SCET is reported in Appendix~\ref{app:SCET_w_delta}, where we comment on the structure of the theory when a IRC regulator $\eps \tau$ is included.

\subsection{SCET with a UV regulator for real radiation}
\label{sec:SCETNoUV}

In this section we perform the calculation of the one-loop soft and jet functions by using an additional UV regulator for the phase-space integrals of the real radiation. This can be compared directly with Section~\ref{sec:SCET_NLLThrust}, where the same calculations were performed without the additional UV regulator. The infrared and collinear divergences, as well as the UV divergences of the virtual corrections, are regularized by conventional dimensional regularization as before. One has some freedom in choosing the form of the UV regulator. In what follows we employ a cutoff $\Lambda$ on the light-cone components of the emissions' momenta which is assumed to be larger than any other scale in the problem, which implies
\begin{align}
\Lambda \ge Q
\,.\end{align}
This mimics what happens in the full theory where the upper bound is set by the center-of-mass energy of the reaction. As a cross check, we have performed the calculations shown below using the exponential regulator proposed in Ref.~\cite{Li:2016axz}, and found analogous results.

As we will see, introducing this new regulator moves UV divergences between the soft and the jet functions, but of course does not affect the result after soft and jet functions have been combined into the total cross section. Since the UV divergences determine the RG equations, and therefore the logarithmic structure, this also implies that logarithmic contributions are moved between the soft and the jet functions. That is of course not a problem, since the separation into the logarithms of contributions from the various ingredients of the factorization theorem is to some extent arbitrary. Even in standard SCET one can move contributions between the different ingredients by changing the choice of the common renormalization scale $\mu$. 

\subsubsection{The soft and jet functions at one loop}
\label{sec:SoftJetOneLoopLambda}
Consider the soft function of the factorization theorem given in \eq{SigmaAddSCET} or  \eq{SigmaMaxSCET}. As before, the virtual contribution (plus its conjugate) is given by
\begin{align}
\label{eq:QCD_SoftV}
S_{\rm bare}^{\rm (V)}(\tau_s;\mu) &= - 2 g_s^2 C_F n \cdot \bar n \mu^{2\epsilon}\int \! \frac{{\rm d}^d k}{(2\pi)^d} \, \frac{1}{n \cdot k}\frac{1}{{\bar n} \cdot k}\frac{1}{k^2} \, \delta(\tau_s)= 0
\,.
\end{align}
This integral is scaleless and therefore vanishes, hence setting $\epsilon_{\rm UV} = \epsilon_{\rm IR}$. 

The real contribution to the soft function is obtained by cutting the gluon propagator and imposing that the contribution to thrust from the real emission is smaller than $\tau_s$. This gives (remember that we impose $k^+,k^- < \Lambda$)
\begin{align}
\label{eq:QCD_Soft_R0}
S_{\rm bare}(\tau_s;\mu,\Lambda) &= \delta(\tau_s)+
2 g_s^2 \mu^{2\epsilon} C_F \, n \mcdot \bn\, Q \int \! \frac{{\rm d}^d k}{(2\pi)^d} \,(2 \pi) \delta(k^2) \,  \frac{1}{n \cdot k}\frac{1}{{\bar n} \cdot k}  \, \delta\left({\rm min}(k^+, k^-) - Q \tau_s\right) 
\nn
&=\delta(\tau_s)+C_F \, \frac{\alpha_s}{\pi} \, Q (\tau_s Q)^{-1-\epsilon} \mu^{2\epsilon} \, \frac{(4\pi)^\epsilon}{\Gamma(1-\epsilon)}\left[\int_{\tau_s Q}^{\Lambda} \frac{\df k^-}{(k^-)^{1+\epsilon}}  + \{k^-\to k^+\}\right]
\nn
&=\delta(\tau_s)+ 2 \, C_F \, \frac{\alpha_s}{\pi} \left(\frac{\mu}{Q}\right)^{2\epsilon} (\tau_s)^{-1-\epsilon} \left[(\tau_s)^{-\epsilon} - \left(\frac{\Lambda}{Q}\right)^{-\epsilon}\right] \, \frac{(4\pi)^\epsilon}{\Gamma(1-\epsilon)} \, \frac{1}{\epsilon} 
\,.
\end{align}
After renormalization, we take the Laplace transform and expand in $\alpha_s(\mu)$. We obtain
\begin{align}
\label{eq:QCD_Soft_R0_Laplace}
\tilde S_{\rm bare}(u;\mu,\Lambda) &= 1+ C_F \frac{\alpha_s(\mu)}{\pi} \left[ \frac{1}{\epsilon^2} + 2\left(\ln\frac{Q}{\Lambda} + \ln\frac{\mu}{Q}\right)\frac{1}{\epsilon} - 2 \ln\frac{Q}{\Lambda} \ln\frac{u_0}{u} - \ln^2\frac{u_0}{u}   \right.\nn
&\qquad
\left. + \left(-\frac{\pi^2}{4} +  \ln^2\frac{Q}{\Lambda} + 4 \ln\frac{Q}{\Lambda} \ln\frac{\mu}{Q} + 2 \ln^2\frac{\mu}{Q}\right)\right]
\nn
&= 1+ C_F \frac{\alpha_s(\mu)}{\pi} \left[ \frac{1}{\epsilon^2} + \frac{2}{\epsilon} \ln\frac{\mu}{\Lambda} -  \ln^2\frac{Q\, u_0}{\Lambda \, u} + 2 \ln^2\frac{\mu}{\Lambda} - \frac{\pi^2}{4}\right]
\,.
\end{align}
From the above expression one can see that the soft function does not contain any logarithmically enhanced terms at the characteristic scales
\begin{align}
\label{eq:soft_characteristic_Lambda}
\mu_S = \Lambda_S = \frac{Q \, u_0}{u}
\,.
\end{align}

Next, we consider the jet function along the direction $n^\mu$. The virtual contribution is again scaleless (also for the zero-bin subtraction) and thus vanishes, so the only non-zero contribution is obtained from the real radiation. The collinear diagrams are unaffected by the extra UV regulator, since their integrals are cut off by the scale $Q$. Thus, we find the same result as in \eq{Jnbare}, which we repeat here for convenience
\begin{align}
\tilde{J}_{n \,{\rm bare}}^{\rm coll}(u;\mu) &= 1 + C_F\frac{\alpha_s(\mu)}{\pi}\left[ \frac{1}{\epsilon^2}+\frac{\frac{3}{4}+\ln\frac{\mu^2u}{Q^2u_0}}{\epsilon}+ \frac{1}{4}\left(3+2\ln\frac{\mu^2u}{Q^2u_0}\right)\ln\frac{\mu^2u}{Q^2u_0}+ \frac{7}{4}-\frac{\pi^2}{6}\right]
\,.
\end{align}
However, unlike in common dimensional regularization, in the zero-bin subtraction corresponding to the diagrams ${\rm (a)}$ and ${\rm (c)}$ of Figure~\ref{fig:jetfunction} (obtained by taking the limit $k\ll l$) the $k^-$ component is cut off by $\Lambda$, hence giving
\begin{align}
\label{eq:JR0bin}
J_{n \,{\rm bare}}^{\rm (0-bin)}(\tau_n;\mu,\Lambda) &= 2 g_s^2\mu^{2\epsilon} C_F n\cdot\bn \, Q \int \df l^+\int \frac{\df k^+ \df k^- \df^{d-2}k_\perp}{2(2\pi)^{d-1}}\frac{1}{l^+}\frac{l^-}{k^-}\delta(k^2)\Theta[k^->0]
\nn
&\qquad \times\delta(l^-(l^+-k^+)-k_\perp^2)\Theta[\Lambda>k^-]\delta(l^+-\tau_n Q)
\nn
&=C_F\,\frac{\alpha_s}{\pi} (4 \pi)^\epsilon\mu^{2\epsilon} \tau_n^{-1-\epsilon} Q^{-\epsilon} \frac{1}{\Gamma(1-\epsilon)}\int_0^\Lambda \! \frac{\df k^-}{(k^-)^{1+\epsilon}} 
\nn
&=-C_F\,\frac{\alpha_s}{\pi} (4 \pi)^\epsilon\left(\frac{\mu^{2}}{Q\Lambda}\right)^\epsilon \tau_n^{-1-\epsilon} \frac{1}{\epsilon \Gamma(1-\epsilon)}
\,.\end{align}
Putting everything together and renormalizing the strong coupling, we obtain the following result for the one-loop jet function in Laplace space
\begin{align}
\label{eq:JRSE_Laplace}
\tilde J_{n \,{\rm bare}}(u;\mu,\Lambda) &= 1+ C_F \frac{\alpha_s(\mu)}{\pi} \left[ \frac{3}{4\epsilon} -\frac{1}{\epsilon}\ln\frac{Q}{\Lambda}+\frac{3}{2}\ln\frac{\mu}{Q} - \frac{1}{2}\ln^2\frac{Q}{\Lambda} \right.\\
&\qquad\left.
- 2\ln\frac{Q}{\Lambda}\ln\frac{\mu}{Q}
+\ln\frac{Q}{\Lambda}\ln\frac{u_0}{u} - \frac{3}{4}\ln\frac{u_0}{u}  + \frac{7}{4}-\frac{\pi^2}{6} \right]
\nn
& = 1+ C_F \frac{\alpha_s(\mu)}{\pi} \left[ \frac{3}{4\epsilon} -\frac{1}{\epsilon}\ln\frac{Q}{\Lambda} +\ln\frac{\mu^2 \, u}{Q^2 \, u_0} \left( \frac{3}{4} - \ln\frac{Q}{\Lambda}\right)
- \frac{1}{2}\ln^2\frac{Q}{\Lambda} 
+ \frac{7}{4}-\frac{\pi^2}{6}\right]
\notag
\,.\end{align}
The jet function does not contain any logarithmically enhanced terms at the characteristic scales
\begin{align}
\label{eq:jet_characteristic_Lambda}
\mu_J = \frac{Q \sqrt{u_0}}{\sqrt{u}}\, \qquad \Lambda_J = Q
\,.
\end{align}

By combining the soft function~\eq{QCD_Soft_R0_Laplace} and two jet functions (\eq{JRSE_Laplace} plus the analogous contribution for the direction $\bn^\mu$) one sees that the dependence on the cutoff $\Lambda$ cancels, and that the result coincides with the usual SCET result obtained in pure dimensional regularization. The new regularization scheme that we have introduced, therefore, only changes the expression of the soft and jet function while leaving their combination in the physical cross section unchanged. 

One can now proceed to write the RG equations for the soft and jet function. Since there are now two scales characterizing the UV structure of the theory, one needs to write two separate evolution equations for each subprocess, the first of which describes the evolution in the dimensional regularization scale $\mu$ and the second one describes the dependence on the UV cutoff $\Lambda$. This is in spirit similar to what happens in SCET$_{\rm II}$ problems~\cite{Bauer:2002aj} where a rapidity regulator is introduced to regularize the additional UV divergence of the real radiation~\cite{Manohar:2006nz,Chiu:2007dg,Becher:2011dz,Chiu:2011qc,Chiu:2012ir,Li:2016axz}. In fact, the same conclusions that follow would apply in that case.
One finds for the soft and jet functions\footnote{We have renormalized the $\Lambda$ anomalous dimension by using the fact that the derivatives in $\mu$ and $\Lambda$ commute.}
\begin{align}
\label{eq:RGE_UV_cutoff}
\frac{\df \ln \tilde{S}(u;\mu, \Lambda)  }{\df\ln\mu} &= 4 C_F \frac{\alpha_s(\mu)}{\pi}\ln\frac{\mu}{\Lambda} & \frac{\df \ln \tilde{S}(u;\mu, \Lambda)  }{\df\ln\Lambda} &= -\int^\mu_{\sqrt{\frac{u_0 Q \Lambda}{u}}} \frac{\df \mu'}{\mu'}4 C_F \frac{\alpha_s(\mu')}{\pi}
\,,
\nn
\frac{\df \ln \tilde{J}_n(u;\mu, \Lambda)  }{\df\ln\mu} &=  \left(\frac{3}{2}+2\ln\frac{\Lambda}{Q}\right) C_F \frac{\alpha_s(\mu)}{\pi}& \frac{\df \ln \tilde{J}_n(u;\mu, \Lambda)  }{\df\ln\Lambda} &= \int_{\sqrt{\frac{u_0 Q \Lambda}{u}}}^\mu \frac{\df \mu'}{\mu'}2 C_F \frac{\alpha_s(\mu')}{\pi}
\,,
\end{align}
where the lower bound of the $\Lambda$ RGE arises from the fact that the corresponding anomalous dimension vanishes at this scale, which occurs at 
\begin{equation}
\mu=\sqrt{\frac{\Lambda}{\Lambda_F}}\mu_F\,,\qquad {\rm with~}F=S,J
\,.\end{equation}
 From the system of equations~\eqref{eq:RGE_UV_cutoff} one can easily see that the $\Lambda$ dependence cancels in the combination of the soft and two jet functions at a given $\mu$.

One can now solve the RG equations by evolving the jet and soft functions simultaneously from their characteristic scales in $\mu$ and $\Lambda$ to a common scale. Since the order of taking the derivatives with respect to $\ln\mu$ and $\ln \Lambda$ commutes, this evolution is independent of the path chosen in the 2-dimensional $\mu - \Lambda$ plane. We can therefore write the evolution kernels relating the soft and jet functions at the characteristic scales $\mu_F$ and $\Lambda_F$ to the common scales $\mu$ and $\Lambda$ as
\begin{align}
\tilde U_F(u; \mu, \Lambda,  \mu_F, \Lambda_F) &= \tilde U^{(\mu)}_F(u; \mu, \mu_F; \Lambda) \, \tilde U^{(\Lambda)}_F(u; \Lambda, \Lambda_F; \mu_F)
\,,
\end{align}
with
\begin{align}
\label{eq:softKernels_Lambda}
\tilde U^{(\mu)}_S(u; \mu, \mu_S; \Lambda) &= \exp\left\{\int_{\mu_S}^{\mu} \frac{\df\mu'}{\mu'}4 C_F \frac{\alpha_s(\mu')}{\pi} \ln\frac{\mu'}{\Lambda}\right\}
\,,\\
\tilde U^{(\Lambda)}_S(u; \Lambda, \Lambda_S; \mu) &= \exp\left\{-\int_{\Lambda_S}^{\Lambda} \frac{\df\Lambda'}{\Lambda'} \int_{\sqrt{\frac{\Lambda'}{\Lambda_S}}\mu_S}^\mu \frac{\df\mu'}{\mu'}\, 4 C_F \frac{\alpha_s(\mu')}{\pi}\right\}
\,,\\
\label{eq:collKernels_Lambda}
\tilde U^{(\mu)}_J(u; \mu, \mu_J; \Lambda) &= \exp\left\{\int_{\mu_J}^{\mu} \frac{\df\mu'}{\mu'}\left(\frac{3}{2}+2\ln\frac{\Lambda}{Q}\right) C_F \frac{\alpha_s(\mu')}{\pi}\right\}
\,,\\
\tilde U^{(\Lambda)}_J(u; \Lambda, \Lambda_S; \mu) &= \exp\left\{\int_{\Lambda_J}^{\Lambda} \frac{\df\Lambda'}{\Lambda'} \int_{\sqrt{\frac{\Lambda'}{\Lambda_J}}\mu_J }^\mu \frac{\df \mu'}{\mu'}\, 2 C_F \frac{\alpha_s(\mu')}{\pi} \right\}
\,,\end{align}

From~\eq{collKernels_Lambda} we observe that the evolution of the jet function now starts at NLL. All double logarithms are entirely contained in the soft function, contrary to the case of standard SCET, where both the soft and jet function contained double logarithmic terms when evolved to the hard scale. However, as shown in Appendix~\ref{app:equivalence} once the soft and jet functions are combined into a physical cross section, the logarithmic terms in the two formulations of SCET agree to all orders in perturbation theory, and one reproduces again the result given in \eq{SigmaNLLResummedSCET}. The logarithmic structure of the soft and jet radiation in this formulation of SCET reproduce exactly the physical structure in full QCD, allowing us to establish a one-to-one correspondence between the two formulations. Notably, this makes it possible to formulate the resummation in SCET via a Monte-Carlo approach, as it will be described in the next section.

\subsection{Monte-Carlo resummation of the transfer functions at NLL}
\label{sec:MC_transfer}
Using the decomposition in \eqs{scet-F}{transfer-soft-jet}, we first consider the soft transfer function. We start from the definition of the thrust soft function as the
following expectation value
\begin{equation}
\label{eq:MC_soft-function}
S(\tau_s,\mu) = \sum_{|k\rangle}|\langle
k|Y_n\bar{Y}_\bn|0\rangle|^2\delta(\tau_s - V_{\rm soft}(p_n,p_\bn)),
\end{equation}
where $|k\rangle$ denotes a generic state with a fixed number of soft
particles, e.g. $|k\rangle = |k_1\rangle$ for a single real emission,
$|k\rangle = |k_1,k_2\rangle$ for two real emissions and so on. In Eq.~\eqref{eq:MC_soft-function}, $V_{\rm soft}(p_n,p_\bn)$ 
denotes the expression of thrust in the soft limit, and $p_n$
($p_{\bar n}$) denotes the sum of the momenta of the soft particles in
the $n$ ($\bar n$) hemisphere. Our goal is to use
Eq.~\eqref{eq:MC_soft-function} such that the soft transfer function $\fullF_S(\tau, \tau_s, \mu)$ required in~\eq{Sigma_MC_SCET} can be 
computed via a MC algorithm. 

The first ingredient to evaluate Eq.~\eqref{eq:MC-soft-jet} for the soft function is $\Sigma_{S}^{\rm max}$. To the order we are working, the result for $\Sigma^{\rm max}$ both for soft (jet) function is trivially obtained from the Laplace space results reported in the previous section by simply evaluating the soft (jet) function with the $\Lambda$ regulator directly in thrust space, i.e.
\begin{align}
\label{eq:thrustLaplaceRel}
S^{\max}(\tau; \mu, \Lambda) &= \tilde S(u=u_0 / \tau; \mu, \Lambda)
\nn
J^{\max}(\tau; \mu, \Lambda) &= \tilde J(u=u_0 / \tau; \mu, \Lambda)
\,.
\end{align}
At higher orders the initial conditions in Laplace space are different than they are in thrust space, such that the \eq{thrustLaplaceRel} is no longer exactly correct. To obtain the correct expression requires to perform the calculation of $S^{\max}$ and $J^{\max}$ directly in thrust space according to the factorization theorem~\eqref{eq:SigmaMaxSCET} with the additional UV regulator $\Lambda$. Eq.~\eqref{eq:thrustLaplaceRel} leads to
\begin{align} 
\label{eq:Sigma_max_MC}
\Sigma_{S}^{\rm max, NLL}(\tau)&=\Sigma^{\rm max, NLL}_{S}(\tau ,\mu=Q,\Lambda=Q)\nn
 &= \exp\left\{\int_{\tau Q}^{Q} \frac{\df\mu}{\mu} 4\Gamma_{\rm cusp}[\alpha_s)] \ln\frac{\mu}{Q}\right\}\exp\left\{\int_{\tau Q}^{Q} \frac{\df\Lambda}{\Lambda} \int^{\sqrt{\tau \Lambda Q} }_{\tau Q} \frac{\df\mu}{\mu}4\Gamma_{\rm cusp}[\alpha_s]\right\}
 \,.
\end{align}

We now consider the soft transfer function of \eq{transfer-soft-jet}, defined as
\begin{align}
\fullF_{S}^{\rm NLL}(\tau_s, \tau, \mu) &= \frac{\Sigma_{S}^{\rm max, LL}(\eps\tau, \mu)}{\Sigma_{S}^{\rm max, LL}(\tau, \mu)} \frac{\Sigma_{S}^{\rm LL}(\tau_s, \mu)}{\Sigma_{S}^{\rm max, LL}(\eps\tau, \mu)}
\,.
\end{align}
The first ingredient is the ratio $\Sigma_{S}^{\rm max}(\eps\tau, \mu) / \Sigma_{S}^{\rm max}(\tau, \mu)$, which is obtained using \eq{Sigma_max_MC}. Since the leading logarithms cancel in the ratio of the $\Sigma_{S}^{\rm max}(\eps\tau)$ and $\Sigma_{S}^{\rm max}(\tau)$, one needs the resummed expression only to LL, which is given by Eq.~\eqref{eq:Sigma_max_MC} where only the one-loop cusp anomalous dimension is considered.

The second ratio $\Sigma_{S}(\tau_s)/\Sigma_{S}^{\rm max}(\eps\tau)$ can be now computed numerically, as it is both IRC
and UV finite. Indeed, the UV finiteness is guaranteed by the presence
of the cutoff $\Lambda$ in the real radiation, while the IRC
finiteness is due to the rIRC safety of the observable that ensures
that the radiation below the resolution scale $\eps\tau$ cancels out
completely in the ratio. 
To achieve this, we introduce the
decomposition for the squared amplitude in Eq.~\eqref{eq:MC_soft-function}, similar to what was done in Section~\ref{sec:reviewCAESAR}
\begin{align}
\label{eq:soft_nPC}
|{M}_S(k_1)|^2 &= |\langle
k_1|Y_n\bar{Y}_\bn|0\rangle|^2,\notag\\
|{\tilde M}_S(k_1,k_2)|^2 &=  |\langle
k_1,k_2|Y_n\bar{Y}_\bn|0\rangle|^2 - 
  |M_S(k_1)|^2 |M_S(k_2)|^2 ,\notag\\
|{\tilde M}_S(k_1,k_2,k_3)|^2 &=  |\langle
k_1,k_2,k_3|Y_n\bar{Y}_\bn|0\rangle|^2 - 
|{\tilde M}_S(k_1,k_2)|^2 |M_S(k_3)|^2 - |{\tilde M}_S(k_3,k_1)|^2 |M_S(k_2)|^2 \notag\\
&\qquad -|{\tilde M}_S(k_2,k_3)|^2 |M_S(k_1)|^2 -|M_S(k_1)|^2
  |M_S(k_2)|^2 |M_S(k_3)|^2, \notag\\
& \dots
\end{align}
We recall that the reason for the above decomposition is that squared
amplitudes $\tilde M_S$ with $n$ correlated real emissions start contributing at
N$^{n-1}$LL to the evolution of the soft function for all rIRC safe
observables. Using SCET with a UV regulator for the real emissions, as discussed in Section~\ref{sec:SCETNoUV}, these are now in clear correspondence with the QCD counterparts discussed in Section~\ref{sec:reviewCAESAR}. Just as before, each of the squared amplitudes in the
r.h.s. of Eq.~\eqref{eq:soft_nPC} admits a perturbative expansion in powers
of $\alpha_s$ due to virtual corrections
\begin{equation}
\label{eq:soft_nPC-def}
|{\tilde M}_S(k_1, \dots,k_n)|^2~\equiv~ \sum_{j=0}^{\infty}\left(\frac{\alpha_s(\mu)}{2\pi}\right)^{n+j}n\mbox{PC}_S^{(j)}(k_1, \dots,k_n)
\,.
\end{equation}
The notation $n\mbox{PC}_S$ in Eq.~\eqref{eq:soft_nPC} denotes the
soft $n$-particle correlated blocks. In order to compute the transfer function to NLL accuracy, we only require the $1\mbox{PC}^{(0)}_S$ block, or in other words the 
squared amplitude $|{M}_S(k_1)|^2$ at tree level. 

Putting all this together one finds
\begin{align}
\label{eq:FS-fact}
\fullF_{S}^{\rm NLL}(\tau_s,\tau, Q)=\frac{\Sigma_{S}^{\rm max}(\eps\tau)}{\Sigma_{S}^{\rm
  max}(\tau)}\sum_{n=1}^{\infty}\frac{1}{n!}\prod_{i=1}^{n}\int_{\eps \tau}
  [\df k_i]|M_S(k_i)|^2 \Theta[\tau_s > V_{\rm soft}(k_1,\dots,k_n)]
  \,,
\end{align}
which can be evaluated with the same MC algorithm described in Section~\ref{sec:simplifiedResummation}.

Next we consider the jet function. As for the soft function, the expression for $\Sigma_J^{\max}(\tau)$ is immediately obtained from the results of Section~\ref{sec:SoftJetOneLoopLambda} [see Eq.~\eqref{eq:thrustLaplaceRel}]
\begin{align}
\label{eq:Sigma_max_MC_Jet}
\Sigma_{J_n}^{\rm max}(\tau) &= \Sigma^{\rm max}_{J_n}(\tau ,\mu=Q,\Lambda=Q)= \exp\left\{\int_{\sqrt{\tau} Q}^{Q} \frac{\df\mu}{\mu} \frac{3}{2} C_F\frac{\alpha_s(\mu)}{\pi}\right\}
\,.
\end{align}
The computation of the jet transfer function is trivial at NLL order. As we have seen in Section~\ref{sec:SoftJetOneLoopLambda}, the jet function is only single logarithmic once the additional UV regulator $\Lambda$ has been introduced, since the only kinematic region of phase space giving rise to large logarithms is of hard-collinear origin. Given that the collinear sensitivity is the same in $\Sigma_J(v)$ and $\Sigma_J^{\max}(v)$, the resulting logarithmic dependence due to the phase space bounds cancels in their ratio, and the only logarithmic sensitivity in the jet transfer function comes from the running coupling constant. This implies that each additional emission is suppressed by an additional power of $\alpha_s$, such that only a finite number of emissions need to be taken into account at a given order N$^k$LL. 
In particular, to NLL accuracy, the jet transfer function does not contribute for the reasons stated above, and one has the trivial result
\begin{equation}
\fullF_{J_n}^{\rm NLL}(\tau_n,\tau, Q) = \Theta[\tau_n>0]
\,.
\end{equation}

We can now combine the result for the two jet functions just computed with the NLL soft function as in Eq.~\eqref{eq:scet-F} obtaining
\begin{align}
\label{eq:sigma_final_MC}
\Sigma^{\rm NLL}(\tau)  & =  \exp\left\{\int_{\tau Q}^{Q} \frac{\df\mu}{\mu} 4\Gamma_{\rm cusp}[\alpha_s(\mu)] \ln\frac{\mu}{Q}\right\}\exp\left\{\int_{\tau Q}^{Q} \frac{\df\Lambda}{\Lambda} \int^{\sqrt{\tau \Lambda Q} }_{\tau Q} \frac{d\mu}{\mu}4\Gamma_{\rm cusp}[\alpha_s(\mu)]\right\}\notag\\
& \qquad \times \exp\left\{\int_{\sqrt{\tau} Q}^{Q} \frac{\df\mu}{\mu} 3 C_F\frac{\alpha_s(\mu)}{\pi}\right\}\int\df \tau_s \ \fullF^{\rm NLL}_{S}(\tau_s, \tau,Q) \, \Theta[\tau>\tau_s]\notag\\
& \qquad =\Sigma_{\rm max}(\tau) \fullF^{\rm NLL}_{S}(\tau, \tau,Q)
\,,
\end{align} 
where we have performed the trivial integrations over $\tau_n$ and $\tau_\bn$. We note that the prefactor given by the product $\Sigma_{\rm max}=\Sigma^{\rm max}_S \Sigma^{\rm max}_{J_n}\Sigma^{\rm max}_{J_\bn}$ can be directly computed in the standard SCET without the need for the $\Lambda$ regulator, whose dependence will completely cancel in the product of the three terms.

Using the same steps as in Section~\ref{sec:reviewCAESAR}, one can neglect terms that only contribute to order NNLL and higher, such that one can write the above result in a way that allows for a simpler MC implementation. To this end, we define
\begin{align}
R'_{\rm LL}(\tau) &\equiv \tau\int [\df k] |M_S(k)|^2 \delta(\tau - V_{\rm soft}(k)) \nn
&=  \int \frac{\df k_t}{k_t}\int_0^{\ln\frac{Q}{k_t}} \df\eta\, \frac{\df\phi}{2\pi}4 C_F \frac{\alpha_s(k_t)}{\pi}\delta\left[\ln (k_t/Q) - \eta - \ln(\tau)\right] 
\nn
& = \int_{\tau Q}^{\sqrt{\tau}Q} \frac{\df k_t}{k_t}4 C_F \frac{\alpha_s(k_t)}{\pi}
\,,
\end{align}
where we have evaluated the scale of the running coupling constant at $k_t = \sqrt{k^+ k^-}$. This is the only available choice in the soft function differential in the two light-cone components, since it is the only possible scale which is invariant under a rescaling of the directions of the Wilson lines. $\fullF^{\rm NLL}_{S}(\tau, \tau,Q)$ becomes
\begin{align}
\fullF^{\rm NLL}_{S}(\tau, \tau,Q)
& =\delta^{R'_{\rm LL}(\tau)} \sum_{n=0}^{\infty}
  \left(\frac{1}{n!}\prod_{{\substack{i=1\\\phantom{x}}}}^{n} \int_{\eps \tau}^\tau\frac{\df \tau_i}{\tau_i} \, R'_{\rm LL}(\tau)\right)\Theta\left[\sum_i \tau_i  <\tau\right]
\,,
\end{align}
which can be solved with the following MC procedure:
\begin{enumerate}
\item Start with $i=0$ and $v_0 = \tau$
\item Increase $i$ by one
\item Generate $\tau_i$ randomly according to $(\tau_{i-1}/\tau_i)^{-R'_{\rm LL}(\Phi_B; \tau)} = r$, with $r \in [0,1]$
\item If $\tau_i < \eps \tau$ exit the algorithm, otherwise go back to step 2
\end{enumerate}
If the sum over all generated $\tau_i$ are less than $\tau$, accept the event, otherwise reject it. The value of $\fullF^{\rm NLL}_{S}(\tau, \tau, Q)$ is equal to the fraction of the accepted events.

One can compare the result obtained in \eq{sigma_final_MC} using the MC algorithm above to determine the transfer function $\fullF^{\rm NLL}_{S}(\tau, \tau, Q)$ to the analytical expression, given in \eq{SigmaNLLResummedSCET}. We show this comparison in Figure.~\ref{fig:MonteCarlo}, where we observe a perfect agreement between the two predictions. 

\FIGURE[h]{
 \centering
  \includegraphics[scale=0.5]{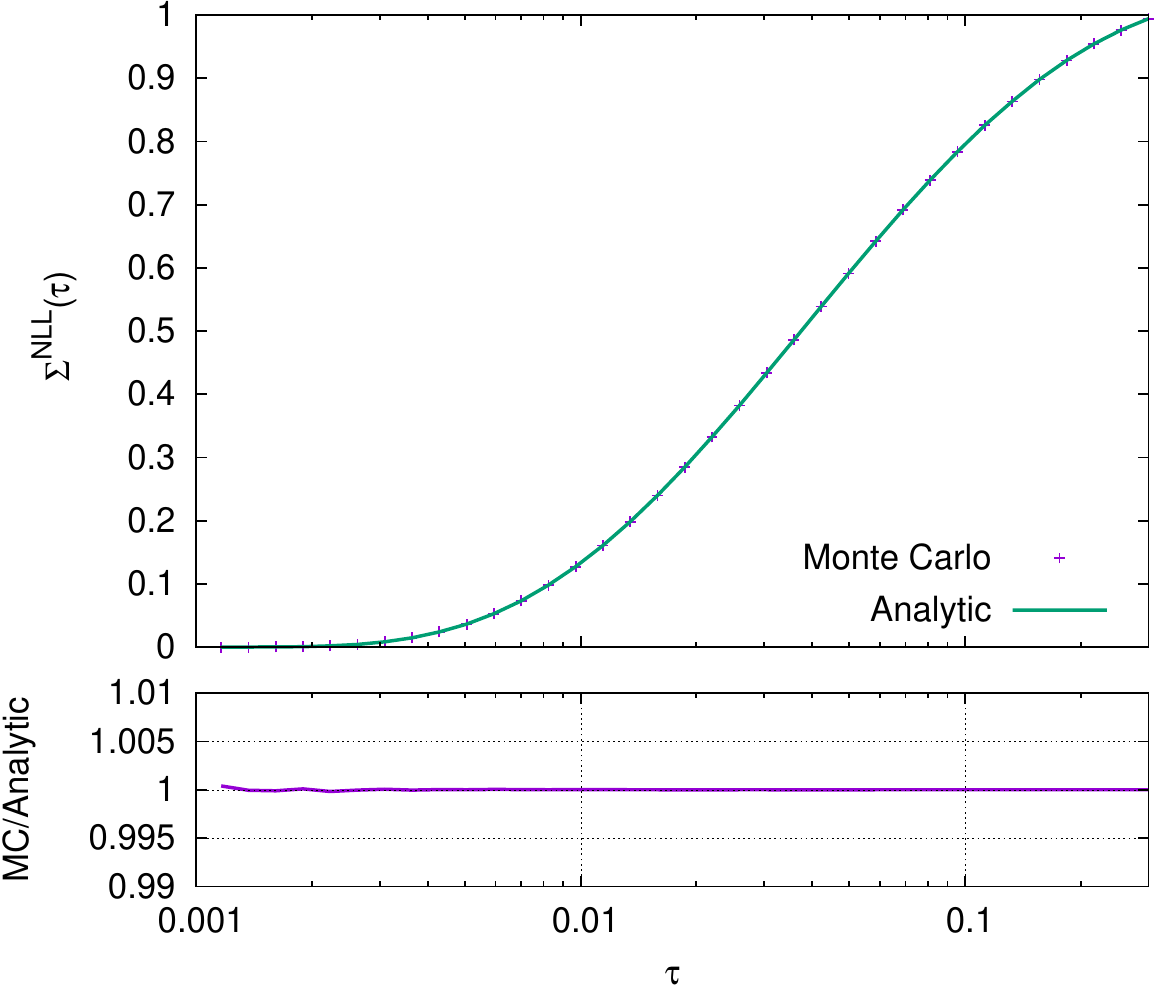}\hspace{1cm}
  \includegraphics[scale=0.5]{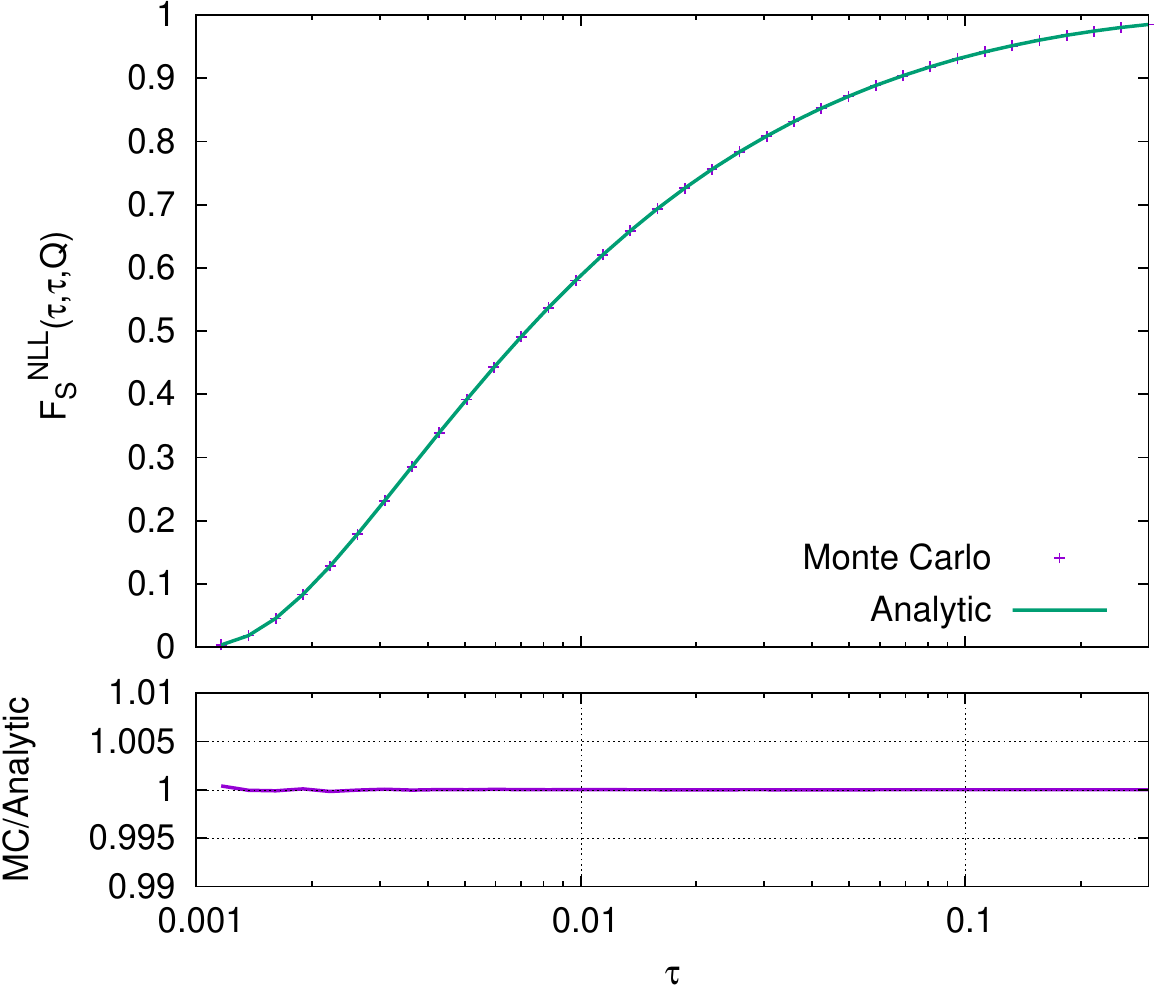}
\caption{\label{fig:MonteCarlo}%
The left figure shows the thrust cross section at NLL obtained with the Monte-Carlo algorithm given in the text (crosses in the plot). The analytic result is reported as a solid line for comparison. The right plot reports the comparison between numerical and analytical solutions for the soft transfer function at the same order. The numerical results have been obtained with $\ln(\eps)=-20$. 
}}

Although the extension to the general case is beyond the scope of this article, we do want to mention that it is possible to apply the above method to a more complicated observable than thrust. In general, if one is able to find an SCET Lagrangian for the simple observable and define $\Sigma{\rm max}$ which by definition contains the same LL as the full observable $v$, then the resummation for $v$ can be obtained by means of a transfer function that is defined in terms of the fields of the same Lagrangian, and can be computed via Monte Carlo methods.

\section{Conclusions and Outlook}
\label{sec:conclusions}

In this work we have shown how to formulate a numerical approach to resummation in SCET using the example of NLL resummation of the thrust distribution. This was achieved by combining the automated {\tt CAESAR/ARES} approach to resummation with the factorization of the long distance degrees of freedom in SCET. 

In SCET, resummation is obtained by first factorizing the required cross section such that a process independent hard function ($H$) multiplies the convolution over jet ($J$) and soft ($S$) functions
\begin{align}
\Sigma(\tau) = H \, J \otimes J \otimes S
\,.\end{align}
The jet and soft functions describe the long distance physics of the process and therefore contain all observable dependence. Each of the factorization ingredients depend on only a single scale, and logarithms can be resummed by solving RG equations for each of the factorization ingredients separately. This general approach makes resummation relatively straightforward once the appropriate factorization formula has been obtained, and simply requires the computation of anomalous dimensions at a given order in perturbation theory. 

In the numerical approach introduced in this paper, we identify a simplified observable, which has the same leading logarithmic structure as the thrust distribution, and for which a factorization theorem can be built in SCET. This simplified ($\max$) observable is constructed such that it has a very simple multiplicative factorization theorem, which is just the product of the same hard function ($H$) multiplied by jet ($J^{\rm max}$) and soft ($S^{\rm max}$) functions
\begin{align}
\Sigma^{\max}(\tau) = H \, J^{\max} \, J^{\max} \, S^{\max}
\,.\end{align}
Due to this simple multiplicative form of the factorization theorem, resummation is achieved in a straightforward manner by solving multiplicative renormalization group equations. The ratio between the full and simplified jet and soft functions defines a transfer function
\begin{align}
\fullF_F \equiv \frac{F}{F^{\max}}
\,,
\end{align}
where $F = J, S$. The main result of this work was to show how to compute this transfer function by performing the phase space integration over real emission diagrams to all orders in perturbation theory. To NLL accuracy, this was shown to result in a rather simple expression, which can be numerically implemented into a straightforward MC algorithm. 

In order to compute the phase space integrals numerically, we needed to ensure that they are finite in 4 dimensions. This is not the case in regular SCET, where the multipole expansion of the phase space limits of the soft function (and the 0-bin of the jet functions) leads to UV divergences in the real integration. In order to overcome this, we introduced an additional regulator to control the UV divergences in the soft real phase space integrations. While this modifies the UV structure of the theory, and requires to perform the RG evolution in two different variables, we showed that the results obtained in SCET with and without this extra regulator are in fact equivalent.

While we have focused for simplicity only on the NLL resummation of the thrust distribution, our results are very general and are readily extended to higher orders in resummation accuracy and to more general observables as long as one can find a simple observable that has the same LL as the full observable, and it is factorizable in SCET.  Using the general definition of $\Sigma^{\rm max}$ given in this article, this is an almost trivial task for most observables. In the approach discussed in this paper, the degrees of freedom required, and hence the effective Lagrangian, are determined by the simple observable which can be resummed analytically through a factorization theorem. The required transfer function, that relates the simple observable to the desired observable, can then be computed numerically using the Feynman rules of the above Lagrangian. Moreover, owing to the fact that the UV limit is now separately regularized,  SCET$_{\rm II}$ problems can be formulated exactly on the same footing as SCET$_{\rm I}$ ones. 

Higher-logarithmic accuracy can be obtained in a relatively straightforward manner by keeping subleading terms in the expansions performed in this work. Furthermore, the numerical approach to resummation in SCET applies even to observables for which a factorization theorem is not known. This opens the door to a systematic resummation for a wide class of observables by combining the analytical power of SCET with numerical MC integrations, which can be automated in an algorithmic way. The details of the general formulation are discussed in a forthcoming paper~\cite{futureWork}.

\acknowledgments 
We are very grateful to A.~Banfi, F.~Caola, A.~Manohar, G.~Salam, I.~Stewart, and G.~Zanderighi for constructive comments on the manuscript, and stimulating conversations on the topics of this article. PFM wishes to thank D.~Y.~Shao for useful discussions about resummation in SCET$_{\rm II}$.  The work of CWB was supported by the Director, Office of Science, Office of High Energy Physics of the U.S. Department of Energy under the Contract No. DE-AC02-05CH11231 (CWB), and (PFW). PFM has been supported by a Marie Sk\l{}odowska Curie Individual Fellowship of the European Commission's Horizon 2020 Programme under contract number 702610 Resummation4PS.

\appendix
\section{Sudakov radiator for Thrust at NLL}
\label{app:g_functions}
In this appendix we report the analytic expressions for the radiator used in the text. The NLL radiator is generally expressed as
\begin{equation}
R(v) = - L g_1(\alpha_s L) - g_2(\alpha_s L) = -\sum_{\ell=1}^2\left(L g^{(\ell)}_1(\alpha_s L) + g^{(\ell)}_2(\alpha_s L) \right)
\,,
\end{equation}
where $L=\ln \frac{1}{v}$, and the $g_i$ functions read (we define $\lambda=\alpha_s \beta_0 L$):
\begin{equation}
\label{eq:g1-b/=0}
\begin{split}
g^{(\ell)}_1(\alpha_s L)=\frac{\Gamma_{\rm cusp}^{(1)} \left(2(1-\lambda ) \ln \left(1-\lambda\right)-(1-2 \lambda ) \ln \left(1-2 \lambda \right)\right)}{4 \pi \beta_0 \lambda }\,,
\end{split}
\end{equation}

\begin{equation}
\label{eq:g2-b/=0}
\begin{split}
g^{(\ell)}_2(\alpha_s L)&=\frac{\Gamma_{\rm cusp}^{(2)} \left( \ln \left(1-2 \lambda\right)-2 \ln \left(1- \lambda\right)\right)}{8 \pi ^2 \text{$\beta_0
 $}^2}+\frac{\gamma_J^{(1)}\ln \left(1- \lambda\right)}{2
\pi  \text{$\beta_0 $}}\\
&+\frac{\Gamma_{\rm cusp}^{(1)} \left(\text{$\beta_1 $} 2 \ln ^2\left(1- \lambda\right)+2 \text{$\beta_1 $} 2\ln \left(1- \lambda\right)\right)}{8 \pi \text{$\beta_0 $}^3}\\
&-\Gamma_{\rm cusp}^{(1)}\frac{\ln \left(1-2 \lambda\right) \left(\text{$\beta_1 $} \ln \left(1-2 \lambda\right)+2 \text{$\beta
  _1 $}\right)}{8 \pi  \text{$\beta_0 $}^3}\,.
\end{split}
\end{equation}
The coefficients of the QCD beta function are given by
\begin{align}
     &\beta_0 = \frac{11 C_A - 2 n_f}{12\pi}\,,\quad 
     \beta_1 = \frac{17 C_A^2 - 5 C_A n_f - 3 C_F n_f}{24\pi^2}\,,
\end{align}
while the anomalous dimensions appearing in the $g_i$ functions read
\begin{equation}
\Gamma_{\rm cusp}^{(1)}= 2C_F\,,~~~ \Gamma_{\rm cusp}^{(2)}= C_F\left[C_A\left(\frac{67}{9}-\frac{\pi^2}{3}\right)-\frac{10}{9}\,n_f\right]\,,~~~\gamma_J^{(1)}= -\frac{3}{2}C_F\,.
\end{equation}

\section{Equivalence of SCET with and without additional UV regulator}
\label{app:equivalence}
The resummation of the large logarithms in the Laplace transform of the cross section in standard SCET is given by
\begin{align}
\Sigma_{\rm NLL}(u) &= \tilde S\left(u; \mu_S \right) \, \tilde J^2\left(u;\mu_J \right) \, \tilde U_S\left(u; \mu_H, \mu_S \right) \, \tilde U^2_J\left(u; \mu_H, \mu_J \right)
\,,
\end{align}
with
\begin{align}
\tilde U_S\left(u; \mu_H, \mu_S \right) &=  \exp\left\{\int_{\mu_S}^{\mu_H} \frac{d\mu}{\mu}4 \Gamma_{\rm cusp}[\alpha_s] \ln\frac{\mu_S}{\mu}- 2\gamma_S[\alpha_s]\right\} 
\nn
\tilde U_J\left(u; \mu_H, \mu_J \right) &=  \exp\left\{\int_{ \mu_J}^{\mu_H} \frac{d\mu}{\mu}\left( 2\Gamma_{\rm cusp}[\alpha_s] \ln\frac{\mu^2}{\mu_J^2} - 2\gamma_J[\alpha_s]\right) \right\}
\,,
\end{align}
with
\begin{align}
\label{eq:mu_characteristic}
\mu_S = \frac{u_0}{u} Q \,, \qquad \mu_J =\sqrt{ \frac{u_0}{u}} Q \,, \qquad \mu_H = Q
\,.
\end{align}
Here we have chosen the common renormalization scale to be $\mu = \mu_H $, such that we need to include the RG evolution of the jet function from $\mu_J $ to $\mu_H$ (given by the first line) and the RG evolution of the soft function from $\mu_S$ to $\mu_H$ (given by the second line). Taking the inverse Laplace transform and dropping the matching coefficients, one reproduces the result given in \eq{SigmaNLLResummedSCET}. 

In SCET with an explicit regulator for the UV divergences in real radiation one finds
\begin{align}
\Sigma^{\Lambda}_{\rm NLL}(u) &= \tilde S_{\Lambda}\left(u;\mu_S, \Lambda_S \right) \, \tilde J_{\Lambda}^2\left(u; \mu_J, \Lambda_J \right) \, \tilde U_S\left(u; \mu_H, \Lambda_H,  \mu_S, \Lambda_S\right) \, \tilde U^2_J\left(u;  \mu_H, \Lambda_H,  \mu_J, \Lambda_J\right)
\,.\end{align}
with $\mu_F$ being the same as in \eq{mu_characteristic} and 
\begin{align}
\Lambda_S = \frac{u_0}{u} Q \,, \qquad \Lambda_J = \Lambda_H = Q
\,.
\end{align}
In the evolution Kernels $\tilde U^{(\Lambda)}_F\left(u;  \mu_H, \Lambda_H,  \mu_F, \Lambda_F\right)$ one has to evolve both $\mu$ and $\Lambda$ from their characteristic scale to the corresponding hard scale. Since the derivatives in $\Lambda$ and $\mu$ commute, one can choose any path in this 2-dimensional evolution, and we choose here to first evolve in $\Lambda$ from $\Lambda_F$ to $\Lambda_H$ holding $\mu$ fixed at $\mu_F$, and then evolve in $\mu$ from $\mu_F$ to $\mu_H$ holding $\Lambda$ fixed at $\Lambda_H$. This allows us to write
\begin{align}
\tilde U_S(u; \mu_H, \Lambda_H,  \mu_S, \Lambda_S) &\equiv \tilde U^{(\mu)}_S(u; \mu_H, \mu_S; \Lambda_H) \, \tilde U^{(\Lambda)}_S(u; \Lambda_H, \Lambda_S; \mu_S)
\nn
\tilde U_J(u;  \mu_H, \Lambda_H,  \mu_J, \Lambda_J) &\equiv  \tilde U^{(\mu)}_J(u; \mu_H, \mu_J; \Lambda_H) 
\,,
\end{align}
where we have used that $\Lambda_J = \Lambda_H$, such that one does not need any $\Lambda$ evolution for the jet function, and we have defined
\begin{align}
\tilde U^{(\mu)}_S(u; \mu_H, \mu_S; \Lambda_H) &= \exp\left\{\int_{\mu_S}^{\mu_H} \frac{d\mu}{\mu}\left[4 \Gamma_{\rm cusp}[\alpha_s] \ln\frac{\mu}{\Lambda_H} - 2 \gamma'_S[\alpha_s]\right]\right\}
\,,\nn
\tilde U^{(\Lambda)}_S(u; \Lambda_H, \Lambda_S; \mu_H) &= \exp\left\{\int_{\Lambda_S}^{\Lambda_H} \frac{d\Lambda}{\Lambda} \int^{\sqrt{\frac{\Lambda}{\Lambda_S}} \mu_S}_{\mu_S} \frac{d\mu}{\mu}4 \Gamma_{\rm cusp}[\alpha_s] \right\}
\,,\\
\tilde U^{(\mu)}_J(u; \mu_H, \mu_J; \Lambda_H) &= \exp\left\{\int_{\mu_J}^{\mu_H} \frac{d\mu}{\mu} \left[ - 2 \gamma'_J[\alpha_s]\right]\right\}
\,,\end{align}

The two anomalous dimensions $\gamma'_S$ and $\gamma'_J$ are different from the usual SCET ones starting from their NLO expression. However, they satisfy $\gamma_S' + 2 \gamma_J' = \gamma_S + 2 \gamma_J$. The integration over $\Lambda'$ in the $\Lambda$ evolution kernels $\tilde U^{(\Lambda)}_F$ can be performed analytically by changing the order of integration. For this, we write
\begin{align}
\int_{\Lambda_F}^{\Lambda_H} \frac{d\Lambda}{\Lambda} \int_{\mu_F}^{\sqrt{\frac{\Lambda}{\Lambda_F}} \mu_F} \frac{\df \mu}{\mu} f(\mu) &= \int_{\mu_F}^{\sqrt{\frac{\Lambda_H}{\Lambda_F}} \mu_F} \frac{\df \mu}{\mu} f(\mu) \int_{\frac{\mu^2}{\mu_F^2} \Lambda_F}^{\Lambda_H} \frac{d\Lambda}{\Lambda}
\nn
&= \int_{\mu_F}^{\sqrt{\frac{\Lambda_H}{\Lambda_F}} \mu_F} \frac{\df \mu}{\mu} f(\mu) \, \ln \frac{\mu_F^2 \, \Lambda_H}{\mu^2 \, \Lambda_F}
\,.
\end{align}
This gives for the $\tilde U^{(\Lambda)}_S$ evolution kernel, using $\Lambda_H = \mu_H$ and $\mu_S = \Lambda_S$
\begin{align}
\tilde U^{(\Lambda)}_S(u; \Lambda_H, \Lambda_S; \mu_H) &= \exp\left\{\int_{\mu_S}^{\sqrt{\mu_H \mu_S}} \frac{\df \mu}{\mu} 4 \Gamma_{\rm cusp}[\alpha_s] \, \ln \frac{\mu_H \mu_S}{\mu^2} \right\}
\nn&= \exp\left\{\int_{\mu_S}^{\mu_J} \frac{\df \mu}{\mu} 4 \Gamma_{\rm cusp}[\alpha_s] \, \ln \frac{\mu_H \mu_S}{\mu^2} \right\}
\nn&= \exp\left\{\int_{\mu_S}^{\mu_H} \frac{\df \mu}{\mu} 4 \Gamma_{\rm cusp}[\alpha_s] \,\ln \frac{\mu_H \mu_S}{\mu^2} \right\} 
\nn
& \qquad \times 
\exp\left\{\int_{\mu_J}^{\mu_H} \frac{\df \mu}{\mu} 4 \Gamma_{\rm cusp}[\alpha_s] \,\ln \frac{\mu^2}{\mu_H \mu_S} \right\}
\end{align}
Putting this together with the $\tilde U^{(\mu)}_S$ one finds for the combined soft evolution factor
\begin{align}
\label{eq:US_tilde}
\tilde U_S(u; \mu_H, \Lambda_H,  \mu_S, \Lambda_S) & = \exp\left\{\int_{\mu_S}^{\mu_H} \frac{\df \mu}{\mu} \left[ 4 \Gamma_{\rm cusp}[\alpha_s] \,\ln \frac{\mu_S}{\mu} - 2\gamma_S'[\alpha_s] \right]\right\} 
\nn
& \qquad \times 
\exp\left\{\int_{\mu_J}^{\mu_H} \frac{\df \mu}{\mu} 4 \Gamma_{\rm cusp}[\alpha_s] \,\ln \frac{\mu^2}{\mu_J^2} \right\}
\end{align}
where we have used in the last line $\mu_J ^2 = \mu_S \mu_H$. This combined evolution factor therefore contains the complete evolution due to the cusp anomalous dimension, which is usually split between the soft and the jet evolution kernels, as well as the non-cusp part of the soft evolution. The evolution factor of the jet function only contains the non-cusp part of the collinear evolution which reads
\begin{align}
\label{eq:UJ_tilde}
\tilde U_J(u; \mu_H, \Lambda_H, \mu_J, \Lambda_J) &= \exp\left\{\int_{\mu_J}^{\mu_H} \frac{\df\mu}{\mu} \left[ - 2 \gamma'_J[\alpha_s]\right]\right\},
\end{align} 
from which it is easy to see that the product of Eqs.~\eqref{eq:US_tilde} and~\eqref{eq:UJ_tilde} fulfills the following equality
\begin{equation}
\tilde U_S(u; \mu_H, \Lambda_H,  \mu_S, \Lambda_S) \tilde U^2_J(u; \mu_H, \Lambda_H, \mu_J, \Lambda_J) = \tilde U_S\left(u; \mu_H, \mu_S \right) \tilde U^2_J\left(u; \mu_H, \mu_J \right),
\end{equation}
which shows that the physical combination of the evolution factors is identical to the standard SCET one at all orders.

\section{RGE of the thrust soft function with a IRC resolution scale}
\label{app:SCET_w_delta}
In this appendix we wish to comment more on the logarithmic structure of the decomposition~\eqref{eq:Sigma_MC_SCET} for the resummed cross section. It is instructive to consider the soft function as a case study, although the same conclusions apply to the two jet functions. We express the soft function as in Eq.~\eqref{eq:MC-soft-jet}, namely
\begin{align}
\label{eq:soft_deco}
S(\tau_s;\mu) &= \frac{\df\Sigma_{S}(\tau_s)}{\df\tau_s}\notag\\
\Sigma_{S}(\tau_s, \mu) &\equiv \Sigma_{S}^{\rm max}(\tau, \mu)\fullF_S(\tau_s, \tau, \mu) = \Sigma_{S}^{\rm max}(\tau, \mu)\frac{\Sigma_{S}^{\rm max}(\eps\tau, \mu)}{\Sigma_{S}^{\rm max}(\tau, \mu)} \frac{\Sigma_{S}(\tau_s, \mu)}{\Sigma_{S}^{\rm max}(\eps\tau, \mu)}
\,.
\end{align}
In this paper we have shown how $\Sigma_{S}^{\rm max}$ is computed analytically, while the transfer function $\fullF_S$ is obtained via MC methods. The latter task requires the introduction of an explicit UV regulator in the theory, which led to the results in Section~\ref{sec:MC_transfer}.  The goal of this appendix is to study how the decomposition~\eqref{eq:soft_deco} modifies the soft function's RGE if it were used in the standard SCET formulation, i.e. without the $\Lambda$ cutoff. This is a useful exercise to understand from a different viewpoint the method proposed in this article.

We start by simplifying, without lost of generality, Eq.~\eqref{eq:soft_deco} by getting rid of the $\Sigma_{S}^{\rm max}(\tau, \mu)$ factor and recast it as
\begin{align}
S(\tau_s;\mu)  &= \Sigma_{S}^{\rm max}(\eps\tau, \mu)\frac{\Sigma'_{S}(\tau_s, \mu)}{\Sigma_{S}^{\rm max}(\eps\tau, \mu)}
\,.
\end{align}
The full soft function for thrust in momentum space fulfills the non-local RGE [see \eqs{RGE_J}{RGE_S}]
\begin{align}
\label{eq:RGE-soft-full}
\mu \frac{\df}{\df \mu} S(\tau_s; \mu) &= \left\{2\Gamma_{\rm cusp}[\alpha_s(\mu)] \ln\frac{\tau_s^2 Q^2}{\mu^2}- 2 \gamma_{S}[\alpha_s(\mu)]\right\} S(\tau_s; \mu) \notag\\
&\qquad 
- 4\Gamma_{\rm cusp}[\alpha_s(\mu)]\int_0^{\tau_s}\!  {\rm d}\tau_s' \, \frac{S(\tau_s; \mu)- S(\tau_s'; \mu)}{\tau_s-\tau_s'}
\,,
\end{align}
where the precise observable dependence (in particular its additive nature) is reflected in the second term in the right-hand-side of the above equation, that essentially shows how an extra real emission modifies the existing value of the observable (i.e. in an additive way).

Now we study how the RGE is modified by the introduction of the IR resolution scale $\eps \tau_s$. We first consider the contribution $\Sigma_{S}^{\rm max}(\eps\tau, \mu)$, which 
fulfills the following local RG equation in momentum space [see \eqs{laplace_RGE_soft}{thrustLaplaceRel}]
\begin{equation}
\label{eq:RGE-soft-resolved-0}
\frac{d \Sigma_{S}^{\rm max}(\eps\tau, \mu)}{d\ln\mu} = \left[4\Gamma_{\rm
  cusp}[\alpha_s(\mu)]\ln\frac{\eps \tau Q}{\mu} -2\gamma_s[\alpha_s(\mu)]\right]\Sigma_{S}^{\rm max}(\eps\tau, \mu)
  \,,
\end{equation}
which can be easily solved by setting the initial conditions at $\mu=\eps \tau Q$.

Next we consider the remaining ratio $\Sigma'_{S}(\tau_s, \mu)/\Sigma_{S}^{\rm max}(\eps\tau, \mu) $ which, by means of \eqs{RGE-soft-full}{RGE-soft-resolved-0}, fulfills the non-local RGE
\begin{align}
\label{eq:RGE-soft-resolved-1}
\mu \frac{\df}{\df \mu} \frac{\Sigma'_{S}(\tau_s, \mu)}{\Sigma_{S}^{\rm max}(\eps\tau, \mu)} &= \frac{\Sigma'_{S}(\tau_s, \mu)}{\Sigma_{S}^{\rm max}(\eps\tau, \mu)}4\Gamma_{\rm
  cusp}[\alpha_s(\mu)]\ln\frac{\tau_s}{\eps\tau}\notag\\
&\qquad - \frac{4 \Gamma_{\rm
    cusp}[\alpha_s(\mu)]}{\Sigma_{S}^{\rm max}(\eps\tau, \mu)} \int_0^{\tau_s}\df \tau'\frac{\Sigma'_{S}(\tau_s, \mu) - \Sigma'_{S}(\tau', \mu)}{\tau_s-\tau'}
    \,.
\end{align}
Since by definition the resolution scale is small ($\eps \tau \ll \tau$), one can write
\begin{align}
&\int_0^{\tau_s}\df \tau' \, \frac{\Sigma'_{S}(\tau_s, \mu) - \Sigma'_{S}(\tau', \mu)}{\tau_s-\tau'} 
\nn
&\qquad = \int_0^{\tau_s}\frac{\df u}{u}\left[\Sigma'_{S}(\tau_s;\mu) - \Sigma'_{S}(\tau_s-u;\mu)\right]
\nn
&\qquad=\int_{\eps \tau}^{\tau_s}\frac{\df u}{u}\left[\Sigma'_{S}(\tau_s;\mu) - \Sigma'_{S}(\tau_s-u;\mu)\right]+\int^{\eps \tau}_0\frac{\df u}{u}\left[\Sigma'_{S}(\tau_s;\mu) - \Sigma'_{S}(\tau_s;\mu)\right]
\nn
&\qquad=\int_{\eps \tau}^{\tau_s}\frac{\df u}{u}\left[\Sigma'_{S}(\tau_s;\mu) - \Sigma'_{S}(\tau_s-u;\mu)\right] + {\cal O}(\eps\tau)
\,,
\end{align}
where the power suppressed ${\cal O}(\eps\tau)$ corrections can be ignored in the limit $\eps \to 0$.
The remaining integral can be split as
\begin{align}
  \int_{\eps \tau}^{\tau_s}\frac{\df u}{u}\left[\Sigma'_{S}(\tau_s;\mu) - \Sigma'_{S}(\tau_s-u;\mu)\right] = \ln\frac{\tau_s}{\eps\tau} \Sigma'_{S}(\tau_s;\mu) - \int_{\eps \tau_s}^{\tau_s}\frac{\df u}{u}\Sigma'_{S}(\tau_s-u;\mu)
  \,.
\end{align}
By plugging the above expression into \eq{RGE-soft-resolved-1} we obtain
\begin{align}
\label{eq:RGE-soft-resolved-2}
\mu \frac{\df}{\df \mu} \frac{\Sigma'_{S}(\tau_s, \mu)}{\Sigma_{S}^{\rm max}(\eps\tau, \mu)}  &= 4 \Gamma_{\rm
    cusp}[\alpha_s(\mu)] \int_{\eps \tau}^{\tau_s}\frac{\df u}{u}\frac{\Sigma'_{S}(\tau_s-u;\mu)}{\Sigma_{S}^{\rm max}(\eps\tau, \mu)}
    \,.
\end{align}
Thus, the decomposition of \eqref{eq:soft_deco} allowed us to separate the initial evolution equation into a local piece~\eqref{eq:RGE-soft-resolved-0} and a non-local piece ~\eqref{eq:RGE-soft-resolved-2}. The precise definition of the non-local piece depends on the form of the observable, and the result given here holds for an additive observable. 
The decomposition of \eq{soft_deco} allows to separate the RGE into a local piece, which is independent of the definition of the observable and can therefore easily be solved analytically, and a purely non-local piece, which contains all the observable dependence. 

While we have discussed how to compute the non-local piece via a MC algorithm, for an additive observable it can easily be calculated analytically, as we now show. We take the Laplace transform ${\cal L}$ of the ratio defined as
\begin{equation}
\frac{\Sigma'_{S}(\tau_s, \mu)}{\Sigma_{S}^{\rm max}(\eps\tau, \mu)}=\frac{1}{2\pi i}\int \df u \,e^{u \tau_s}\Pi^{(\eps)}(u,\mu)
\,,
\end{equation}
and one finds the RG equation
\begin{equation}
\label{eq:RGE-resolved-2}
\frac{d \Pi^{(\eps)}(u,\mu)}{d\ln\mu} = 4 \Gamma_{\rm
  cusp}[\alpha_s(\mu)]\Gamma(0,\eps
\tau_su)\Pi^{(\eps)}(u,\mu)
\,,
\end{equation}
where we used the result 
\begin{equation}
{\cal L}\left(\frac{\Theta[u>\eps \tau]}{u}\right) =
\int_u^\infty \df u\frac{e^{-\eps \tau u}}{u} = \Gamma(0,\eps
\tau u) \simeq \ln\frac{u_0}{u \tau \eps} + {\cal O}(\eps)
\,,
\end{equation}
with $u_0=e^{-\gamma_E}$. The above anomalous dimension can be easily obtained from an explicit calculation of the soft function by fixing the thrust value to $\tau_s$, while requiring that it be larger than $\eps \tau$. The solution to Eq.~\eqref{eq:RGE-resolved-2} reads
\begin{equation}
\Pi^{(\eps)}(u,\mu)  = \Pi^{(\eps)}(u,\mu_0)
\left(\frac{u_0}{u \eps \tau}\right)^{\eta(\mu,\mu_0)}
\,,
\end{equation}
where
\begin{equation}
\eta(\mu,\mu_0) = \int_{\mu_0}^\mu\frac{d\mu'}{\mu'}4\Gamma_{\rm cusp}[\alpha_s(\mu')]
\,.
\end{equation}
The solution in momentum space can be obtained by performing the
inverse Laplace transform which, at NLL, yields
\begin{align}
\frac{\Sigma'_{S}(\tau_s, \mu)}{\Sigma_{S}^{\rm max}(\eps\tau, \mu)} =\frac{1}{\tau_s Q} \left(\frac{\tau_s
   }{\eps \tau}\right)^{\eta(\mu,\mu_0)}\frac{e^{-\gamma_E \,\eta(\mu,\mu_0)}}{\Gamma(\eta(\mu,\mu_0))}
   \,.
\end{align}
This can be combined with the solution to Eq.~\eqref{eq:RGE-soft-resolved-0} to obtain the full NLL result (at this order we set $\gamma_s=0$)
\begin{align}
S(\tau_s;\mu)&= \exp\left\{\int_{\mu_0}^{\mu} \frac{\df\mu'}{\mu'}4\Gamma_{\rm
  cusp}(\alpha_s(\mu'))\ln\frac{\eps \tau Q}{\mu'} \right\} \frac{1}{\tau_s Q} \left(\frac{\tau_s
    }{\eps \tau }\right)^{\eta(\mu,\mu_0)}\frac{e^{-\gamma_E \eta(\mu,\mu_0)}}{\Gamma(\eta(\mu,\mu_0))}
    \,.
\end{align}
By writing the logarithm in the exponential function as $\ln\frac{\eps \tau Q}{\mu'} = \ln\frac{\tau_s Q}{\mu'}+\ln\frac{\eps \tau}{\tau_s}$ we obtain
\begin{align}
\label{eq:thrust-final}
S(\tau_s;\mu)&= \exp\left\{\int_{\mu_0}^{\mu} \frac{\df\mu'}{\mu'}4\Gamma_{\rm
  cusp}(\alpha_s(\mu'))\ln\frac{\tau_s Q}{\mu'} \right\} \frac{1}{\tau_s Q} \frac{e^{-\gamma_E \eta(\mu,\mu_0)}}{\Gamma(\eta(\mu,\mu_0))}
  \,,
\end{align}
which reproduces the NLL result for the thrust soft function with $\mu_0=\tau_s Q$.

\addcontentsline{toc}{section}{References}
\bibliographystyle{JHEP}
\bibliography{CAESAR_SCET}

\providecommand{\href}[2]{#2}\begingroup\raggedright\begin{thebibliography}{10}

\bibitem{Collins:1984kg}
J.~C. Collins, D.~E. Soper and G.~F. Sterman, \emph{{Transverse Momentum
  Distribution in Drell-Yan Pair and W and Z Boson Production}},
  \href{http://dx.doi.org/10.1016/0550-3213(85)90479-1}{\emph{Nucl. Phys.} {\bf
  B250} (1985) 199--224}.

\bibitem{Bauer:2000ew}
C.~W. Bauer, S.~Fleming and M.~E. Luke, \emph{{Summing Sudakov logarithms in B
  ---> X(s gamma) in effective field theory}},
  \href{http://dx.doi.org/10.1103/PhysRevD.63.014006}{\emph{Phys. Rev.} {\bf
  D63} (2000) 014006}, [\href{http://arxiv.org/abs/hep-ph/0005275}{{\tt
  hep-ph/0005275}}].

\bibitem{Bauer:2000yr}
C.~W. Bauer, S.~Fleming, D.~Pirjol and I.~W. Stewart, \emph{{An Effective field
  theory for collinear and soft gluons: Heavy to light decays}},
  \href{http://dx.doi.org/10.1103/PhysRevD.63.114020}{\emph{Phys. Rev.} {\bf
  D63} (2001) 114020}, [\href{http://arxiv.org/abs/hep-ph/0011336}{{\tt
  hep-ph/0011336}}].

\bibitem{Bauer:2001ct}
C.~W. Bauer and I.~W. Stewart, \emph{{Invariant operators in collinear
  effective theory}},
  \href{http://dx.doi.org/10.1016/S0370-2693(01)00902-9}{\emph{Phys. Lett.}
  {\bf B516} (2001) 134--142}, [\href{http://arxiv.org/abs/hep-ph/0107001}{{\tt
  hep-ph/0107001}}].

\bibitem{Bauer:2001yt}
C.~W. Bauer, D.~Pirjol and I.~W. Stewart, \emph{{Soft collinear factorization
  in effective field theory}},
  \href{http://dx.doi.org/10.1103/PhysRevD.65.054022}{\emph{Phys. Rev.} {\bf
  D65} (2002) 054022}, [\href{http://arxiv.org/abs/hep-ph/0109045}{{\tt
  hep-ph/0109045}}].

\bibitem{Bauer:2008dt}
C.~W. Bauer, S.~P. Fleming, C.~Lee and G.~F. Sterman, \emph{{Factorization of
  e+e- Event Shape Distributions with Hadronic Final States in Soft Collinear
  Effective Theory}},
  \href{http://dx.doi.org/10.1103/PhysRevD.78.034027}{\emph{Phys. Rev.} {\bf
  D78} (2008) 034027}, [\href{http://arxiv.org/abs/0801.4569}{{\tt
  0801.4569}}].

\bibitem{Bauer:2008jx}
C.~W. Bauer, A.~Hornig and F.~J. Tackmann, \emph{{Factorization for generic jet
  production}}, \href{http://dx.doi.org/10.1103/PhysRevD.79.114013}{\emph{Phys.
  Rev.} {\bf D79} (2009) 114013}, [\href{http://arxiv.org/abs/0808.2191}{{\tt
  0808.2191}}].

\bibitem{Catani:1990rr}
S.~Catani, B.~R. Webber and G.~Marchesini, \emph{{QCD coherent branching and
  semiinclusive processes at large x}},
  \href{http://dx.doi.org/10.1016/0550-3213(91)90390-J}{\emph{Nucl. Phys.} {\bf
  B349} (1991) 635--654}.

\bibitem{Catani:1991kz}
S.~Catani, G.~Turnock, B.~R. Webber and L.~Trentadue, \emph{{Thrust
  distribution in e+ e- annihilation}},
  \href{http://dx.doi.org/10.1016/0370-2693(91)90494-B}{\emph{Phys. Lett.} {\bf
  B263} (1991) 491--497}.

\bibitem{Banfi:2004yd}
A.~Banfi, G.~P. Salam and G.~Zanderighi, \emph{{Principles of general
  final-state resummation and automated implementation}},
  \href{http://dx.doi.org/10.1088/1126-6708/2005/03/073}{\emph{JHEP} {\bf 03}
  (2005) 073}, [\href{http://arxiv.org/abs/hep-ph/0407286}{{\tt
  hep-ph/0407286}}].

\bibitem{Abbate:2010xh}
R.~Abbate, M.~Fickinger, A.~H. Hoang, V.~Mateu and I.~W. Stewart, \emph{{Thrust
  at N$^3$LL with Power Corrections and a Precision Global Fit for
  alphas(mZ)}}, \href{http://dx.doi.org/10.1103/PhysRevD.83.074021}{\emph{Phys.
  Rev.} {\bf D83} (2011) 074021}, [\href{http://arxiv.org/abs/1006.3080}{{\tt
  1006.3080}}].

\bibitem{Banfi:2001bz}
A.~Banfi, G.~P. Salam and G.~Zanderighi, \emph{{Semi-numerical resummation of
  event shapes}},
  \href{http://dx.doi.org/10.1088/1126-6708/2002/01/018}{\emph{JHEP} {\bf 01}
  (2002) 018}, [\href{http://arxiv.org/abs/hep-ph/0112156}{{\tt
  hep-ph/0112156}}].

\bibitem{Banfi:2014sua}
A.~Banfi, H.~McAslan, P.~F. Monni and G.~Zanderighi, \emph{{A general method
  for the resummation of event-shape distributions in $e^{+} e^{−}$
  annihilation}}, \href{http://dx.doi.org/10.1007/JHEP05(2015)102}{\emph{JHEP}
  {\bf 05} (2015) 102}, [\href{http://arxiv.org/abs/1412.2126}{{\tt
  1412.2126}}].

\bibitem{Banfi:2016zlc}
A.~Banfi, H.~McAslan, P.~F. Monni and G.~Zanderighi, \emph{{The two-jet rate in
  $e^+e^-$ at next-to-next-to-leading-logarithmic order}},
  \href{http://dx.doi.org/10.1103/PhysRevLett.117.172001}{\emph{Phys. Rev.
  Lett.} {\bf 117} (2016) 172001}, [\href{http://arxiv.org/abs/1607.03111}{{\tt
  1607.03111}}].

\bibitem{Bizon:2017rah}
W.~Bizon, P.~F. Monni, E.~Re, L.~Rottoli and P.~Torrielli,
  \emph{{Momentum-space resummation for transverse observables and the Higgs
  p$_{\perp}$ at N$^{3}$LL+NNLO}},
  \href{http://dx.doi.org/10.1007/JHEP02(2018)108}{\emph{JHEP} {\bf 02} (2018)
  108}, [\href{http://arxiv.org/abs/1705.09127}{{\tt 1705.09127}}].

\bibitem{Becher:2008cf}
T.~Becher and M.~D. Schwartz, \emph{{A precise determination of $\alpha_s$ from
  LEP thrust data using effective field theory}},
  \href{http://dx.doi.org/10.1088/1126-6708/2008/07/034}{\emph{JHEP} {\bf 07}
  (2008) 034}, [\href{http://arxiv.org/abs/0803.0342}{{\tt 0803.0342}}].

\bibitem{Chien:2010kc}
Y.-T. Chien and M.~D. Schwartz, \emph{{Resummation of heavy jet mass and
  comparison to LEP data}},
  \href{http://dx.doi.org/10.1007/JHEP08(2010)058}{\emph{JHEP} {\bf 08} (2010)
  058}, [\href{http://arxiv.org/abs/1005.1644}{{\tt 1005.1644}}].

\bibitem{Becher:2012qc}
T.~Becher and G.~Bell, \emph{{NNLL Resummation for Jet Broadening}},
  \href{http://dx.doi.org/10.1007/JHEP11(2012)126}{\emph{JHEP} {\bf 11} (2012)
  126}, [\href{http://arxiv.org/abs/1210.0580}{{\tt 1210.0580}}].

\bibitem{Hoang:2014wka}
A.~H. Hoang, D.~W. Kolodrubetz, V.~Mateu and I.~W. Stewart,
  \emph{{$C$-parameter distribution at N$^3$LL′ including power
  corrections}},
  \href{http://dx.doi.org/10.1103/PhysRevD.91.094017}{\emph{Phys. Rev.} {\bf
  D91} (2015) 094017}, [\href{http://arxiv.org/abs/1411.6633}{{\tt
  1411.6633}}].

\bibitem{Becher:2015lmy}
T.~Becher, X.~Garcia~i Tormo and J.~Piclum, \emph{{Next-to-next-to-leading
  logarithmic resummation for transverse thrust}},
  \href{http://dx.doi.org/10.1103/PhysRevD.93.054038,
  10.1103/PhysRevD.93.079905}{\emph{Phys. Rev.} {\bf D93} (2016) 054038},
  [\href{http://arxiv.org/abs/1512.00022}{{\tt 1512.00022}}].

\bibitem{Frye:2016aiz}
C.~Frye, A.~J. Larkoski, M.~D. Schwartz and K.~Yan, \emph{{Factorization for
  groomed jet substructure beyond the next-to-leading logarithm}},
  \href{http://dx.doi.org/10.1007/JHEP07(2016)064}{\emph{JHEP} {\bf 07} (2016)
  064}, [\href{http://arxiv.org/abs/1603.09338}{{\tt 1603.09338}}].

\bibitem{Moult:2018jzp}
I.~Moult and H.~X. Zhu, \emph{{Simplicity from Recoil: The Three-Loop Soft
  Function and Factorization for the Energy-Energy Correlation}},
  \href{http://arxiv.org/abs/1801.02627}{{\tt 1801.02627}}.

\bibitem{Baron:2018nfz}
J.~Baron, S.~Marzani and V.~Theeuwes, \emph{{Soft-Drop Thrust}},
  \href{http://arxiv.org/abs/1803.04719}{{\tt 1803.04719}}.

\bibitem{Bauer:2002nz}
C.~W. Bauer, S.~Fleming, D.~Pirjol, I.~Z. Rothstein and I.~W. Stewart,
  \emph{{Hard scattering factorization from effective field theory}},
  \href{http://dx.doi.org/10.1103/PhysRevD.66.014017}{\emph{Phys. Rev.} {\bf
  D66} (2002) 014017}, [\href{http://arxiv.org/abs/hep-ph/0202088}{{\tt
  hep-ph/0202088}}].

\bibitem{Banfi:2004nk}
A.~Banfi, G.~P. Salam and G.~Zanderighi, \emph{{Resummed event shapes at hadron
  - hadron colliders}},
  \href{http://dx.doi.org/10.1088/1126-6708/2004/08/062}{\emph{JHEP} {\bf 08}
  (2004) 062}, [\href{http://arxiv.org/abs/hep-ph/0407287}{{\tt
  hep-ph/0407287}}].

\bibitem{Bozzi:2005wk}
G.~Bozzi, S.~Catani, D.~de~Florian and M.~Grazzini, \emph{{Transverse-momentum
  resummation and the spectrum of the Higgs boson at the LHC}},
  \href{http://dx.doi.org/10.1016/j.nuclphysb.2005.12.022}{\emph{Nucl. Phys.}
  {\bf B737} (2006) 73--120}, [\href{http://arxiv.org/abs/hep-ph/0508068}{{\tt
  hep-ph/0508068}}].

\bibitem{Becher:2010tm}
T.~Becher and M.~Neubert, \emph{{{Drell-Yan} Production at Small $q_T$,
  Transverse Parton Distributions and the Collinear Anomaly}},
  \href{http://dx.doi.org/10.1140/epjc/s10052-011-1665-7}{\emph{Eur. Phys. J.}
  {\bf C71} (2011) 1665}, [\href{http://arxiv.org/abs/1007.4005}{{\tt
  1007.4005}}].

\bibitem{Stewart:2010pd}
I.~W. Stewart, F.~J. Tackmann and W.~J. Waalewijn, \emph{{The Beam Thrust Cross
  Section for Drell-Yan at NNLL Order}},
  \href{http://dx.doi.org/10.1103/PhysRevLett.106.032001}{\emph{Phys. Rev.
  Lett.} {\bf 106} (2011) 032001}, [\href{http://arxiv.org/abs/1005.4060}{{\tt
  1005.4060}}].

\bibitem{Banfi:2011dx}
A.~Banfi, M.~Dasgupta and S.~Marzani, \emph{{QCD predictions for new variables
  to study dilepton transverse momenta at hadron colliders}},
  \href{http://dx.doi.org/10.1016/j.physletb.2011.05.028}{\emph{Phys. Lett.}
  {\bf B701} (2011) 75--81}, [\href{http://arxiv.org/abs/1102.3594}{{\tt
  1102.3594}}].

\bibitem{Berger:2010xi}
C.~F. Berger, C.~Marcantonini, I.~W. Stewart, F.~J. Tackmann and W.~J.
  Waalewijn, \emph{{Higgs Production with a Central Jet Veto at NNLL+NNLO}},
  \href{http://dx.doi.org/10.1007/JHEP04(2011)092}{\emph{JHEP} {\bf 04} (2011)
  092}, [\href{http://arxiv.org/abs/1012.4480}{{\tt 1012.4480}}].

\bibitem{Jouttenus:2011wh}
T.~T. Jouttenus, I.~W. Stewart, F.~J. Tackmann and W.~J. Waalewijn, \emph{{The
  Soft Function for Exclusive N-Jet Production at Hadron Colliders}},
  \href{http://dx.doi.org/10.1103/PhysRevD.83.114030}{\emph{Phys. Rev.} {\bf
  D83} (2011) 114030}, [\href{http://arxiv.org/abs/1102.4344}{{\tt
  1102.4344}}].

\bibitem{Becher:2012qa}
T.~Becher and M.~Neubert, \emph{{Factorization and NNLL Resummation for Higgs
  Production with a Jet Veto}},
  \href{http://dx.doi.org/10.1007/JHEP07(2012)108}{\emph{JHEP} {\bf 07} (2012)
  108}, [\href{http://arxiv.org/abs/1205.3806}{{\tt 1205.3806}}].

\bibitem{Zhu:2012ts}
H.~X. Zhu, C.~S. Li, H.~T. Li, D.~Y. Shao and L.~L. Yang,
  \emph{{Transverse-momentum resummation for top-quark pairs at hadron
  colliders}},
  \href{http://dx.doi.org/10.1103/PhysRevLett.110.082001}{\emph{Phys. Rev.
  Lett.} {\bf 110} (2013) 082001}, [\href{http://arxiv.org/abs/1208.5774}{{\tt
  1208.5774}}].

\bibitem{Banfi:2012jm}
A.~Banfi, P.~F. Monni, G.~P. Salam and G.~Zanderighi, \emph{{Higgs and Z-boson
  production with a jet veto}},
  \href{http://dx.doi.org/10.1103/PhysRevLett.109.202001}{\emph{Phys. Rev.
  Lett.} {\bf 109} (2012) 202001}, [\href{http://arxiv.org/abs/1206.4998}{{\tt
  1206.4998}}].

\bibitem{Becher:2013xia}
T.~Becher, M.~Neubert and L.~Rothen, \emph{{Factorization and
  $N^{3}LL_{p}$+NNLO predictions for the Higgs cross section with a jet veto}},
  \href{http://dx.doi.org/10.1007/JHEP10(2013)125}{\emph{JHEP} {\bf 10} (2013)
  125}, [\href{http://arxiv.org/abs/1307.0025}{{\tt 1307.0025}}].

\bibitem{Stewart:2013faa}
I.~W. Stewart, F.~J. Tackmann, J.~R. Walsh and S.~Zuberi, \emph{{Jet $p_T$
  resummation in Higgs production at $NNLL'+NNLO$}},
  \href{http://dx.doi.org/10.1103/PhysRevD.89.054001}{\emph{Phys. Rev.} {\bf
  D89} (2014) 054001}, [\href{http://arxiv.org/abs/1307.1808}{{\tt
  1307.1808}}].

\bibitem{Procura:2014cba}
M.~Procura, W.~J. Waalewijn and L.~Zeune, \emph{{Resummation of
  Double-Differential Cross Sections and Fully-Unintegrated Parton Distribution
  Functions}}, \href{http://dx.doi.org/10.1007/JHEP02(2015)117}{\emph{JHEP}
  {\bf 02} (2015) 117}, [\href{http://arxiv.org/abs/1410.6483}{{\tt
  1410.6483}}].

\bibitem{Li:2016ctv}
Y.~Li and H.~X. Zhu, \emph{{Bootstrapping Rapidity Anomalous Dimensions for
  Transverse-Momentum Resummation}},
  \href{http://dx.doi.org/10.1103/PhysRevLett.118.022004}{\emph{Phys. Rev.
  Lett.} {\bf 118} (2017) 022004}, [\href{http://arxiv.org/abs/1604.01404}{{\tt
  1604.01404}}].

\bibitem{Monni:2016ktx}
P.~F. Monni, E.~Re and P.~Torrielli, \emph{{Higgs Transverse-Momentum
  Resummation in Direct Space}},
  \href{http://dx.doi.org/10.1103/PhysRevLett.116.242001}{\emph{Phys. Rev.
  Lett.} {\bf 116} (2016) 242001}, [\href{http://arxiv.org/abs/1604.02191}{{\tt
  1604.02191}}].

\bibitem{Becher:2006mr}
T.~Becher, M.~Neubert and B.~D. Pecjak, \emph{{Factorization and Momentum-Space
  Resummation in Deep-Inelastic Scattering}},
  \href{http://dx.doi.org/10.1088/1126-6708/2007/01/076}{\emph{JHEP} {\bf 01}
  (2007) 076}, [\href{http://arxiv.org/abs/hep-ph/0607228}{{\tt
  hep-ph/0607228}}].

\bibitem{Kelley:2011ng}
R.~Kelley, M.~D. Schwartz, R.~M. Schabinger and H.~X. Zhu, \emph{{The two-loop
  hemisphere soft function}},
  \href{http://dx.doi.org/10.1103/PhysRevD.84.045022}{\emph{Phys. Rev.} {\bf
  D84} (2011) 045022}, [\href{http://arxiv.org/abs/1105.3676}{{\tt
  1105.3676}}].

\bibitem{Monni:2011gb}
P.~F. Monni, T.~Gehrmann and G.~Luisoni, \emph{{Two-Loop Soft Corrections and
  Resummation of the Thrust Distribution in the Dijet Region}},
  \href{http://dx.doi.org/10.1007/JHEP08(2011)010}{\emph{JHEP} {\bf 08} (2011)
  010}, [\href{http://arxiv.org/abs/1105.4560}{{\tt 1105.4560}}].

\bibitem{futureWork}
C.~W. Bauer and P.~F. Monni \href{http://arxiv.org/abs/in preparation}{{\tt in
  preparation}}.

\bibitem{Dixon:2008gr}
L.~J. Dixon, L.~Magnea and G.~F. Sterman, \emph{{Universal structure of
  subleading infrared poles in gauge theory amplitudes}},
  \href{http://dx.doi.org/10.1088/1126-6708/2008/08/022}{\emph{JHEP} {\bf 08}
  (2008) 022}, [\href{http://arxiv.org/abs/0805.3515}{{\tt 0805.3515}}].

\bibitem{Korchemsky:1999kt}
G.~P. Korchemsky and G.~F. Sterman, \emph{{Power corrections to event shapes
  and factorization}},
  \href{http://dx.doi.org/10.1016/S0550-3213(99)00308-9}{\emph{Nucl. Phys.}
  {\bf B555} (1999) 335--351}, [\href{http://arxiv.org/abs/hep-ph/9902341}{{\tt
  hep-ph/9902341}}].

\bibitem{Fleming:2007qr}
S.~Fleming, A.~H. Hoang, S.~Mantry and I.~W. Stewart, \emph{{Jets from massive
  unstable particles: Top-mass determination}},
  \href{http://dx.doi.org/10.1103/PhysRevD.77.074010}{\emph{Phys. Rev.} {\bf
  D77} (2008) 074010}, [\href{http://arxiv.org/abs/hep-ph/0703207}{{\tt
  hep-ph/0703207}}].

\bibitem{Schwartz:2007ib}
M.~D. Schwartz, \emph{{Resummation and NLO matching of event shapes with
  effective field theory}},
  \href{http://dx.doi.org/10.1103/PhysRevD.77.014026}{\emph{Phys. Rev.} {\bf
  D77} (2008) 014026}, [\href{http://arxiv.org/abs/0709.2709}{{\tt
  0709.2709}}].

\bibitem{Bauer:2002ie}
C.~W. Bauer, A.~V. Manohar and M.~B. Wise, \emph{{Enhanced nonperturbative
  effects in jet distributions}},
  \href{http://dx.doi.org/10.1103/PhysRevLett.91.122001}{\emph{Phys. Rev.
  Lett.} {\bf 91} (2003) 122001},
  [\href{http://arxiv.org/abs/hep-ph/0212255}{{\tt hep-ph/0212255}}].

\bibitem{Bauer:2003di}
C.~W. Bauer, C.~Lee, A.~V. Manohar and M.~B. Wise, \emph{{Enhanced
  nonperturbative effects in Z decays to hadrons}},
  \href{http://dx.doi.org/10.1103/PhysRevD.70.034014}{\emph{Phys. Rev.} {\bf
  D70} (2004) 034014}, [\href{http://arxiv.org/abs/hep-ph/0309278}{{\tt
  hep-ph/0309278}}].

\bibitem{Becher:2006qw}
T.~Becher and M.~Neubert, \emph{{Toward a NNLO calculation of the anti-B --->
  X(s) gamma decay rate with a cut on photon energy. II. Two-loop result for
  the jet function}},
  \href{http://dx.doi.org/10.1016/j.physletb.2006.04.046}{\emph{Phys. Lett.}
  {\bf B637} (2006) 251--259}, [\href{http://arxiv.org/abs/hep-ph/0603140}{{\tt
  hep-ph/0603140}}].

\bibitem{Sterman:1986aj}
G.~F. Sterman, \emph{{Summation of Large Corrections to Short Distance Hadronic
  Cross-Sections}},
  \href{http://dx.doi.org/10.1016/0550-3213(87)90258-6}{\emph{Nucl. Phys.} {\bf
  B281} (1987) 310--364}.

\bibitem{Bauer:2002aj}
C.~W. Bauer, D.~Pirjol and I.~W. Stewart, \emph{{Factorization and endpoint
  singularities in heavy to light decays}},
  \href{http://dx.doi.org/10.1103/PhysRevD.67.071502}{\emph{Phys. Rev.} {\bf
  D67} (2003) 071502}, [\href{http://arxiv.org/abs/hep-ph/0211069}{{\tt
  hep-ph/0211069}}].

\bibitem{Manohar:2006nz}
A.~V. Manohar and I.~W. Stewart, \emph{{The Zero-Bin and Mode Factorization in
  Quantum Field Theory}},
  \href{http://dx.doi.org/10.1103/PhysRevD.76.074002}{\emph{Phys. Rev.} {\bf
  D76} (2007) 074002}, [\href{http://arxiv.org/abs/hep-ph/0605001}{{\tt
  hep-ph/0605001}}].

\bibitem{Bosch:2003fc}
S.~W. Bosch, R.~J. Hill, B.~O. Lange and M.~Neubert, \emph{{Factorization and
  Sudakov resummation in leptonic radiative B decay}},
  \href{http://dx.doi.org/10.1103/PhysRevD.67.094014}{\emph{Phys. Rev.} {\bf
  D67} (2003) 094014}, [\href{http://arxiv.org/abs/hep-ph/0301123}{{\tt
  hep-ph/0301123}}].

\bibitem{Bosch:2004th}
S.~W. Bosch, B.~O. Lange, M.~Neubert and G.~Paz, \emph{{Factorization and shape
  function effects in inclusive B meson decays}},
  \href{http://dx.doi.org/10.1016/j.nuclphysb.2004.07.041}{\emph{Nucl. Phys.}
  {\bf B699} (2004) 335--386}, [\href{http://arxiv.org/abs/hep-ph/0402094}{{\tt
  hep-ph/0402094}}].

\bibitem{Almeida:2014uva}
L.~G. Almeida, S.~D. Ellis, C.~Lee, G.~Sterman, I.~Sung and J.~R. Walsh,
  \emph{{Comparing and counting logs in direct and effective methods of QCD
  resummation}}, \href{http://dx.doi.org/10.1007/JHEP04(2014)174}{\emph{JHEP}
  {\bf 04} (2014) 174}, [\href{http://arxiv.org/abs/1401.4460}{{\tt
  1401.4460}}].

\bibitem{Li:2016axz}
Y.~Li, D.~Neill and H.~X. Zhu, \emph{{An Exponential Regulator for Rapidity
  Divergences}}, {\emph{Submitted to: Phys. Rev. D} (2016) },
  [\href{http://arxiv.org/abs/1604.00392}{{\tt 1604.00392}}].

\bibitem{Chiu:2007dg}
J.-y. Chiu, F.~Golf, R.~Kelley and A.~V. Manohar, \emph{{Electroweak
  Corrections in High Energy Processes using Effective Field Theory}},
  \href{http://dx.doi.org/10.1103/PhysRevD.77.053004}{\emph{Phys. Rev.} {\bf
  D77} (2008) 053004}, [\href{http://arxiv.org/abs/0712.0396}{{\tt
  0712.0396}}].

\bibitem{Becher:2011dz}
T.~Becher and G.~Bell, \emph{{Analytic Regularization in Soft-Collinear
  Effective Theory}},
  \href{http://dx.doi.org/10.1016/j.physletb.2012.05.016}{\emph{Phys. Lett.}
  {\bf B713} (2012) 41--46}, [\href{http://arxiv.org/abs/1112.3907}{{\tt
  1112.3907}}].

\bibitem{Chiu:2011qc}
J.-y. Chiu, A.~Jain, D.~Neill and I.~Z. Rothstein, \emph{{The Rapidity
  Renormalization Group}},
  \href{http://dx.doi.org/10.1103/PhysRevLett.108.151601}{\emph{Phys. Rev.
  Lett.} {\bf 108} (2012) 151601}, [\href{http://arxiv.org/abs/1104.0881}{{\tt
  1104.0881}}].

\bibitem{Chiu:2012ir}
J.-Y. Chiu, A.~Jain, D.~Neill and I.~Z. Rothstein, \emph{{A Formalism for the
  Systematic Treatment of Rapidity Logarithms in Quantum Field Theory}},
  \href{http://dx.doi.org/10.1007/JHEP05(2012)084}{\emph{JHEP} {\bf 05} (2012)
  084}, [\href{http://arxiv.org/abs/1202.0814}{{\tt 1202.0814}}].

\end{thebibliography}\endgroup

\end{document}